\documentclass[aps,onecolumn,floats,prd,nofootinbib,superscriptaddress,10pt,longbibliography]{revtex4-1}
\usepackage[utf8]{inputenc}
\usepackage{graphicx}
\usepackage{dcolumn}
\usepackage{bm}
\usepackage{lipsum}
\usepackage{xcolor}
\usepackage{calc}
\usepackage{accents}
\usepackage{comment}
\usepackage{float}
\usepackage{diagbox}
\usepackage{ulem}
\usepackage{hyperref}

\usepackage{soul}

\newcommand{\lcdm}{$\Lambda$CDM}
\makeatletter
\newcommand\footnoteref[1]{\protected@xdef\@thefnmark{\ref{#1}}\@footnotemark}
\makeatother
\newcommand{\ie}{{\it i.e.~}}

\newcommand{\Planck}{{\it Planck}}

\begin{document}

\title{The Ups and Downs of Early Dark Energy solutions to the Hubble tension: \\ a review of models, hints and constraints circa 2023}

\author{Vivian Poulin}
\affiliation{Laboratoire Univers \& Particules de Montpellier (LUPM), CNRS \& Universit\'{e} de Montpellier (UMR-5299), Place Eug\`{e}ne Bataillon, F-34095 Montpellier Cedex 05, France}

\author{Tristan L.~Smith}
\affiliation{Department of Physics and Astronomy, Swarthmore College, \\ 500 College Ave., Swarthmore, PA 19081, USA}
\affiliation{Center for Cosmology and Particle Physics, Department of Physics, New York University, New York, NY 10003, USA}
\author{Tanvi Karwal}
\affiliation{Center for Particle Cosmology, Department of Physics \& Astronomy, University of Pennsylvania, Philadelphia, PA 19104, USA}

\begin{abstract}We review the current status of Early Dark Energy (EDE) models proposed to resolve the ``Hubble tension'', the discrepancy between ``direct'' measurements of the current expansion rate of the Universe and ``indirect measurements'' for which the values inferred rely on the \lcdm{} cosmological model calibrated on early-universe data.
EDE refers to a new form of dark energy active at early times (typically a scalar-field), that quickly dilutes away at a redshift close to matter-radiation equality.  
The role of EDE is to decrease the sound horizon by briefly contributing to the Hubble rate in the pre-recombination era. 
We summarize the results of several analyses of EDE models suggested thus far in light of recent cosmological data, including constraints from the canonical \Planck\ data, baryonic acoustic oscillations and Type Ia supernovae, and the more recent hints driven by cosmic microwave background observations using the Atacama Cosmology Telescope. 
We also discuss potential challenges to EDE models, from theoretical ones (a second ``cosmic coincidence'' problem in particular) to observational ones, related to the amplitude of clustering on scales of $8h$/Mpc as measured by weak-lensing observables (the so-called $S_8$ tension) and the galaxy power spectrum from BOSS analyzed through the effective field theory of large-scale structure. 
We end by reviewing recent attempts at addressing these shortcomings of the EDE proposal. 
While current data remain inconclusive on the existence of an EDE phase, we stress that given the signatures of EDE models imprinted in the CMB and matter power spectra, next-generation experiments can firmly establish whether EDE is the mechanism responsible for the Hubble tension and distinguish between the various models suggested in the literature. 
\end{abstract}

\maketitle

\tableofcontents

\section{Introduction to the ``Hubble tension''}

Over the last two decades, our physical description of the Universe on the largest scales has made tremendous progress. The application of General Relativity to the Universe as a whole under the assumption of statistical homogeneity and isotropy has led to the establishment of a standard cosmological model known as the $\Lambda$ cold dark matter ($\Lambda$CDM) model. The $\Lambda$CDM model -- which includes a cosmological constant $\Lambda$ and cold dark matter (CDM), along with baryons, photons, and neutrinos -- has made predictions that can be tested with cosmological observables up to a very high degree of accuracy. This is especially true for our observations of the cosmic microwave background (CMB), primordial element abundances from big bang nucleosynthesis (BBN), baryon acoustic oscillations (BAO) and uncalibrated luminosity distances to Type Ia supernovae (SNIa). In less than 50 years, these measurements have  reached percent-level precision, allowing us to enter an ``era of precision cosmology'' \cite{Turner:2022gvw}. However, despite its many successes,  our cosmological model remains parametric and the natures of its dominant components -- dark matter and dark energy --, as well as the mechanism at the origin of fluctuations -- usually assumed to be inflation -- , are yet to be understood.

\par  As measurement precision improves, several tensions have emerged between probes of the early and late universe in recent years, possibly hinting at the underlying nature 
of  these components (for a recent review, see Ref.~\cite{Abdalla:2022yfr}).
Loosely speaking, the ``Hubble tension'' refers to the inconsistency between ``direct'' measurements of the current expansion rate of the Universe, \ie the Hubble constant $H_0$, made with a variety of probes in the late-universe and ``indirect measurements'' for which the values inferred rely on the \lcdm{} cosmological model calibrated on early universe data.
More precisely, it is now understood that the 
 most statistically significant tension arises between measurements which are calibrated using the so-called ``sound horizon'', imprints of acoustic waves propagating in the primordial plasma until recombination -- under the assumption of the \lcdm{} model-- and those that rely on different calibration methods\footnote{Note that some  high-accuracy ``direct" measurements that find high $H_0$ also rely on the assumption of \lcdm{} (e.g. the `H0LiCOW' strongly lensed quasars \cite{Wong:2019kwg}) but not on the early-universe calibration.}.
This tension is predominantly driven by the {\it Planck} collaboration's observation of the cosmic microwave background (CMB), which predicts a value in \lcdm{} of $H_0 = (67.27 \pm 0.60)$ km/s/Mpc \cite{Planck:2018vyg}, and the value measured by the SH0ES collaboration using the Cepheid-calibrated cosmic distance ladder, whose latest measurement yields $H_0 = 73.04\pm1.04$ km/s/Mpc \cite{Riess:2021jrx}. Taken at face value, these observations alone result in a $5\sigma$ tension.

Great efforts have been mounted to search for any systematic causes of the tension in either the direct or indirect measurements, or both \cite{Rigault:2014kaa,NearbySupernovaFactory:2018qkd,Addison:2017fdm,CSP:2018rag,Jones:2018vbn,Efstathiou:2020wxn,Brout:2020msh,Mortsell:2021nzg,Mortsell:2021tcx,Freedman:2021ahq,Garnavich:2022hef,Kenworthy:2022jdh,Riess:2022mme,Feeney:2017sgx,Breuval:2020trd,Javanmardi:2021viq,Wojtak:2022bct} (for a review, see Refs.~\cite{DiValentino:2021izs,Abdalla:2022yfr}). 
 Yet, the SH0ES team recently provided a comprehensive measurement of the $H_0$ parameter to 1.3\% precision, with attempts at addressing these potential systematic errors, and concluded that there is ``{\it  no indication that the discrepancy arises
from measurement uncertainties or [over 70] analysis variations considered to date}'' \cite{Riess:2021jrx}. 
Moreover, the appearance of this discrepancy across an array of probes suggests that a single systematic effect is unlikely to be sufficient to resolve it.  

\begin{figure}[b]
    \centering
    \includegraphics[scale=0.33]{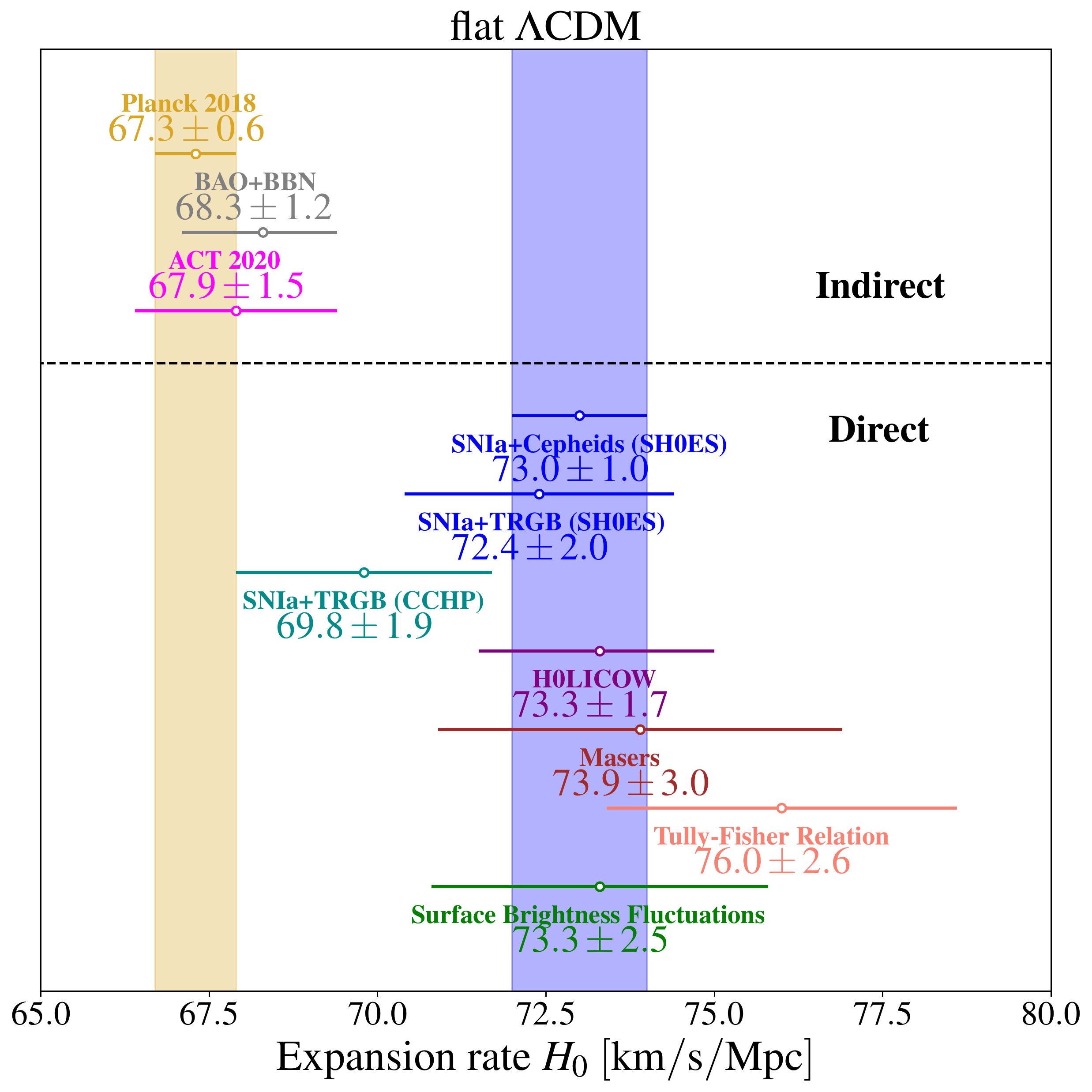}
    \caption{A summary of a representative selection of the indirect and direct measurements of the current expansion rate $H_0$, with errors below $3$ km/$s$/Mpc, based on Ref.~\cite{Verde:2019ivm}. We refer to Ref.~\cite{Abdalla:2022yfr} for a complete compilation.}
    \label{fig:compilations}
\end{figure}

Indeed, on the one hand there exists a variety of different techniques for calibrating $\Lambda$CDM at high-redshifts and subsequently inferring the value of $H_0$, which do not involve {\it Planck} data. For instance, one can use alternative CMB data sets such as WMAP, ACT, or SPT, or even remove observations of the CMB altogether and combine measurements of BBN  with data from BAO \cite{Addison:2017fdm,Schoneberg:2019wmt}, resulting in $H_0$ values in good agreement with {\it Planck}.
On the other hand, alternative methods for measuring the local expansion rate have been proposed in the literature, in an attempt to remove any biases introduced from Cepheid and/or SN1a observations. The Chicago-Carnegie Hubble program (CCHP), which calibrates SNIa using the tip of the red giant branch (TRGB), obtained a value of $H_0=(69.8 \pm 0.6~\mathrm{(stat)} \pm 1.6~\mathrm{(sys)})$ km/s/Mpc \cite{Freedman:2019jwv,Freedman:2021ahq}, in between the {\it Planck} CMB prediction and the SH0ES calibration measurement, and a re-analysis of the CCHP data by Anand et al. yields $H_0=71.5 \pm1.9$km/s/Mpc \cite{Anand:2021sum}. The SH0ES team, using the parallax measurement of $\omega-$Centauri from GAIA DR3 to calibrate the TRGB, obtained $H_0=(72.1 \pm2.0)$km/s/Mpc~\cite{Yuan:2019npk,Soltis:2020gpl}. Additional methods intended to calibrate SNIa at large distances include: surface brightness fluctuations of galaxies \cite{Khetan:2020hmh}, MIRAS \cite{Huang:2019yhh}, or the Baryonic Tully Fisher relation \cite{Schombert:2020pxm}. Bypassing the SN1a altogether and relying on Cepheids alone, the SH0ES team recently obtained $H_0=(73.1 \pm2.5)$km/s/Mpc \cite{Kenworthy:2022jdh}. 
There also exists a variety of observations which do not rely on observations of SNIa -- these include for e.g. time-delay of strongly-lensed quasars \cite{Wong:2019kwg,Birrer:2020tax,Shajib:2023uig}, maser distances \cite{Pesce:2020xfe}, cosmic chronometers \cite{Moresco:2022phi}, the age of old stars \cite{Jimenez:2019onw,Moresco:2022phi,Cimatti:2023gil} or gravitational waves as ``standard sirens'' \cite{Abbott:2019yzh}. 
A \footnote{Some notable measurements not represented here include the TDCOSMO results, which update the H0LiCOW results by constraining lens profiles from stellar kinematic data rather than assuming a specific parameterization.  The latest analysis including spatially-resolved stellar kinematics by Ref.~\cite{Shajib:2023uig} finds $H_0=77.1^{+7.3}_{-7.1}$ km/s/Mpc, while
earlier work that included data from a different sample of lenses to constrain the population of lens galaxies led to $H_0= 67.4^{+4.1}_{-3.2}$ km/s/Mpc \cite{Birrer:2020tax}. Since the TDCOSMO collaboration concludes that their later results corroborate the methodology of time-delay cosmography, we choose to represent the ``optimistic'' number in this summary, but note that there is a level of subjectivity in this decision. 
The choice of including this optimistic measurement is further supported by Ref.~\cite{Pandey:2019yic}, which disfavored the possibility that mass-modelling of strong lenses introduces an $O(10\%)$ bias in measurements of $H_0$. 
This conclusion relies on the excellent agreement found when comparing distances measured with strong-lensing time delays to distance-redshift relations from both calibrated and uncalibrated supernovae, implying that residual systematics in either measurement of $H_0$ cannot account for the current level of tension with the CMB determination. 
} 
summary of current determinations of the Hubble constant is provided in fig.~\ref{fig:compilations}, with errors less than $3$ km/s/Mpc, following Ref.~\cite{DiValentino:2021izs}. While not all measurements are in strong tension with {\it Planck}, these direct probes tend to yield values of $H_0$ systematically larger than the value inferred by {\it Planck}. 
Depending on how one chooses to combine the various measurements, the tension oscillates between the $4-6 \sigma$ level \cite{Abdalla:2022yfr,Riess:2021jrx,DiValentino:2021izs}.

Observational evidence for or against the Hubble tension is poised to change significantly within the next several years. The use of multi-messenger gravitational-wave measurements of binary systems (i.e., `standard sirens') is projected to reach 2\% error bars within the next five years \cite{Chen:2017rfc}, providing an independent check of the local value of $H_0$. 
We are also only a few years away from significant improvements in our measurements of galaxy clustering (e.g., from DESI \cite{DESI:2016fyo} and Euclid \cite{EUCLID:2011zbd}) and in our measurements of CMB anisotropies (e.g., from Simons Observatory \cite{SimonsObservatory:2018koc} and CMB Stage 4 \cite{CMB-S4:2016ple}). 
These upcoming measurements will be decisive in determining if physics beyond \lcdm\ is necessary to account for any discrepancy between the direct and indirect estimations of $H_0$. 
This relatively short timeline provides a strong impetus to develop a range of possible extensions to \lcdm\ which address the Hubble tension. 

\begin{figure}
    \centering
    \includegraphics[width=1\columnwidth]{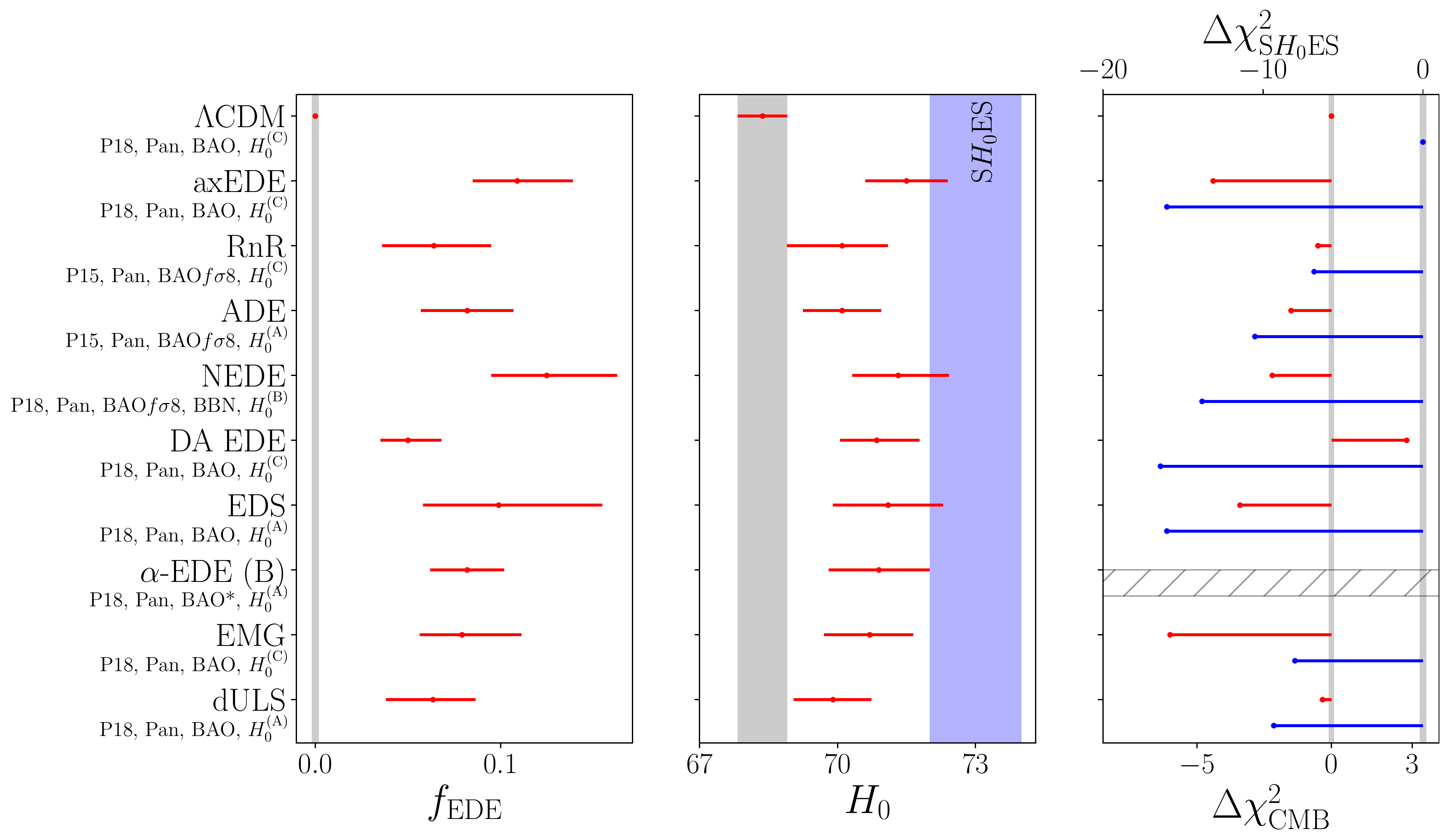}
    \caption{Comparison of the reconstructed 1D posteriors of $\{f_{\rm EDE}, H_0\}$ at 68\% C.L. in a representative sample of EDE models suggested in the literature, compared with $\Lambda$CDM.  
    Note that all models were not analyzed using the exact same datasets (see legends for details), but all analyses include (at least) {\it Planck} data (2015 or 2018), BOSS BAO DR12, Pantheon and SH0ES. 
    The label $H_0^{(A)}$ refers to the value from Ref.~\cite{Riess:2019cxk},  $H_0^{(B)}$ from Ref.~\cite{Riess:2020fzl} and  $H_0^{(C)}$ from Ref.~\cite{Riess:2021jrx}. In the rightmost plot, we show $\Delta \chi^2_{\rm CMB}$ in red and $\Delta \chi^2_{\rm SH_0ES}$ in blue. We do not have access to the  $\Delta \chi^2$ for the $\alpha$-EDE model.}
    \label{fig:EDE_whis_intro}
\end{figure}

In the following, we review attempts involving dark energy in the pre-recombination universe, loosely named ``Early Dark Energy'' (EDE). 
We show in Fig.~\ref{fig:EDE_whis_intro}, a summary of the reconstructed maximum fractional contribution of EDE  to the energy density of the universe taken from the literature, and the associated value of $H_0$, when the models are fit  to a combination of cosmological data involving \Planck{}, a compilation of BAO data, and SN1a measurements calibrated using the SH0ES value. One can see that this data combination consistently yields EDE contribution around $\sim 10\%$ and values of $H_0$ that are within $\sim 2\sigma$ of the SH0ES measurement\footnote{Note that one cannot directly compare the posterior of $H_0$ reconstructed in each model to that of SH0ES to assess the tension level, given that SH0ES is included in the analysis. This estimate of the tension level comes from computing a tension metric presented later.}. We also provide the $\Delta \chi^2\equiv\chi^2({\rm EDE})-\chi^2(\Lambda{\rm CDM})$ of \Planck\ and SH0ES data to gauge the relative success of the models in addressing the Hubble tension: compared to $\Lambda$CDM, most models can significantly improve the fit to SH0ES data, while providing a slightly better fit to \Planck\ data.
The rest of the review is organized as follows: after a brief introduction in Sec.~\ref{sec:IntroResolveTension} to the underlying general mechanism leading to high-$H_0$ in cosmology with EDE, we present the various models suggested in the literature in Sec.~\ref{sec:EDE_models}. We discuss in details the phenomenology of EDE within CMB and LSS data (at linear order in perturbations) in Sec.~\ref{sec:EDE_pheno}, and present the results of analyses of EDE models in light of \Planck\ data in Sec.~\ref{sec:EDE_Planck}. We then review in Sec.~\ref{sec:EDE_ACT_SPT} the recent hints of EDE found when analyzing ACT data, and the impact of SPT data on the preference over $\Lambda$CDM. Sec.~\ref{sec:EDE_challenges} is devoted to discussing challenges to EDE cosmologies, including how probes of matter clustering at late-times (galaxy weak-lensing and clustering in particular) can constrain the EDE solution to the Hubble tension. We eventually summarize our discussion in Sec.~\ref{sec:concl}. 
We mention that another review was recently produced on the same topic \cite{Kamionkowski:2022pkx}, which covers part of the material presented here. While \cite{Kamionkowski:2022pkx} presents in great detail the current status of the Hubble tension, the purpose of the present review is instead to dive deeper into details of the EDE phenomenology and the analysis results of the various models suggested in the literature, and therefore provide a complementary view on the very active topic of Early Dark Energy and the Hubble tension.


\section{General considerations for models that attempt to address the Hubble tension}
\label{sec:IntroResolveTension}

The parameter that drives constraints on $H_0$ from observations of physics in the early universe is the angular scale of the sound horizon $\theta_s(z)\equiv r_s(z_*)/D_A(z)$, where $r_s(z_*)$ is the comoving size of the sound horizon at the redshift $z_*$ at which the oscillating material (i.e., photons and baryons) decouples (i.e, $z_*$ refers to recombination or baryon drag), and $D_A(z)$ is the angular diameter distance to the observation (CMB or galaxy surveys):
\begin{eqnarray}
    r_s(z_*) &=&  \int_{z_*}^\infty \frac{c_s(z')}{H(z')}dz'\,, \label{eq:r_s}\\
    D_A(z) &=& \int_0^{z} \frac{1}{H(z')}dz'\,.\label{eq:dA}
\end{eqnarray}
Here $H(z) \equiv \dot a/a$ is the Hubble parameter, the sound speed of acoustic oscillations in the tightly coupled photon-baryon fluid is given by $c_s(z) \equiv c[3(1+R)]^{-1/2}$, $R \equiv 3\rho_b/(4 \rho_{\gamma})$, $\rho_X$ is the energy density in baryons ($X=b$) and photons ($X=\gamma$). It is in this way that $r_s$ provides a standardizable ruler: it is a fixed length scale which, when projected on our sky, can be used to measure distances ($D_A(z)$) in the Universe.

Resolutions to the Hubble tension can be split between {\it late}- and {\it early}-universe solutions\footnote{We note that it has also been suggested that a ``local'' modification may play a role in the Hubble tension (e.g. a void \cite{Huterer:2023ldv}, or a screened fifth force \cite{Desmond:2019ygn}).}. 
Late-universe solutions modify the $z \lesssim 2$ expansion history to raise the value of $H_0$ without changing the value of the angular diameter distance to last-scattering $D_A(z_*)$. As a result, these theories predict the same value of $r_s(z_*)$ as in  \lcdm. 
A strong constraint on these models comes from the fact that the SH0ES determination of $H_0$ is not a determination of the expansion rate at $z=0$. Instead, it is an absolute calibration of SNIa, which can be used to turn the sample of SN1a measured at $0.001 \leq z \leq 2.26$ (Pantheon+ \cite{Scolnic:2021amr}) into a  measurement of the luminosity distance (and therefore the expansion rate) between these redshifts. 
This constrains large deviations away from $\Lambda$CDM across these redshifts which would be necessary to explain the tension. 
In addition to this, the assumption of \lcdm\ in the early universe allows us to infer a value of $r_s(z_d)$ from observations of the CMB, calibrating measurements of the BAO which also extend to $z\simeq 2$ \cite{eBOSS:2020yzd}, and provide an independent measurements of the angular diameter distance (and therefore also of the expansion rate). 
A remarkable result is that, if one uses the measurements from SH0ES to calibrate SN1a on the one hand, and the value of $r_s(z_d)$ predicted in \lcdm\ to calibrate BAO on the other hand, the reconstructed angular/luminosity distances are in tension with one another \cite{Bernal:2016gxb,Poulin:2018zxs,Camarena:2021jlr,Efstathiou:2021ocp,Pogosian:2021mcs}. 
As a result, measurements of $H(z)$ along with $\theta_s(z)$ significantly constrain late-universe solutions\footnote{It has been suggested that the Pantheon+ sample and strongly lensed QSOs may indicate that at high redshifts, the determination of $\Omega_m$ deviates from $\Lambda$CDM \cite{Krishnan:2020obg,Dainotti:2021pqg,Dainotti:2022bzg,Krishnan:2021dyb,Malekjani:2023dky}, although these deviations are still under the $2\sigma$ level \cite{Brout:2022vxf}.} \cite{Poulin:2018zxs, Benevento:2020fev,Efstathiou:2021ocp}.

\begin{figure}[t]
    \centering
    \includegraphics[scale=0.66]{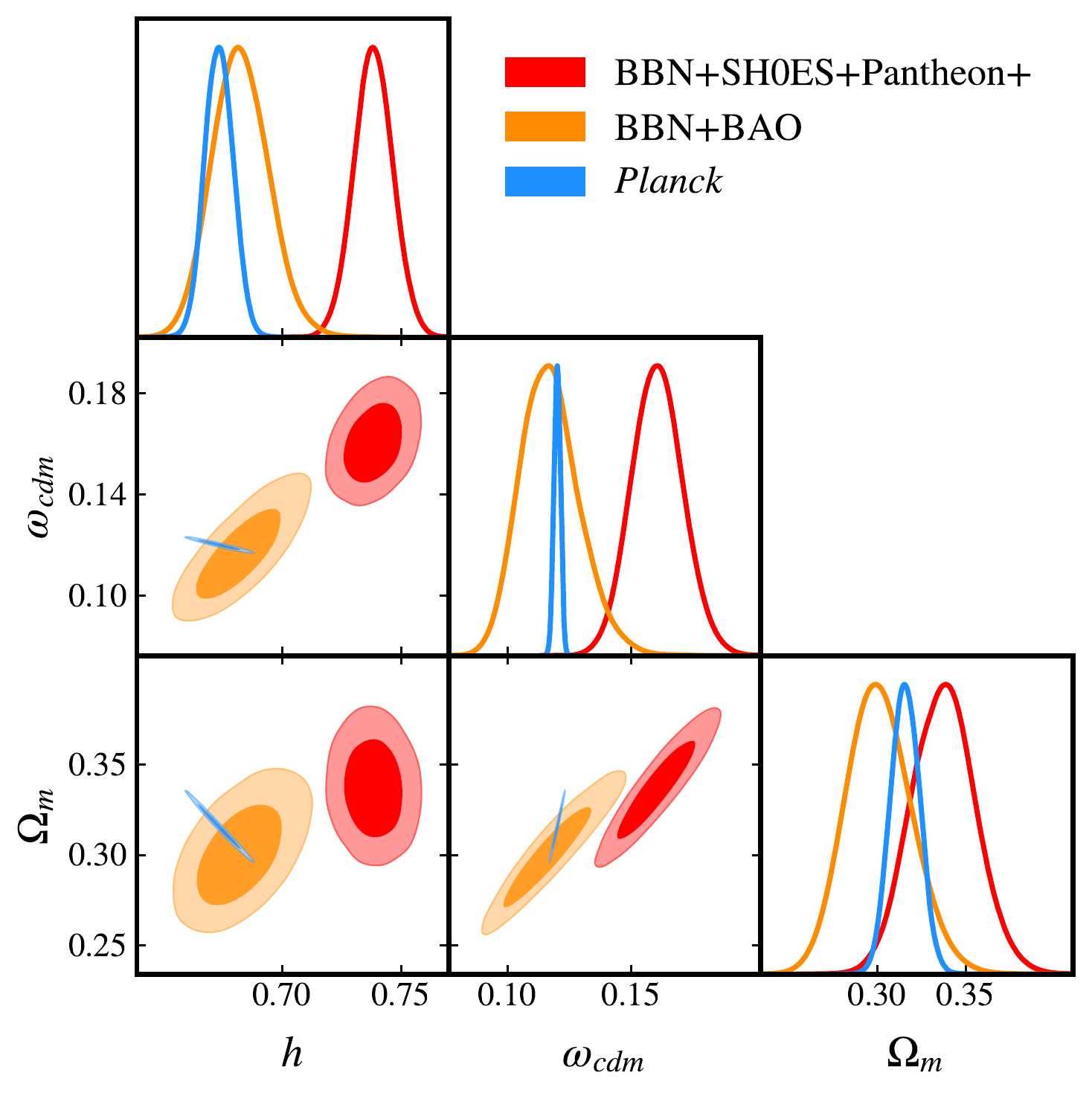}
    \caption{A triangle plot showing constraints (within $\Lambda$CDM) to the Hubble constant, $h$, the physical cold dark matter density, $\omega_{cdm}$, and the total matter density in units of the critical energy density, $\Omega_m$ using a variety of data sets.}
    \label{fig:late-time-constraints}
\end{figure}

 We can visualize how measurements of Type Ia supernovae and BAO give consistent expansion histories but different calibrations in Fig.~\ref{fig:late-time-constraints}. In that Figure, the red contours show constraints using SH0ES, Pantheon+ \cite{Brout:2020msh}, and a prior on $\omega_b$ from big bang nucleosynthesis, $\omega_b = 0.0226 \pm 0.00034$; the orange contours show constraints using a variety of BAO measurements \cite{Beutler:2011hx,Alam:2016hwk,Blomqvist:2019rah} calibrated within $\Lambda$CDM and the same prior on $\omega_b$; the blue contours show constraints using \textit{Planck}. Within $\Lambda$CDM, both the uncalibrated supernovae and BAO measurements are sensitive to the dimensionless Hubble parameter, $H(z)/H_0 = \sqrt{\Omega_m (1+z)^3 + (1-\Omega_m)}$, and therefore provide constraints to $\Omega_m$. Fig.~\ref{fig:late-time-constraints} shows that both SH0ES+Pantheon+ and BAO measurements place consistent constraints on $\Omega_m$. On the other hand, once either data set is calibrated (using Cepheid variables for the SNeIa and the comoving sound horizon in $\Lambda$CDM for the BAO) we gain information on dimensionful quantities such as $h$ and $\omega_{\rm cdm}$. This figure makes it clear that the Hubble tension is in fact a tension in both the Hubble constant \emph{and} $\omega_{cdm}$ and that it is present in data that does not include measurements of the CMB. This fact places significant pressure on any attempt to build a model which just modifies late-time dynamics to address the Hubble tension. 

Early-universe solutions reduce the size of the sound horizon (Eq. (\ref{eq:r_s})) and change the CMB/BAO calibrator \cite{Bernal:2016gxb,Aylor:2018drw}, resulting in a larger range of allowed dynamics. Resolutions using EDE decrease the size of the sound horizon by proposing the existence of additional pre-recombination energy density which, in turn, increases the Hubble parameter around the time of photon decoupling. 
Since $H(z)$ is a decreasing function, both integrals in Eqs.~(\ref{eq:r_s}) and (\ref{eq:dA}) are dominated by their behaviors at their lower limits, so we can roughly write 
\begin{equation}
\theta_s(z) = \frac{r_s(z_*)}{D_A(z)} \sim H_0 \frac{c_s(z_*)}{H(z_*)}F(\Omega_m;z)~~ \Rightarrow ~~   H_0 \sim \theta_s(z)\frac{H(z_*)}{c_s(z_*)}F^{-1}(\Omega_m;z)\label{eq:thetas}
\end{equation}
where $F(\Omega_m;z)$ asymptotes at high redshift to a redshift-independent function, $F(\Omega_m;z\gg1) \simeq [2.8 (1/\Omega_m -1)^{1/3}-0.95]/\sqrt{1-\Omega_m}$. 

The approximate expression on the right side of Eq.~(\ref{eq:thetas}) helps us to quickly understand how EDE can address the Hubble tension. Measurements of the late-time expansion history (such as through SNeIa) provide tight constraints on $\Omega_m$, and atomic physics (plus the CMB temperature today) gives tight constraints on $z_*$. 
In addition, assuming standard baryon/photon interactions, $c_s(z_*)$ is fixed from strong constraints on the baryon density obtained either through the CMB power spectra or through measurements of the abundances of light elements formed during BBN. 
Given that the CMB or BAO provide a precise measurement of $\theta_s$, the role of EDE is to enhance $H(z_*)$ so as to increase the inferred value of $H_0$ \cite{Bernal:2016gxb,Addison:2017fdm,Aylor:2018drw,Schoneberg:2019wmt, Evslin:2017qdn}.  
We note that modifications to the physics of recombination can increase the indirect $H_0$ by increasing $z_*$ (see, e.g., Refs.~\cite{Jedamzik:2020krr,Hart:2019dxi,Hart:2021kad,Sekiguchi:2020teg,Schoneberg:2021qvd}) but a discussion of these models is beyond the scope of this review. 

It is important to note that the larger value of $H_0$ from direct measurements predicts a larger physical matter density than is inferred from the CMB assuming \lcdm. In a flat \lcdm\ Universe, measurements of the expansion history from SNIa, BAO, and CMB provide a relatively tight constraint on $\Omega_m \equiv \rho_m/\rho_{\rm crit}$, which, when combined with the SH0ES determination of $H_0$ gives a physical matter density $\omega_m  \equiv \Omega_m h^2 = 0.1603 \pm 0.0063$, which is about $3\sigma$ larger than the value inferred from \textit{Planck} assuming \lcdm: $\omega_m = 0.14228^{+0.00079}_{-0.00091}$ \cite{Planck:2018vyg}\footnote{where $\rho_{\rm crit} \equiv 3 H_0^2/(8 \pi G)$ and $h \equiv H_0/(100\ {\rm km/s/Mpc})$}. 
Remarkably, an increase in $H(z>z_*)$ naturally leads to an increase in $\omega_m$. Indeed, at the background level, the increase in $H(z)$ leads to an increase in Hubble friction which slows the growth of the CDM density contrast.  Within the EDE model, this is compensated for by increasing the DM density, leading to a larger $\omega_m$.

An increase in $H(z_*)$ also leads to increased damping of the small scales of the CMB. When $H(z_*)$ increases, the angular scale of the sound horizon, $\theta_s(z_*)$, is kept fixed by increasing $H_0$. 
The angular scale $\theta_D$ corresponding to damping roughly depends on 
\begin{equation}
    \theta_D(z_*) \sim \frac{H_0}{\sqrt{\dot \tau(z_*) H(z_*)}},
\end{equation}
where $ \dot \tau(z_*)=n_e(z_*) x_e(z_*) \sigma_T/(1+z_*)$, $n_e$ is the electron density, $x_e$ is the ionization fraction, and $\sigma_T$ is the Thomson cross section. If the indirect $H_0$ increases by a fraction $f$, with $\theta_s(z_*)$ fixed, $\theta_D(z_*)$ increases by $f^{1/2}$. 
A larger $\theta_D(z_*)$ leads to greater suppression of power which must be compensated for by an increases in $n_s$ and $\omega_m$. This may have interesting consequences for models of cosmic inflation (see, e.g., Ref.~\cite{Ye:2021nej}) as we discuss later on. 
With these general properties of early-universe solutions, we next detail several models of EDE proposed in the literature.

\section{Overview of Early Dark Energy models}
\label{sec:EDE_models}

Broadly speaking, Early Dark Energy is any component of the Universe which was dynamically relevant at $z \gg 1$ and has equation of state $w\simeq -1$ at some point in its evolution. 
Typically (but not only), it takes the form of a cosmological scalar field which is initially frozen in its potential by Hubble friction.  
The possibility of the presence of dark energy before last-scattering has been studied for over a decade \cite{Doran:2000jt,Wetterich:2004pv,Doran:2006kp}, 
but these models gained acute attention in the context of the Hubble tension, starting with the work of Refs.~\cite{Karwal:2016vyq,Poulin:2018cxd}, followed by several variations on the underlying scalar field model by numerous authors \cite{Smith:2019ihp,Agrawal:2019lmo,Lin:2019qug,Alexander:2019rsc,Sakstein:2019fmf,Das:2020wfe,Niedermann:2019olb,Niedermann:2020dwg,Niedermann:2021vgd,Ye:2020btb,Berghaus:2019cls,Freese:2021rjq,Braglia:2020bym,Sabla:2021nfy,Sabla:2022xzj,Gomez-Valent:2021cbe,Moss:2021obd,Guendelman:2022cop,Karwal:2021vpk,McDonough:2021pdg,Wang:2022nap,Alexander:2022own,McDonough:2022pku,Nakagawa:2022knn,Gomez-Valent:2022bku,MohseniSadjadi:2022pfz,Kojima:2022fgo,Rudelius:2022gyu,Oikonomou:2020qah,
Tian:2021omz,Maziashvili:2021mbm}.

EDE models typically postulate the existence of a scalar field whose background dynamics are driven by the homogeneous Klein-Gordon (KG) equation of motion
\begin{equation}
   \ddot{\phi} + 3 H \dot{\phi} + \frac{dV(\phi)}{d\phi} = 0 \,,
\end{equation}
where dots refer to derivatives with respect to cosmic time.
For most EDE models, the dynamics can be summarized as follows: the field is frozen in its potential, such that the background energy density is constant. 
The fractional contribution of the field to the total energy density, $f_{\rm EDE}(z)\equiv \rho_{\rm EDE}(z)/\rho_{\rm tot}(z)$, therefore increases over time, in a manner similar to that of DE. 
Eventually, some mechanism (e.g., Hubble friction dropping below a critical value, or a phase transition changing the shape of the potential) releases the scalar, at which point the field becomes dynamical and the background energy density dilutes away faster than matter. 
The contribution of EDE to the Hubble rate is hence localized in redshift, typically within the decade before recombination, and acts to reduce the size of the sound horizon as previously described. A successful solution usually requires a fractional contribution $f_{\rm EDE}\sim 10\%$ that reaches a maximum at $z\sim10^3-10^4$ and subsequently dilutes at a rate greater than or equal to that of radiation.
The various proposed models can (broadly speaking) be characterised by (1) the shape of the scalar field potential, (2) the mechanism through which they become dynamical, 
and (3) whether or not the scalar field is minimally coupled. 

With these characteristics, we describe several models in the literature, briefly discuss their theoretical motivation and delineate their parameters.

\subsection{Axion-like Early Dark Energy}

\textbf{Model:} The axion-like Early Dark Energy (axEDE) proposes a potential of the form
\begin{equation}\label{eq:potential}
    V(\theta) = m^2 f^2[1-\cos (\theta)]^n,
\end{equation} where $m$ represents the axion mass, $f$ the axion decay constant, and $\theta \equiv \phi/f$ is a re-normalized field variable defined such that $-\pi \leq \theta \leq \pi$. The background field is held fixed at some initial value until $H\sim \partial^2_\theta V(\theta)$,  at which point the field becomes dynamical and oscillates about the minimum of its potential. 
A more detailed description of this model can be found in Refs.~\cite{Poulin:2018cxd,Smith:2019ihp}. 
It was implemented in a publicly available modified version of the \texttt{CLASS} code \cite{Blas:2011rf}\footnote{\url{https://github.com/PoulinV/AxiCLASS}}.

\textbf{Motivation:} Rather than being tied to a fundamental theory, this toy potential was introduced to provide a flexible EDE model with a rate of dilution set by the value of the exponent $n$.  
Taken at face-value, this potential may be generated by higher-order instanton corrections \cite{Abe:2014xja,Choi:2015aem,Kappl:2015esy} that require fine-tuning to cancel the leading contributions \cite{Rudelius:2022gyu,McDonough:2022pku}, or through the interaction of an axion and a dilaton \cite{Alexander:2019rsc} and is therefore non-generic \cite{Kaloper:2019lpl}. 
Related constructions of this potential in the context of `Natural Inflation' were also proposed in Refs.~\cite{Czerny:2014qqa,Croon:2014dma,Higaki:2015kta}. Interestingly, recent works have paved the way to embed this potential in a string-theory framework \cite{McDonough:2022pku,Cicoli:2023qri}, showing that the theory challenges for building an EDE with such a potential are not particularly different from that of other cosmological models, such as inflation, dark energy, or fuzzy dark matter \cite{Cicoli:2023qri}. 
The general behavior of this EDE component is well-described by cycle-averaging the evolution of the background and perturbative field dynamics \cite{Poulin:2018dzj,Poulin:2018cxd}, as we describe in more detail in sec.~\ref{sec:EDE_pheno}.

\textbf{Parameters:} It is now conventional to trade the `theory parameters', $m$ and $f$, for `phenomenological parameters', namely the critical redshift $z_c$ at which the field becomes dynamical and the fractional energy density $f_{\rm EDE}(z_c)$ contributed by the field at the critical redshift.  Additionally, the parameter $\theta_i$ controls the effective sound speed $c_s^2$ and thus the dynamics of the perturbations (mostly). Moreover, it is assumed that the field always starts in slow-roll (as enforced by the very high value of the Hubble rate at early times), and without loss of generality one can restrict $0\leq \theta_i \leq \pi$. Note that running MCMCs on the theory parameters can change the effective priors for the phenomenological parameters, and impacts constraints \cite{Hill:2020osr}.
 
\subsection{Rock 'n' Roll Early Dark Energy}

{\bf Model:} The Rock 'n' Roll Early Dark Energy model \cite{Agrawal:2019lmo} (RnR EDE) refers to a rolling scalar field with a potential of the form
\begin{eqnarray}
V(\phi)=V_0\bigg(\frac{\phi}{M_{\rm pl}}\bigg)^{2n}+V_\Lambda\,,
\end{eqnarray}
where $V_\Lambda$ is a constant. The behavior of this model is very similar to the axion-like early dark energy model, as the $(1-\cos(\phi/f))^n$ term can be approximated by a $(\phi/f)^{2n}$ in the small $\phi/f$ limit. This EDE field also begins in slow-roll and eventually rolls down and oscillates in its potential with a time-averaged constant equation of state $w=(n-1)/(n+1)$ when the Hubble parameter drops below the effective mass $\partial_\phi^2 V(\phi)$.

{\bf Motivation:} Such potentials naturally arise from the requirement of a constant equation of state, and have been studied in the context of axiverse-motivated models. However, they too suffer a similar tuning issue as the axion-like EDE model, as in realistic UV completions, terms with exponents $k<n$ are generated and impact the dynamics, and must therefore be made sub-dominant through fine-tuned cancellations.
 
{\bf Parameters:} Similar to the axion-like EDE, one can trade the two theory parameters, the initial field value $\phi_i$ and $V_0$, for phenomenological parameters $f_{\rm EDE}(z_c)$ and $1+z_c$. The power-law exponent $n$ can be fixed or be free to vary, with data favoring $n\simeq 2$. We again note that RnR EDE is identical to axEDE in the limit $\theta_i \ll 1$.

\subsection{Acoustic Early Dark Energy}

\textbf{Model:} Acoustic Early Dark Energy (ADE) is a phenomenological fluid description of EDE suggested in Ref.~\cite{Lin:2019qug}. It is characterized by its background equation of state $w_{\rm ADE} = p_{\rm ADE} / \rho_{\rm ADE}$ and rest-frame sound speed $c_s^2$, which in general may be different from the adiabatic sound speed $\dot p_{\rm ADE}/\dot \rho_{\rm ADE}$. The ADE equation of state is parameterized as 
\begin{equation}
    1+w_{\rm ADE}(a)=\frac{1+w_f}{[1+(a_c/a)^{3(1+w_f)/p}]^p}\,,
\end{equation}
where $a_c$ controls the scale factor at which the equation of state transitions from $w_{\rm ADE}=-1$ to $w_{\rm ADE}=w_f$ and $p$ controls the rapidity of the transition. The case $p=1$ approximates the behavior of the axion-like EDE model.

\textbf{Motivation:} Scalar-field EDE models with oscillating potentials require a level of fine-tuning that is unsatisfactory from a theoretical perspective. Additionally, they represent somewhat ad-hoc choices of modelling. 
This phenomenological approach generalizes the modeling (at least at the background level) attempting to optimize the success of the EDE model. 
It also has the advantage of being realized in the K-essence class of dark energy models \cite{Armendariz-Picon:2000nqq}, where the dark component is a perfect fluid represented by a minimally-coupled scalar field $\phi$ with a general kinetic term and a constant sound speed $c_s^2$, as given by the Lagrangian density
\begin{equation}
    P(X,\phi)=\bigg(\frac{X}{A}\bigg)^{\frac{1-c_s^2}{2c_s^2}}X-V(\phi)\,,
\end{equation}
where $X=-\nabla^2\phi/2$ and $A$ is a constant density scale. In this category of models, $w_{\rm ADE}\to c_s^2$ if the kinetic term dominates, whereas $w_{\rm ADE}\to-1$ if the potential $V(\phi)$ dominates.

\textbf{Parameters:} In this model, the parameters are: the ADE maximum fractional energy density contribution $f_{\rm EDE}(a_c)$; the critical scale factor $a_c$ at which the maximum is reached, similar to other EDE models; the equation of state at late-times $w_f$; the sound speed $c_s^2$; and the exponent $p$ which quantifies the rapidity of the transition. 
In practice, the number of parameters is reduced by setting $p=1/2$, and data seem to favor $w_f=c_s^2$. The case of a canonical scalar field corresponds to setting $w_f=c_s^2=1$. While data have a slight preference for $c_s^2=w_f=0.8$, the case $w_f=c_s^2=1$ is perfectly compatible, and in the following we will quote results for this minimal model.

\subsection{ New Early Dark Energy}

\textbf{Model:} The new Early Dark Energy (NEDE) model introduced in Refs.~\cite{Niedermann:2019olb,Niedermann:2020dwg} proposes two cosmological scalar fields, the NEDE field $\psi$ of mass $M$ and the `trigger' (sub-dominant) field $\phi$ of mass $m$, whose potential is written as (with canonically normalized kinetic terms):
\begin{equation}
    V(\psi,\phi)=\frac{\lambda}{4}\psi^4+\frac{1}{2}\beta M^2\psi^2-\frac{1}{3}\alpha M\psi^3+\frac{1}{2}m^2\phi^2+\frac{1}{2}\gamma\phi^2\psi^2.
\end{equation}
where $\lambda$, $\beta$, $\alpha$, $\gamma$ are dimensionless couplings. When $H\lesssim m$, $\phi$ rolls down its potential, eventually dropping below a threshold value for which the field configuration with $\psi=0$ becomes unstable. 
At this point, a quantum tunneling to a true vacuum occurs and the energy density contained in the NEDE field rapidly dilutes.
A modified \texttt{CLASS} version presented in Ref.~\cite{Niedermann:2020dwg} is publicly available\footnote{ \url{https://github.com/flo1984/TriggerCLASS}}. 

\textbf{Motivation:}  The NEDE proposal in a first-order phase-transition that can naturally be triggered in two-scalar-field models, and may therefore be more theoretically appealing than the original axEDE model.  The trigger field has two functions: at the background level,
it ensures that the percolation phase is short on cosmological time scales, preventing the emergence of large anisotropies. 
At the perturbation level, adiabatic perturbations of the trigger field set the initial conditions for the decaying NEDE fluid.
In addition, recently Ref.~\cite{Niedermann:2021ijp} suggested a modification to the model, dubbed `hot' NEDE, where the temperature of a (subdominant) dark sector radiation fluid plays the role of the trigger, thereby removing the need for an additional scalar field. In addition, NEDE naturally introduces interactions between the dark sector radiation fluid and NEDE or DM (with potential implications for the $S_8$ tension), and can connect the phase-transition to the origin of neutrino masses. However a test of the latter model against data has yet to appear in the literature.

\textbf{Parameters:} The NEDE model is specified by the fraction of NEDE before the decay $f_{\rm NEDE}(z_*) \equiv \bar\rho_{\rm NEDE}(z_*)/\bar\rho_{\rm tot}(z_*)$  (where $z_*$ is given by the redshift at which $H=0.2 m$), the mass $m_\phi$ of the trigger field\footnote{In the following, we will use the simpler notation $m_{\rm NEDE}\equiv m_\phi\times$Mpc. } which controls the redshift $z_*$ of the decay, and the equation of state $w_{\rm NEDE}$ after the decay. In Ref.~\cite{Niedermann:2020dwg}, the effective sound speed $c_s^2$ in the NEDE fluid is set equal to the equation of state after the decay, i.e. $c_s^2=w_{\rm NEDE}$.

\subsection{Dissipative Axion Early Dark Energy}

\textbf{Model: }
Thermal friction is a mechanism that gives rise to warm inflation, a class of inflationary models in which the Universe maintains a finite temperature through continuous particle production, but has also been applied to EDE in Refs.~\cite{Berghaus:2019cls, Berghaus:2022cwf}. 
In this scenario a scalar field $\phi$ couples to light degrees of freedom that self-thermalize and make up dark radiation.
It experiences friction $\Upsilon$ through this interaction, which acts to extract energy density from the scalar field into the dark radiation bath. 
This is similar to a drag force acting on an object moving through a fluid - the object slows and loses kinetic energy while the fluid heats up and gains thermal energy. 
The system evolves as 
\begin{eqnarray}
    \ddot{\phi} + (3H + \Upsilon)\dot{\phi} + V_\phi = 0 \\
    \dot{\rho}_{dr} + 4H\rho_{dr} = \Upsilon \dot{\phi}^2 \,.
\end{eqnarray}
The dark radiation has strong self-interactions and rapidly thermalizes. 
The cumulative effect of this coupled system resembles EDE with the frozen scalar field making up the pre-$z_c$ segment, and the dark radiation redshifting as $(1+z)^4$ making up the post-$z_c$ segment. 

\textbf{Motivation: }
This model was proposed to resolve the fine-tuning of the scalar potential of EDE.
Under thermal friction EDE (or dissipative axion EDE as it is called in \cite{Berghaus:2019cls,Berghaus:2022cwf}), the impact and constraints of the model are completely independent of the chosen scalar potential. 
An additional benefit is that dark radiation in the form of extra relativistic degrees of freedom can also alleviate the LSS tension, making it desirable to incorporate into an EDE model. 
A very similar mechanism was also studied in Ref.~\cite{Gonzalez:2020fdy}. 

\textbf{Parameters: }
The parameters of the model are the initial position $\phi_i$ of the scalar, the friction $\Upsilon$ that couples it to dark radiation and a parameter controlling the potential, assumed to be the mass $m$ of the scalar in \cite{Berghaus:2019cls,Berghaus:2022cwf} which employed a quadratic potential. 
Posteriors of this model favour the region of parameter space in which DA EDE asymptotes to an ever-present dark radiation with very high $z_c$, instead of an injection close to matter-radiation equality associated with most EDE models. 
Although DA EDE improves the fit to data over simply varying $N_{\rm eff}$ due to additional degrees of freedom, the improvement in total $\chi^2$ under this model stems exclusively from fitting a higher $H_0$. 
Hence, while this model attempts to address important challenges to EDE, ultimately, it is disfavoured by data - DA EDE cannot increase the Hubble parameter while maintaining the excellent fit to the CMB provided by \lcdm. 

\subsection{Early Dark Energy Coupled to Dark Matter}

\textbf{Model: }
One can introduce a conformal coupling $A(\phi)$ between an EDE scalar field and dark matter \cite{Karwal:2021vpk,McDonough:2021pdg}. 
The impact of the coupling is a modification of the potential $V(\phi)$ of the scalar to an effective potential $V_{\rm eff}$ that includes dark matter contributions
\begin{eqnarray}
    V_{\rm eff}(\phi, a) = V(\phi) + \rho_{\rm DM}(a) \,,
\end{eqnarray}
and a modulation of the particle mass of dark matter
\begin{eqnarray}
    m_{\rm DM}(\phi) = m_0 A(\phi) \,,
    \label{eq:EDS_dm_mass}
\end{eqnarray}
where $m_0$ is its mass at $z=0$, and Eq.~\ref{eq:EDS_dm_mass} modifies the dark matter energy density $\rho_{\rm DM}$. 
Specifically for the choice of an exponential coupling $A(\phi) = e^{c_\theta \phi / M_{\rm pl}}$ between EDE and dark matter, this phenomenology can arise from considering the predictions of the Swampland distance conjecture. 
In this case, axion dark matter is sensitive to super-Planckian field excursions of the EDE scalar field through an exponential coupling, leading to an `early dark sector' (EDS) \cite{McDonough:2021pdg}. 
 This is the simplest coupling in the chameleon scenario, but other couplings can arise from chameleon models, as usually postulated for late-time dark energy \cite{Khoury:2003aq, Khoury:2003rn, Karwal:2021vpk}.

\textbf{Motivation: }
While EDE offers a flexible solution to the Hubble tension, it also raises certain fundamental questions, including a new `why then' problem relating to the injection redshift of EDE, akin to the `why now' problem of late-time dark energy.
In these models, the dynamics of EDE are triggered by the onset of matter domination, when dark matter becomes the dominant energy component in the Universe. 
Another benefit of such a coupling is that it modifies the evolution of dark matter perturbations, with possible implications for the emergent tension in measurements of the growth of structure.

\textbf{Parameters: }
Within this scenario, various forms of the conformal coupling and the potential of the scalar field may be explored. 
EDS \cite{McDonough:2021pdg,Lin:2022phm} explores an exponential coupling and the original axEDE potential \cite{Poulin:2018dzj,Poulin:2018cxd,Smith:2019ihp}. 
The parameters of this model are the three axEDE parameters with the addition of a coupling constant $c_\theta$.

\subsection{$\alpha$-attractors EDE}

\textbf{Model: } The framework of $\alpha$-attractors \cite{Kallosh:2013hoa,Kallosh:2013yoa,Galante:2014ifa} corresponds to an EDE scalar field in which the potential is given by 
\begin{equation}
\label{eq:alphaEDE}
    V(\phi)=\Lambda+V_0\frac{(1+\beta)^{2n}\tanh(\phi/\sqrt{6\alpha}M_{\rm pl})^{2p}}{[1+\beta\tanh(\phi/\sqrt{6\alpha}M_{\rm pl})]^{2n}}\,,
\end{equation}
where $V_0,p,n,\alpha,\beta$ are constants. The normalization factor $(1+\beta)^{2n}$ ensures the same normalization of the plateau at large $\phi$ regardless of the choice of $(p,n)$.  The role of $(p,n)$ is to give flexibility to the form of the potential to reproduce various shapes of the energy injection. The authors of Ref.~\cite{Braglia:2020bym} have studied three specific choices $(p, n) =\{(2, 0), (2, 4), (4, 2)\}$, chosen to imitate the dynamics of the R'n'R, axion-like EDE and cADE models respectively. 

\textbf{Motivation:} Such potentials can naturally arise through a field re-definition that turns a non-canonical kinetic pole-like term into a canonical one \cite{Braglia:2020bym}. 
These $\alpha$-attractor models have been studied in the context of inflation, and lead to predictions for the spectral index $n_s$ and tensor-to-scalar ratio $r$, that are largely independent of the specific functional form of V$(\phi)$, leading to the name of ``attractors''. 
They have also been studied in the context of dark energy \cite{Linder:2015qxa,Garcia-Garcia:2018hlc,LinaresCedeno:2019bgo}, and invoked to connect dark energy and inflation \cite{Dimopoulos:2017zvq,Akrami:2017cir} naturally leading to their application for EDEs as well\footnote{Recently, an alternative, exponential potential was suggested in Ref.~\cite{Brissenden:2023yko}, arguably simpler than the one presented in Eq.~\ref{eq:alphaEDE}, introduced to connect early and late dark energies for suitable choices of parameters in the context of $\alpha$-attractors (with model extensions that can connect with inflation). A dedicated analysis in light of cosmological data is still lacking, though arguments at the background-level suggest that the parameter space might be viable.}.

\textbf{Parameters: } 
Besides $(p,n)$ and $\beta$ which are fixed, the models has three free parameters which are chosen to be the usual maximum fraction of EDE $f_{\rm EDE}(z_c)$, the critical redshift $z_c$ and the initial field value $\theta_i\equiv \phi_i/(\sqrt{6}\alpha M_{\rm pl})$.
In the following, we will report results for the case  $(p, n)=(2, 4)$ and $\beta=1$, as it leads to the best potential resolution of the tension \cite{Braglia:2020bym}.

\subsection{Early Modified Gravity}

\textbf{Model: } Modified gravity (MG) models which deviate from General Relativity at early times have shown promise in explaining the $H_0$ tension \cite{Umilta:2015cta,Rossi:2019lgt,Ballesteros:2020sik,Braglia:2020iik,Zumalacarregui:2020cjh, Abadi:2020hbr,Ballardini:2020iws,Braglia:2020bym}. In particular the early modified gravity (EMG) model from Ref.~\cite{Braglia:2020auw} postulates the existence of a scalar field $\sigma$ with non-minimal coupling $\xi$ to the Ricci scalar of the form $f(\sigma)=(M_{\rm pl}+\xi\sigma^2)R/2$, on top of a simple quartic potential $\lambda\phi^4/4$. In the limit where $\xi\to 0$ this model reduces to the R'n'R model, while in the limit $\lambda\to 0$ it reduces to the case of a non-minimally coupled massless scalar field considered in Refs.~\cite{Ballesteros:2020sik,Braglia:2020iik}.

\textbf{Motivation: } Non-minimally coupled scalar fields were first suggested to circumvent strong constraints on the variation of Newton's constant in laboratory and solar-system experiments. The value of $G_{\rm eff}$ measured in the laboratory by Cavendish-type experiments for a nearly massless scalar tensor theory of gravity is given by 
\begin{equation}
    G_{\rm eff}\equiv \frac{1}{8\pi F}\frac{2F+4F_{\sigma}^2}{2F+3F_{\sigma}^2}\,
\end{equation}
where $F_{\sigma}$ refers to the first derivative of $F(\sigma)$, and deviations away from the locally measured value tend to zero as the scalar field settles into its minimum today. In addition, the non-minimal coupling introduces a new degree of freedom to control the gravitational strength at early times that improves over the result of the standard R'n'R model \cite{Braglia:2020auw}. 

\textbf{Parameters: } This model consists of three extra free parameters, two describing the field, namely $\sigma_i$ the initial field value and $V_0$ which sets the amplitude of the potential and is related to $\lambda$ through
\begin{equation}
    \lambda = 10^{2V_0}/(3.156 \times 10^{109})\,,
\end{equation} 
where $3.156 \times 10^{109}$ is the numerical value of $M_{\rm pl}^4$ in units of $\rm{eV}^4$, while the third one describes the strength of the non-minimal coupling $\xi$. To ease comparison with other EDE models, along with $\xi$, we will report $f_{\rm EDE}(z_c)$ and ${\rm Log}_{10}(z_c)$, derived parameters in the analyses performed in Refs.~\cite{Braglia:2020auw,Schoneberg:2021qvd}, that are defined as the peak of the EDE contribution and the redshift at which it is reached, respectively.

\subsection{Decaying Ultralight Scalar}

\textbf{Model: } The decaying ultralight scalar (dULS) model uses a `standard' axion potential, $V= m^2 f^2 (1-\cos \phi/f)$, coupled to dark radiation. As the axion oscillates, resonant effects pump energy into the dark radiation, leading to an axion energy density that dilutes faster than matter.  

\textbf{Motivation: } Ref.~\cite{Gonzalez:2020fdy} introduced this model in order to address some of the theoretical issues found in the axion-like EDE models. The axion-like EDE potential ($V\sim (1-\cos \phi/f)^3$) phenomenologically leads to a resolution of the Hubble tension since, in large part, the scalar field energy density dilutes faster than matter. However, this form of the potential is also highly fine-tuned since the potential does not have quadratic and quartic contributions (see Sec.~\ref{sec:EDE_challenges} for a discussion). The dULS model addresses these issues, allowing a non-minimally coupled scalar field with a standard axion potential to dilute faster than matter. 

\textbf{Parameters: } Ref.~\cite{Gonzalez:2020fdy} analyzed this model using a fluid approximation tuned to the exact dynamics. The equation of state evolves as 
\begin{equation}
    w_s = -1 +\frac{1}{1+(a_c/a)^3}+\frac{1/3}{1+(g_d a_c/a)^4},
\end{equation}
where $g_d = 1.1$. The dULS background energy density is then specified by $a_c$ and $\Omega_{\rm dULS}$. The perturbative dynamics are modeled using the sound speed
\begin{equation}
    c_s^2 = 1-\frac{1}{1+(a_c/a)^3}+\frac{1/3}{1+(g_d a_c/a)^4}.
\end{equation}
Note that the perturbative dynamics implied by this choice of sound speed may be quite different from the actual perturbative dynamics in the full theory. For example, it is well known that the effective sound speed of an oscillating scalar field varies in both time and scale as described in detail in Sec.~\ref{sec:EDE_pheno} \cite{Hlozek:2017zzf,Poulin:2018dzj}. 
We note that Ref.~\cite{Weiner:2020sxn} showed the gravitational wave production that accompanies the scalar field decay is ruled out by current CMB observations.

\subsection{Other models}

Finally, we mention a selection of other models that have been suggested in the literature but have not directly confronted cosmological data, or are disfavored by the data. 
These include the model of Refs.~\cite{Sakstein:2019fmf,CarrilloGonzalez:2020oac}, which attempt to explain the injection redshift of EDE (\ie., the EDE mass must overcome the Hubble friction right around matter-radiation equality to be successful) by introducing a conformal coupling between EDE and neutrinos. This leads to a modified KG equation, that now counts an additional source term proportional to the trace of the neutrino energy-momentum tensor. The effect of the neutrino coupling is to
kick the scalar out of its minimum and up its potential right when the neutrinos become non-relativistic, $T_\nu\sim m_\nu$, independently of the value of the scalar field mass and its initial condition. 
Although elegant, this mechanism requires neutrinos with individual masses of ${\cal O}(0.1-0.5)$ eV, testable with current and near-future CMB and LSS observations, with current bounds on the sum of neutrino masses $\sum m_\nu \lesssim 0.1$ eV (e.g. Refs.~\cite{Planck:2018vyg,Brieden:2022lsd,Simon:2022csv}) that can be relaxed in scenarios with exotic neutrino interactions (e.g. Refs.~\cite{FrancoAbellan:2021hdb,Esteban:2021ozz}).
While a dedicated analysis including the conformal coupling is still lacking, we mention that Refs.~\cite{Murgia:2020ryi,Reeves:2022aoi} performed analyses of EDE while simultaneously allowing the sum of neutrino masses to vary and did not find any significant relaxation of the neutrino masses constraint. 

Another example is the ``chain EDE'' model of Ref.~\cite{Freese:2021rjq}, where the Universe undergoes a series of first-order phase transitions, starting at a high-energy vacuum in a potential, and tunneling down through a chain of lower-energy metastable minima. 
In general, a single phase transition occurring around matter-radiation equality (as favored to resolve the Hubble tension), and characterized by a constant tunneling rate per volume $\Gamma$, is excluded by CMB and LSS observations due to the non-observation of anisotropies induced by the presence of bubbles of false vacuum on large scales. 
One way out of this constraint is a time-dependent tunneling rate $\Gamma$, e.g. due to an additional trigger field as in the NEDE proposal (an idea originally introduced in the context of double-field inflation \cite{Linde:1990gz,Adams:1990ds}). 
The authors of Ref.~\cite{Freese:2021rjq} suggest a differnt way to evade the CMB constraints. 
They invoke tunneling along a chain of false vacua with decreasing energy at a constant tunneling rate and, using simple scaling arguments, find that a solution to the Hubble tension requires $N>600$ phase transitions to avoid very large anisotropies. 
A specific example of Chain EDE is given, featuring a scalar field in a titled cosine potential, that authors argue to be ubiquitous in axion physics and have strong theoretical motivation. However, a dedicated analysis of this promising model against cosmological data is still lacking.

A variant of RnR EDE is AdS-EDE, originally studied in Ref.~\cite{Ye:2020btb}. 
It consists of a scalar field with a quartic potential ($V \propto \phi^4$) which is modified so that the field goes through an Anti-De Sitter (AdS) (i.e., $V<0$) phase. 
Constraints on this model in the literature are incomplete due to the use of a theory prior that appears to enforce a non-zero lower bound on the EDE fraction\footnote{Evidence for this can be found by noting that in Ref.~\cite{Ye:2021iwa} the AdS-EDE $f_{\rm EDE}$ is non-zero at the 17$\sigma$ level but the overall $\chi^2$ is {\it degraded} by 3.}. 
Yet, in Ref.~\cite{Ye:2020btb}, when fixing the depth of the AdS phase, the model is shown to have a better $\chi^2$ when including the SH0ES prior than RnR, which is encouraging regarding the potential of this model, and deserves further investigation.

Our final example is the model of assisted quintessence studied in Ref.~\cite{Sabla:2021nfy} (see also Ref.~\cite{GARCIA2021101503} for a similar phenomenological model). 
This model, originally introduced two decades ago to resolve the `cosmic coincidence' problem \cite{Dodelson:2001fq,Kim:2005ne}, introduces a scalar-field that exhibits tracking behavior, such that its energy density is a fixed fraction of the dominant background species. As a result of the transition from  radiation- to matter-domination, an era of early dark energy occurring around matter-radiation equality is inevitable. Nevertheless, the authors of Ref.~\cite{Sabla:2021nfy} show that this field leads to an irreducible contribution to DM after the transition, that prevents a resolution of the tension (in fact it slightly exacerbates it). Indeed, as mentioned in the text above, EDE must vanish faster than matter to alleviate the Hubble tension. An additional energy component in the early universe that dilutes like matter worsens the tension as shown in \cite{Poulin:2018dzj}.

\section{Phenomenology of Early Dark Energy}
\label{sec:EDE_pheno}

The basic physics of a minimally-coupled EDE can be captured through the `generalized dark matter' formalism first presented in Ref.~\cite{Hu:1998kj} (see also Ref.~\cite{Poulin:2018dzj}). The dynamics of any cosmological material can be described by specifying an equation of state $w(a)$, an effective sound-speed $c_{s}^2(k,a)$ (defined in the material's local rest-frame), and the anisotropic stress $\sigma(k,a)$. For scalar fields, the anisotropic stress is zero; for the following discussion we will take $\sigma = 0$ (see Ref.~\cite{Sabla:2022xzj} for a discussion of the phenomenology when the anisotropic stress is non-zero). 

\subsection{Background and perturbations evolution}

For an EDE with an equation of state $w_{\rm EDE}(a)$, the continuity equation immediately gives the evolution of the energy density, 
\begin{equation}
\rho_{\rm EDE}(a) = \rho_{\rm EDE,0} e^{3\int_a^1 [1+w_{\rm EDE}(a)]da/a} \,.
\end{equation}
The basic background dynamics of EDEs we review here have $w_{\rm EDE} \rightarrow -1$ at some point in the past, which then transitions to $0<w_{\rm EDE}<1$ at some critical scale factor $a_c$. 
Specifically for a scalar field with a potential of the form $V \propto \phi^{2n}$ around its minimum, we have $w_f=(n-1)/(n+1)$ \cite{Turner:1983he, Poulin:2018dzj}. We can parameterize this through the function 
\begin{equation}
\label{eq:w_EDE_pheno}
w_{\rm EDE}(a) = \frac{1+w_f}{1+(a_c/a)^{3(1+w_f)}} -1\,,
\end{equation}
which describes a fluid that does not dilute with cosmic expansion when $a \ll a_c$, but dilutes as $a^{-3(1+w_f)}$ for $a \gg a_c$. This parameterization also describes the ADE model when $p=1$ as discussed in the previous section. 

\begin{figure}
    \centering
    \includegraphics[width=0.8\columnwidth]{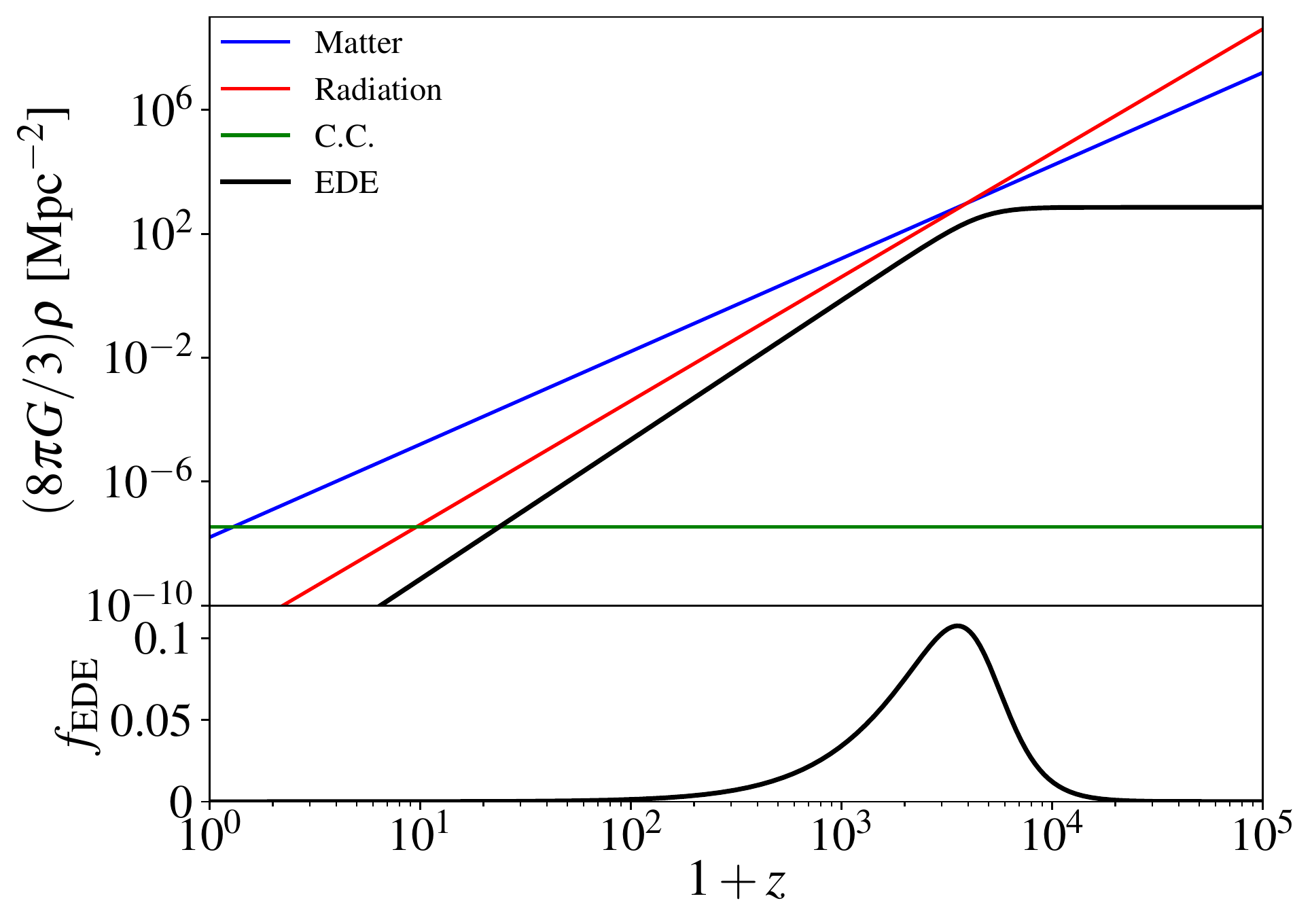}
    \caption{{\it Top panel:} Evolution of the energy densities of the various components of the Universe, including an early dark energy active before recombination. {\it Bottom panel:} Fractional contribution $f_{\rm EDE}\equiv \rho_{\rm EDE}/\rho_{\rm tot}$ of EDE. As a representative example, we chose a model where the maximum contribution $f_{\rm EDE}(z_c)=0.1$ is reached at the critical redshift $z_c \simeq 3500$ subsequently diluting with an equation of state $w=1/2$ afterwards. }
    \label{fig:EDE_background}
\end{figure}

For such an equation of state, the background evolution of $\rho_{\rm EDE}(a)$ is shown in Fig.~\ref{fig:EDE_background}, and is summarized by $f_{\rm EDE} \equiv \rho_{\rm EDE}/\rho_{\rm tot}$, which shows that the contribution of EDE is localized around the time when the field becomes dynamical (denoted by the `critical' redshift, $z_c$).

Perturbations on the other hand evolve according to the continuity and Euler equations \cite{Ma:1995ey,Hu:1998kj}:
\begin{eqnarray}
\frac{d}{d\eta} \left(\frac{\delta_{\rm EDE}}{1+w_{\rm EDE}}\right) &=& -\left(\theta_{\rm EDE}+h_\delta'\right)-3\frac{a'}{a}(c_s^2-c_a^2)\left(\frac{\delta_{\rm EDE}}{1+w_{\rm EDE}} + 3 \frac{a'}{a} \frac{\theta_{\rm EDE}}{k^2}\right)\label{eq:Cont}\\
\theta_{\rm EDE}' &=& -\frac{a'}{a} \left(1-3 c_s^2\right) \theta_{\rm EDE} + c_s^2 k^2 \frac{\delta_{\rm EDE}}{1+w_{\rm EDE}} + k^2 h_v, \label{eq:Euler}
\end{eqnarray}
where primes denote derivatives with respect to conformal time $\eta$, 
the gravitational potentials are $h_\delta = h_s/6$ and $h_v =0$ in synchronous gauge, and $h_\delta = \phi_N$ and $h_v = \psi_N$ in conformal Newtonian gauge \cite{Ma:1995ey}, 
\begin{equation}
    c_a^2 = \frac{\rho'_{\rm EDE}}{P'_{\rm EDE}} = w_{\rm EDE} - \frac{1}{3} \frac{dw_{\rm EDE}/d\ln a}{1+w_{\rm EDE}} 
\end{equation}
is the adiabatic sound speed defined as the gauge-independent linear relation between time variations of the background pressure and energy density of the fluid, $\dot{p}=c_a^2\dot{\rho}$, and $c_s^2$ is the EDE's effective sound speed, defined in its local rest-frame \cite{Hu:1998kj}. 
Note that the choice of $w_{\rm EDE}(a)$ given by Eq.~\ref{eq:w_EDE_pheno} gives $c_a^2(a\ll a_c) = -(2+w_f)$ and $c_a^2(a\gg a_c) = w_f$, which approximates the time variation of the adiabatic sound speed in scalar-field EDE models \cite{Poulin:2018dzj}. 

The choice of $c_{s}^2$ is more complicated, since in general it depends on both $k$ and $a$. In scalar-field models with a potential of the form $V \propto \phi^{2n}$ \cite{Poulin:2018dzj}, 
\begin{equation}
\label{eq:cs2_axEDE}
c_{s}^2(k,a) = \frac{2 a^2(n-1) \varpi^2+k^2}{2 a^2 (n+1) \varpi^2 + k^2} \,,
\end{equation}

where $\varpi$ is the angular frequency of the oscillating background field and is well-approximated by \cite{Johnson:2008se,Poulin:2018dzj,Smith:2019ihp}
\begin{eqnarray}
\varpi(a) &\simeq& m\frac{\sqrt{\pi} \Gamma(\frac{1+n}{2n})}{\Gamma\left(1+\frac{1}{2n}\right)}2^{-(1+n)/2}  \theta^{n-1}_{\rm env}(a)\,,
\label{eq:omega}\\
&\simeq& 3 H(z_c)\frac{\sqrt{\pi} \Gamma(\frac{1+n}{2n})}{\Gamma \left(1+\frac{1}{2n}\right)}2^{-(1+n)/2} \frac{\theta^{n-1}_{\rm env}(a)}{{\sqrt{|E_{n,\theta \theta}(\theta_i)|}}}\,,
\nonumber
\end{eqnarray}
where $\Gamma(x)$ is the Euler Gamma function and the envelope of the background field ($\theta_{\rm env}\equiv \phi_{\rm env}/f)$ once it is oscillating is roughly
\begin{equation}
\phi_{\rm env}(a)=\phi_c\left(\frac{a_c}{a}\right)^{3/(n+1)}\,,\label{eq:phi_env}
\end{equation}
and we have written the scalar field potential as $V_n(\phi) = m^2 f^2 E_n(\theta = \phi/f)$. We recall that $m$ and $f$ are the axion mass and decay constant respectively. 

The  angular frequency $\varpi$ defines a scale that governs the sound speed affecting a given mode: For $a \gg a_c$, $\varpi = 0$, so that $c_s^2 = 1$; when $a \ll a_c$ and $k\ll a \varpi$, $c_{s,{\rm eff}}^2 = (n-1)/(n+1) = w$; when $a \ll a_c$ and $k\gg a \varpi$, $c_{s}^2 =1$. 
Note that this approximate sound speed only applies when $\varpi/H \gg 1$-- i.e. when the field oscillates several times per Hubble time.  
For simplicity in what follows, we will take the effective sound speed to be constant in time and scale. Since, in all cases, $c_{s}^2$ is of order unity, our choice of $c_{s}^2={\rm constant}$ will give a good sense of how the perturbative dynamics impact observables. This approximation also matches the dynamics in the ADE model. We discuss the axion-like case later on.

\subsection{Understanding the dynamics of EDE}

\begin{figure}
    \centering
    \includegraphics[width=\columnwidth]{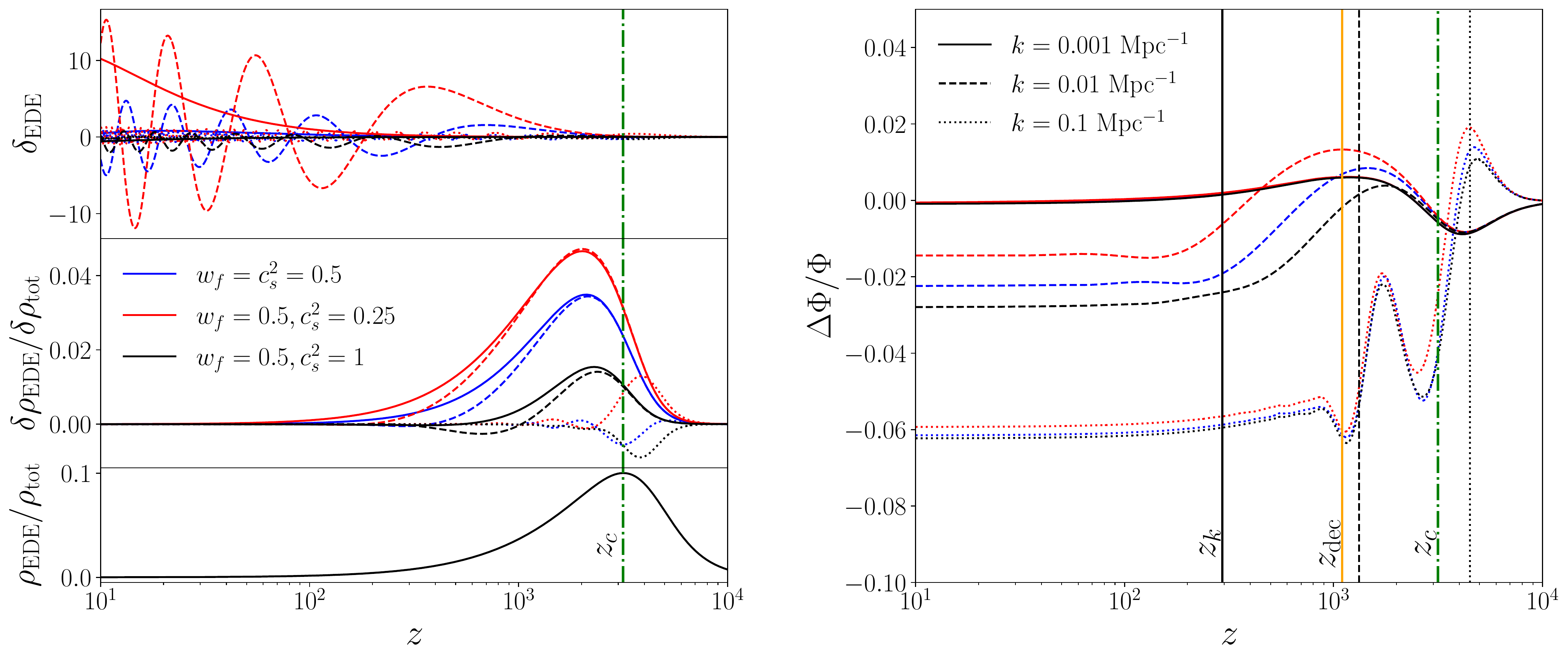}
    
    \caption{{\it Left panel:} The  evolution of EDE perturbations in the phenomenological fluid model for different values of the effective sound speed. In each panel, we show the evolution of the background (bottom) and perturbed EDE (top and middle) energy density for three different wavenumbers. 
    {\it Right panel:} The difference in the evolution of the Weyl potential, relative to  $\Lambda$CDM with the same values of $\{h,\omega_{\rm cdm},\omega_b,A_s,n_s,\tau_{\rm reio}\}$. 
    We also show the redshift of horizon entry for each mode (black), the critical redshift of the EDE (dot-dashed green) and the redshift of decoupling (solid orange).}
    \label{fig:pheno_EDE_comp}
\end{figure}

\begin{figure}
    \centering
    \includegraphics[width=0.8\columnwidth]{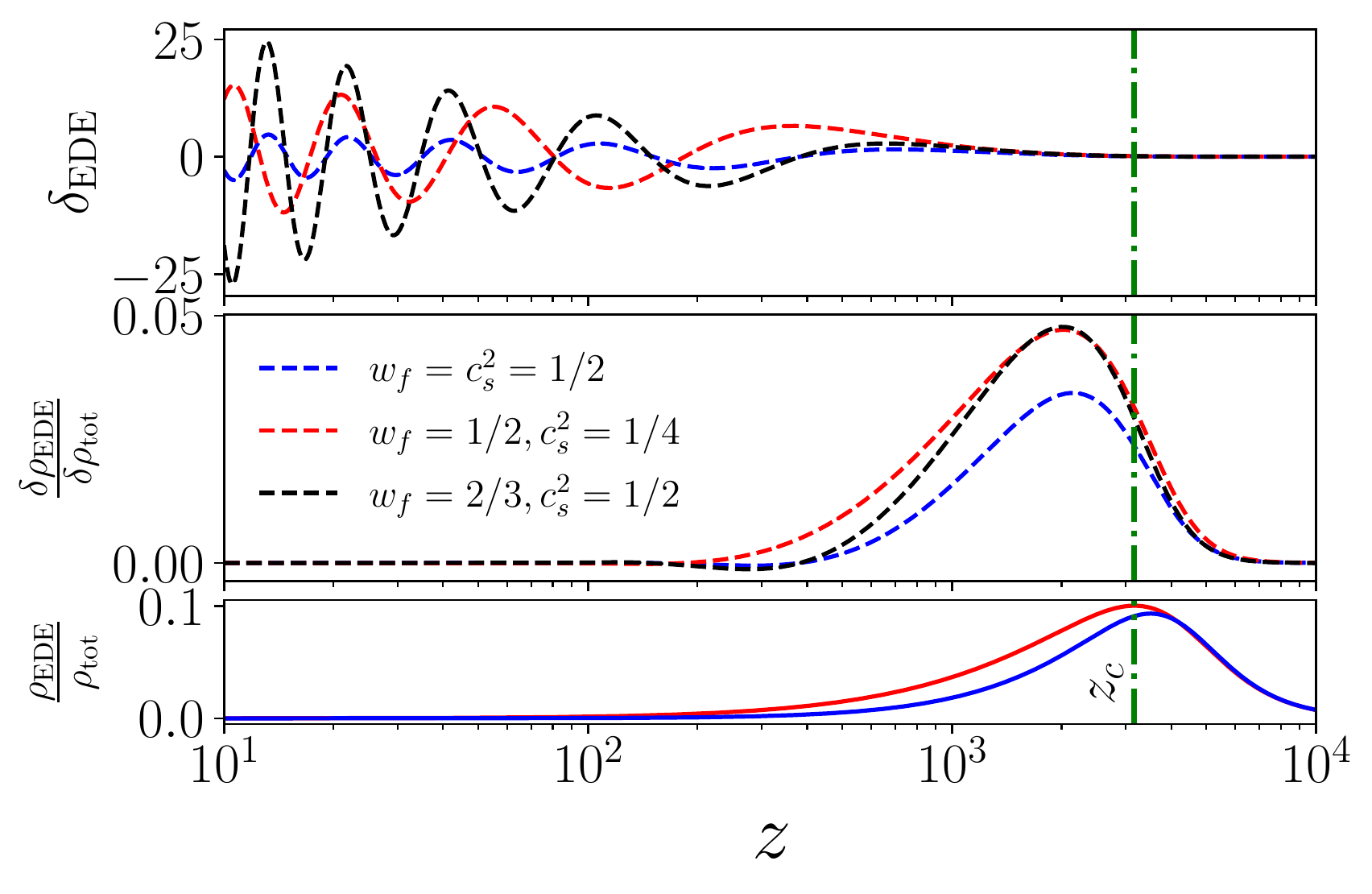}
    \caption{ The evolution of EDE perturbations with $k = 0.1 h/$Mpc in the phenomenological fluid model for different values of $c_s^2$ and $w_f$. 
    The middle panel shows that the amplitude of the first compression in $\delta \rho_{\rm EDE}/\delta \rho_{\rm tot}$ can be increased by either decreasing $c_s^2$ or increasing $w$. }
    \label{fig:phenoEDE_cs2-vs-w}
\end{figure}

We now turn to describing the evolution of EDE density perturbations $\delta_{\rm EDE}$.  The initial conditions are easiest to derive in synchonous gauge (it is straightforward to derive the corresponding initial conditions in conformal Newtonian gauge through the standard gauge transformation \cite{Ma:1995ey}). Adiabatic perturbations are generated dynamically since the gravitational potential, $h_S$, is initially non-zero. 
Specifically we have $h_S =  (k \eta)^2/2$ \cite{Ma:1995ey}, leading to the initial behavior when $w_{\rm EDE} \simeq -1$, on superhorizon scales, and during radiation-domination for the EDE fluid variables\footnote{When $w_{\rm EDE} \simeq -1$, the EDE makes a negligible contribution to the total energy density and so does not affect $h_S$ at this time.}
\begin{eqnarray}
\frac{\delta_{\rm EDE}}{1+w_{\rm EDE}} &=& -\frac{(4-3 c_s^2)/2}{32+6 c_s^2 + 12 w_f} (k \eta)^2, \label{eq:deltaIC}\\
\theta_{\rm EDE} &=& -\frac{c_s^2/2 }{32+6 c_s^2 + 12 w_f}k (k \eta)^3,\label{eq:thetaIC}
\end{eqnarray}
where $\eta$ is the conformal time, and $w_f$ appears because the adiabatic sound speed at $a\gg a_c$ is $c_a^2(a\ll a_c) \simeq -(2+w_f)$.  
The initial conditions show that on superhorizon scales the density perturbation is suppressed for $a< a_c$ since $\delta_{\rm EDE}\propto (1+w_{\rm EDE})$, with $w_{\rm EDE}\to -1$. This statement is also true in conformal Newtonian gauge since the gauge transformation is proportional to $\rho_{\rm EDE}'/\rho_{\rm EDE} \propto (1+w_{\rm EDE})$.

On subhorizon scales where matter perturbations dominate, one has
\begin{eqnarray}
\frac{d^2}{d\eta^2}\left(\frac{\delta_{\rm EDE}}{1+w_{\rm EDE}}\right) &=& -k^2\left(c_s^2 \frac{\delta_{\rm EDE}}{1+w_{\rm EDE}} + \psi_N\right) -(1-3c_a^2) \frac{a'}{a} \frac{d}{d\eta}\left(\frac{\delta_{\rm EDE}}{1+w_{\rm EDE}}\right) \,,
\label{eq:subhorEDE}
\end{eqnarray}
where we have written the fluid equation in conformal Newtonian gauge, since this gauge can help us build intuition for the mode dynamics on subhorizon scales.

The redshift evolution of $\delta_{\rm EDE}$ (and the fractional contribution $\delta\rho_{\rm EDE}/\delta\rho_{\rm tot}$) in a model with\footnote{The choice of parameters is arbitrary, but close to the best-fit EDE model, with $a_c=10^{-3.5}\Rightarrow z_c \simeq 3160$.} $f_{\rm EDE}(a_c) = 0.1$, $a_c = 10^{-3.5}$ and $w=1/2$ is shown in the left panels of Fig.~\ref{fig:pheno_EDE_comp} for three different modes $k=0.001,0.01,0.1$ Mpc$^{-1}$, and three different sound speeds $c_s^2 = 0.25,0.5,1$. 
These wavenumbers are chosen to demonstrate the impact of EDE and the key role played by $c_s^2$ (and the interplay with $w$) on modes entering the horizon after, close to and before the EDE critical redshift, respectively. 
In addition, we show the effect of varying the equation of state $w$ compared to varying $c_s^2$ for the mode  $k=0.01$ Mpc$^{-1}$ in Fig.~\ref{fig:phenoEDE_cs2-vs-w}.
The modes evolve as follows:

\begin{itemize}

 \item[\textbullet]  First, one can see from Eq.~\ref{eq:subhorEDE} that with a non-zero sound speed $c_s^2$, the fluid has significant pressure support, leading to oscillations whose frequency is set by $c_s^2$.  The greater $c_s^2$ is, the more effective the pressure gradients are in preventing the EDE density contrast from collapsing. Accordingly, smaller $c_s^2$ lead to a larger contribution to the overall density perturbation, as shown in the middle panel of the left graphic in Fig.~\ref{fig:pheno_EDE_comp}.

\item[\textbullet]Second, the friction term ($\propto \delta_{\rm EDE}'$) can cause the oscillation amplitude to either decrease or increase, depending on the sign of $(1-3 c_a^2)$. For the case shown here with $w_f=1/2$ at $a>a_c$\,, $1-3c_a^2 = 1-3 w_f = -1/2$, leading to an {\it enhancement} of the oscillation amplitude as the universe expands.

\item[\textbullet]  Third, at the perturbation level, the competition between the effects of pressure inside the horizon (controlled by $c_s^2$) and the growth of EDE modes outside the horizon ($\delta_{\rm EDE}\propto(1+w_{\rm EDE})$) leads to a positive correlation between $c_s^2$ and $w_f$.  This can be seen in Fig.~\ref{fig:phenoEDE_cs2-vs-w}, where we show that reduced pressure support (i.e., $c_s^2=1/4$) can partially mimic an increase in the final equation of state to $w_f = 2/3$, and consequently, a larger $w_{\rm EDE}$ can partly compensate for the impact of greater pressure support $c_s^2$  \cite{Lin:2019qug}. 

\end{itemize}

\subsection{Impact of EDE perturbations on the Weyl potential}

In the case where the EDE is minimally coupled, the effects of its perturbations are only communicated through changes in the gravitational potentials. In conformal Newtonian gauge, changes to the photon perturbations are governed by the Weyl potential, $\Phi \equiv (\phi_N + \psi_N)/2$, which is sourced by 
\begin{equation}
\Phi = -\frac{3 }{4 k^2 }\left(\frac{a'}{a}\right)^2\left(2 \delta + \sum_i [1+w_i] [6 (a'/a)\theta_i/k^2+3\sigma_i]\right),
\end{equation}
where the sums are over all cosmological species and $\sigma_i$ is the anisotropic stress.
The right panel of Fig.~\ref{fig:pheno_EDE_comp} shows the fractional change to the Weyl potential in the phenomenological EDE model.
The changes shown there can be split into the effects due to modifications to the background and due to EDE perturbative dynamics.  
At the background level, the story is simple: the presence of EDE causes the Hubble parameter to increase around $a \sim a_c$, leading to an increase in the Hubble friction experienced by the CDM modes that are within the horizon at $a_c$.  This explains the dominant effect visible in the right panel of Fig.~\ref{fig:pheno_EDE_comp}: a reduction of the amplitude of the Weyl potential, stronger for larger $k$ modes, that spend a longer time in the horizon while the EDE contributes significantly to the energy budget.  \\  

At the perturbation level, the story is more complex. Indeed, one important aspect of the left panel of Fig.~\ref{fig:pheno_EDE_comp} is that the dynamics of a given EDE mode and its impact on observables are tied to whether the mode enters the horizon (i.e. when it satisfies $k=a_k H(a_k)$) around the time when the EDE fluid becomes dynamical $a_c$.  This can be understood as follows:

\begin{itemize} 
\item[\textbullet]    All modes begin outside the horizon, where the density perturbation $\delta_{\rm EDE} \propto 1+w_{\rm EDE}$. As a result, modes which enter the horizon before $a_c$, when $w_{\rm EDE} \simeq -1$ (satisfying $a_k\ll a_c$ with $k=a_k H(a_k)$), will be stabilized by pressure gradients with an overall amplitude suppressed by $\sim 1+w_{\rm EDE}(a_k)$, relative to modes that enter the horizon later. 
This explains why in the left panel of Fig.~\ref{fig:pheno_EDE_comp}, the mode $k=0.1$ Mpc$^{-1}$ has a smaller amplitude (and overall contribution to $\delta \rho_{\rm tot}$) than other modes which enter when $w_{\rm EDE}>-1$.  Due to this suppression, the Weyl potential is basically insensitive to the $c_s^2$-dependent details of the evolution of these early EDE modes in this phenomenological model.\\

\item[\textbullet] In principle, the EDE velocity perturbations also contribute to the gravitational potentials through the resulting heat-flux $(\rho_{\rm EDE}+P_{\rm EDE}) \theta_{\rm EDE} \propto (1+w_{\rm EDE}) \theta_{\rm EDE}$, but are also suppressed when $w_{\rm EDE} \simeq -1$.  \\
   
\item[\textbullet]  Taking the other limit $a_k\gg a_c$ (see for example the mode with $k=0.001$  Mpc$^{-1}$), one may expect that modes entering the horizon later will strongly impact the Weyl potential dynamics, because of their enhanced growth on super-horizon scales.  However, their impact turns out to be minimal because once they enter the horizon (around $z \sim 300$ for this mode), the physical perturbations are suppressed by $\rho_{\rm EDE}/\rho_{\rm tot} \ll 1$, and one can see on the right panel that the Weyl potential with $k=0.001$  Mpc$^{-1}$ is largely unaffected by EDE. 
\\

\end{itemize}

In summary, details of the EDE perturbative dynamics are imprinted on cosmological observables only for modes that enter the horizon around $a_c$.  We can clearly see this in the fractional change to the Weyl potential for each mode in the right panel of Fig.~\ref{fig:pheno_EDE_comp}: the only mode with an appreciable dependence on the value of $c_s^2$ is $k=0.01\ {\rm Mpc}^{-1}$, which is entering the horizon slightly after $a_c= 10^{-3.5}$. 
This is particularly important in determining where the support for specific EDE dynamics (beyond the background evolution) comes from within the data, and shows that one can hope to precisely measure {\it when} and {\it  how} the EDE contribution occurred. \\

\subsection{Impact on the CMB and matter power spectra}

\begin{figure}
    \centering
    \includegraphics[width=0.8\columnwidth]{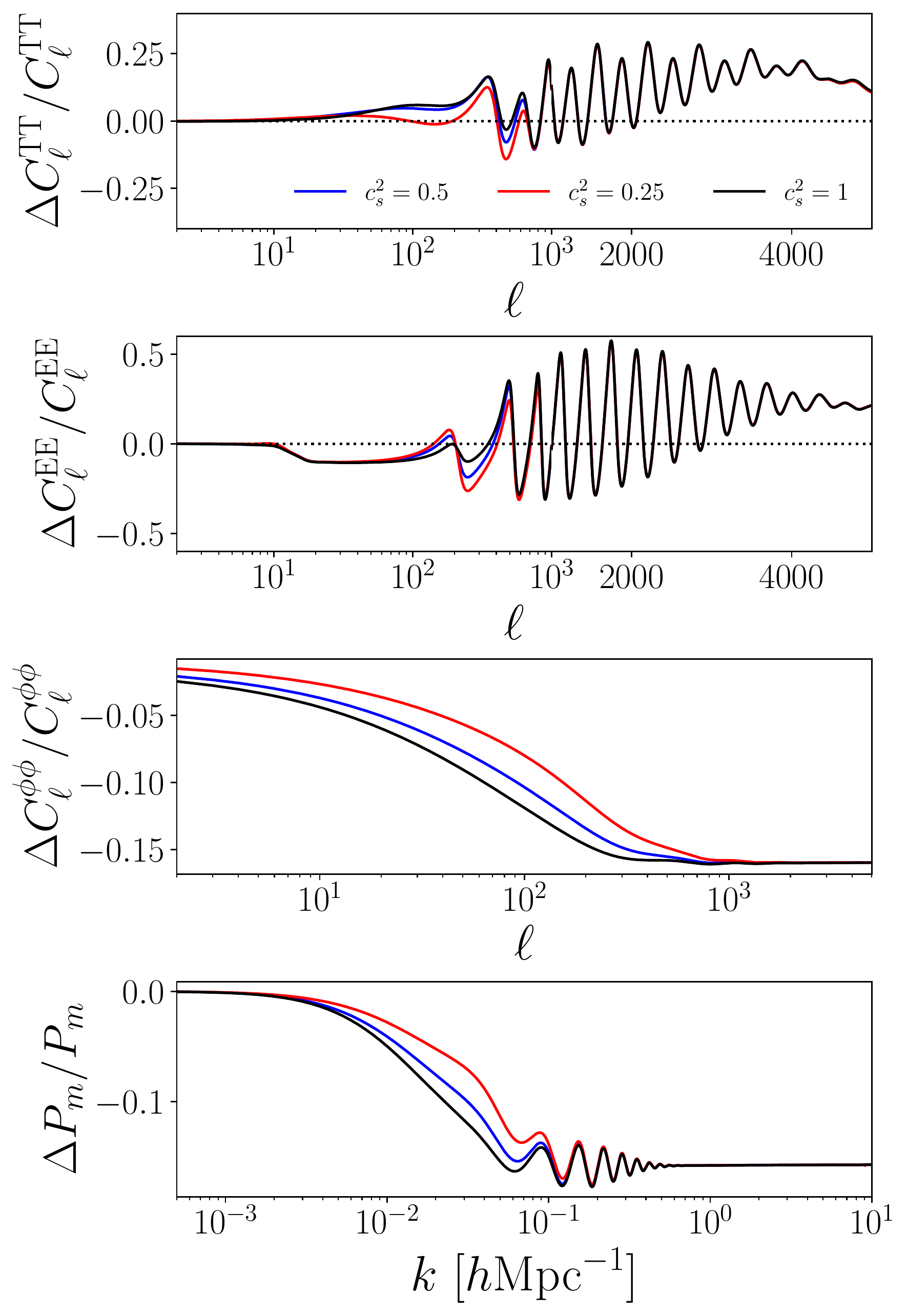}
    \caption{The impact  of EDE  $f_{\rm EDE}(a_c) = 0.1$, $a_c = 10^{-3.5}$ and $w=1/2$ on the CMB and matter power spectra  for three different values of $c_s^2=0.25,0.5,1$. We show the relative difference with respect to  $\Lambda$CDM with the same values of $\{h,\omega_{\rm cdm},\omega_b,A_s,n_s,\tau_{\rm reio}\}$. }
    \label{fig:pheno_EDE_spectra}
\end{figure}

The impact  of EDE  $f_{\rm EDE}(a_c) = 0.1$, $a_c = 10^{-3.5}$ and $w=1/2$ on the CMB and matter power spectra  for three different values of $c_s^2=0.25,0.5,1$ is shown in Fig.~\ref{fig:pheno_EDE_spectra}.  We show the relative difference with respect to  $\Lambda$CDM with the same values of $\{h,\omega_{\rm cdm},\omega_b,A_s,n_s,\tau_{\rm reio}\}$.
We focus on illustrating the impact of $c_s^2$, as the dominant effects of changing $f_{\rm EDE}$, $a_c$ and $w_f$ are fairly straightforward, since they (mostly) come from the changes due to background dynamics. 
For a discussion on the roles of $f_{\rm EDE}$, $a_c$ and $w_f$, we refer to App.~\ref{app:EDE_pheno}.

First,  as discussed in the introduction, the purely background effect of EDE (at fixed $h$) is to increase the early-universe expansion rate, thereby decreasing the angular sound horizon and damping scales, which lead to residual wiggles and a higher amplitude at large $\ell$ in $TT$ and $EE$ that is common to all models, regardless of the value of $c_s^2$. 
The effect of perturbations on the other hand, is localized and tied to the dynamics of modes which enter the horizon right around $z_c$, as discussed previously. Using $k_c(\eta_0 - \eta_c)\simeq l_c$ and $k_c \eta_c = 1$, we see that the perturbative dynamics will impact CMB observations around $l_c \simeq \eta_0/\eta_c \simeq 100$ for $a_c = 10^{-3.5}$, with a width extending about a factor of 10 in redshift, corresponding to a $\delta \ell \simeq 100$. 
This is clearly seen in Fig.~\ref{fig:pheno_EDE_spectra}, where all models lead to the same dynamics at $\ell \gtrsim 1000$, but deviate below this scale. 

\begin{figure}
    \centering
    \includegraphics[width=0.7\columnwidth]{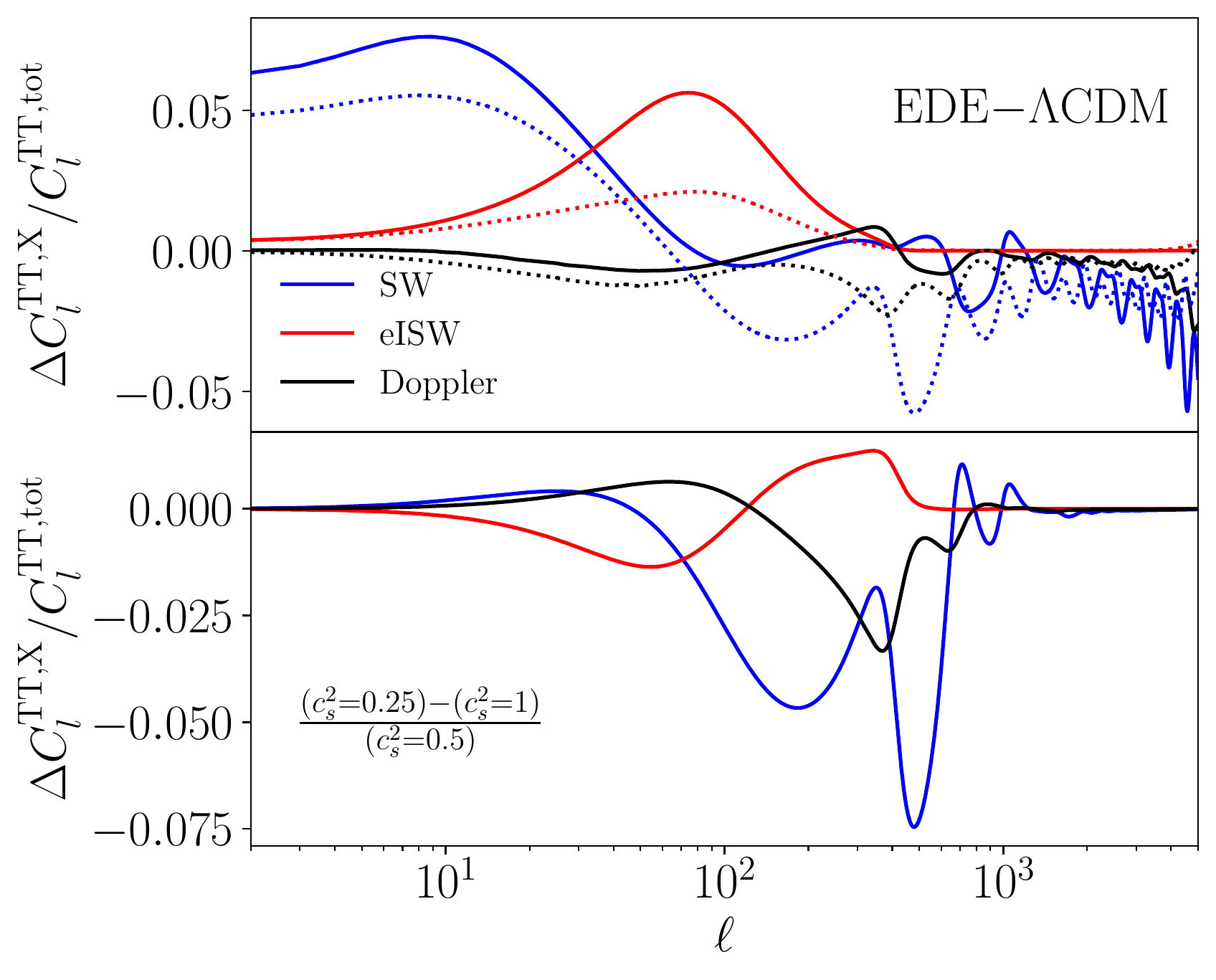}
    \caption{The fractional change to the Sachs Wolfe (SW), early ($z>50$) integrated Sachs Wolfe (early ISW) and Doppler terms in the line-of-sight integral that determines the CMB temperature power spectrum. 
    The top panel shows the fractional change between the phenomenological EDE and $\Lambda$CDM model. 
    The bottom panel shows the fractional change between EDE models with different values of $c_s^2$. 
    Here, unlike in previous figures in this section, we have fixed the angular size of the sound horizon at photon decoupling to $100\theta_s = 1.04$ to remove the dominant background effect induced by EDE. In the top panel the solid curves show the differences at fixed physical CDM density, and in this case the EDE model has $h = 0.80$ and $\Omega_m = 0.225$ while the $\Lambda$CDM model has $h = 0.67$ and $\Omega_m = 0.31$. The dashed curves show thee differences when $\Omega_m$ is held fixed (at 0.31) and in this case EDE model has $h = 0.723$ and $\omega_{\rm cdm} = 0.14$ while the $\Lambda$CDM model has $\omega_{\rm cdm} = 0.12$. Note that the contributions to the temperature anisotropies from photon polarization and anisotropic Thomson scattering only constitute $\sim 0.5\%$ of the total anisotropies and are affected at the $\sim 1-10\%$ level by EDE, making it negligible compared to the other sources we show here.
    }
    \label{fig:sw_eisw_dop}
\end{figure} 

\begin{figure}
    \centering
    \includegraphics[width=0.7\columnwidth]{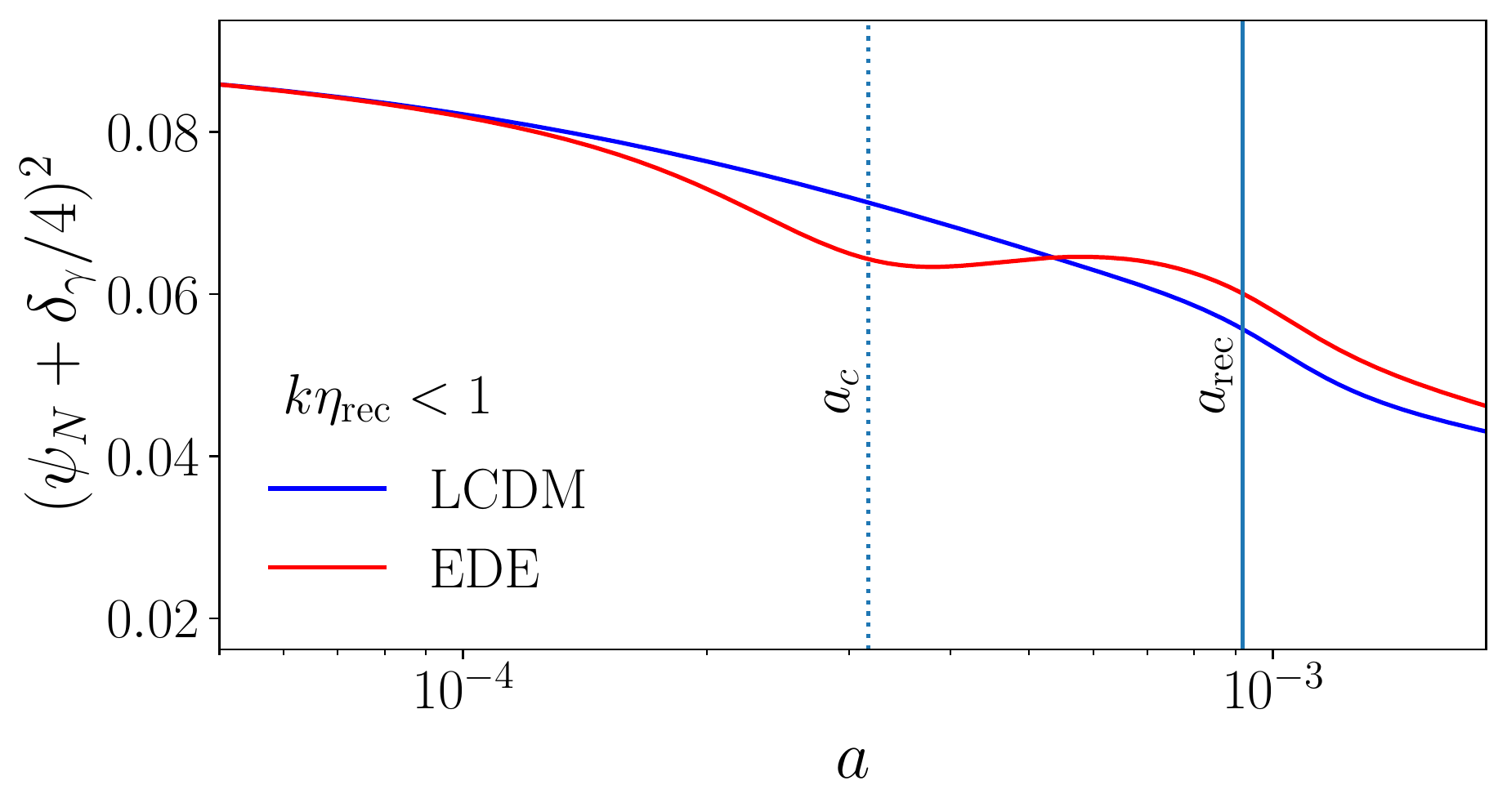}
    \caption{ The time evolution of SW Fourier modes which are superhorizon before recombination (i.e., $k\eta_{\rm rec}<1$) in $\Lambda$CDM and EDE. The presence of non-adiabatic pressure in the EDE causes a time variation in this mode around $a\simeq a_c$ which persists up to recombination, leading to an enhanced Sachs-Wolfe plateau at low multipoles.}
    \label{fig:sw_pert}
\end{figure}

The more subtle effects induced by modifying the sound speed are detailed in the previous section, closely following their first presentation in Ref.~\cite{Lin:2019qug} (see also Ref.~\cite{Poulin:2018zxs} for a related discussion). 
To further explore the role of EDE perturbations,  we show the fractional change to the individual contributions to the CMB TT power spectra, namely the Sachs-Wolfe, (early) integrated Sachs-Wolfe (eISW), and Doppler terms for $c_s^2=0.25$ and 0.5 in the bottom panel of Fig.~\ref{fig:sw_eisw_dop}. 
The top panel of Fig.~\ref{fig:sw_eisw_dop} shows the fractional difference of the phenomenological EDE {\it at fixed $c_s^2=0.5$} and $\Lambda$CDM.
We remove the dominant background effect that is highly correlated with $h$ from these plots by fixing $\theta_s$ instead of $h$. The solid curves shows the fractional differences at fixed physical CDM density (at $\omega_{\rm cdm} = 0.12$), giving $h = 0.67$ and $\Omega_m = 0.31$ in $\Lambda$CDM, while the phenomenological EDE model has $h = 0.80$ and $\Omega_m = 0.22$. The dotted curves show what happens when $\Omega_m$ is held fixed (at $\Omega_m = 0.31$), giving $\omega_{\rm cdm} = 0.14$ and $h=0.723$ in the EDE model. 
The effect of EDE perturbations can then be understood as follows:

\begin{itemize}
    \item[\textbullet] Changes to the Weyl potential due to EDE perturbations affect the ``acoustic driving'' of the CMB photon perturbations - the decay of gravitational potentials can boost the amplitude of acoustic oscillations in the baryon-photon fluid. Acoustic driving impacts the Sachs-Wolfe term in the line-of-sight integral which determines the overall CMB power spectra, and this is primarily responsible for the differences in the CMB power spectra between EDE models with different perturbative dynamics.  
    Concretely, we have seen in the previous section that a larger $c_s^2$ (i.e. greater pressure support) leads to a smaller amplitude of EDE perturbations and therefore a faster decay of the Weyl potential, affecting the driving force acting on CMB perturbations. 
    This means that larger $c_s^2$ will show a larger SW amplitude than smaller $c_s^2$ (and vice versa), leading to the small differences between the EDE models seen in the bottom panel of Fig.~\ref{fig:sw_eisw_dop}. 
    This effect dominates constraints on $c_s^2$ \cite{Li:2019yem}. \\

    \item[\textbullet]
    In addition, differences in the time-dependence of the gravitational potential around recombination affect the amplitude of the eISW term, which is strongly impacted by the presence of EDE \cite{Vagnozzi:2021gjh}. 
    This is most clearly visible in the top panel of Fig.~\ref{fig:sw_eisw_dop}, where we show the fractional difference of the phenomenological EDE (at fixed $c_s^2=0.5$) and $\Lambda$CDM, with a large eISW contribution around $\ell \sim 100$.  
    In fact, as we have adjusted $h$ to match $\theta_s$, one can see that the dominant impact of EDE at $\ell \sim 100$ is through the eISW term, rather than the SW term. 
    The additional time variation in the gravitational potentials in EDE is due to the small residual EDE contribution to the background energy density after recombination.
    The dotted curve shows that when we fix $\Omega_m$, thereby increasing $\omega_{\rm cdm}$, the eISW difference is reduced.  This feature is particularly important to understand the correlation in parameters that comes out of the MCMC analysis, specifically a (perhaps surprising) increase in $\omega_{\rm cdm}$.\\
    
    \item[\textbullet] The top panel of Fig.~\ref{fig:sw_eisw_dop} shows an increase in the SW contribution at low multipoles. This is due to the effect of the time-varying total equation of state while the EDE is a sizable fraction of the total energy density. The SW term for modes that are superhorizon before recombination is shown in Fig.~\ref{fig:sw_pert}. This evolution-- a slight decrement followed by an enhancement-- is also seen in the Weyl potential (see the right panel of Fig.~\ref{fig:pheno_EDE_comp}). Note that the enhancement of the SW contribution is partially canceled by an anticorrelation with the early ISW term. The cancellation is nearly exact when comparing EDE and $\Lambda$CDM models with the same $h$ and $\omega_{\rm cdm}$ and is only partial when fixing $\theta_s$. This can be seen in the low-$\ell$ differences shown in the figures in Appendix \ref{app:EDE_pheno}.
\end{itemize}

Finally, the dynamics of EDE at the background level largely explain the changes to the matter and lensing spectra seen in Fig.~\ref{fig:pheno_EDE_spectra}: modes with larger $k$ are suppressed due to the increased Hubble friction induced by the presence of EDE around $a_c$. 
This choice of $a_c$ implies a suppression of the matter power spectrum for modes $k\gtrsim 0.01\ h {\rm Mpc}^{-1}$, 
while the amount of suppression at $z=0$ increases with $f_{\rm EDE}$, and decreases as $w_f$ increases, since a larger $w_f$ leads to a shorter period of time where the EDE is dynamically relevant. We stress that both {\it how much} and {\it how long} EDE contributes to $H(z)$ matter in setting the overall amplitude of the suppression. 
Yet, one can see that the shape of the suppression strongly depends on $c_s^2$ for the reasons discussed extensively above: one can partly compensate the effect of the Hubble friction by reducing the pressure support of EDE perturbations. 
As a result, the model with smaller $c_s^2$ shows a shallower suppression around $k_c$ than that with larger $c_s^2$. 
The CMB lensing potential power spectrum has a similar dependence, with the rough mapping  $\ell \sim 10^3-10^4 k$, and the important difference that its large-scale amplitude is set by the matter power spectrum around $k_{\rm eq} \simeq 0.02\ h {\rm Mpc}^{-1}$ \cite{Planck:2015mym}.

\subsection{Beyond the phenomenological model: the axion-EDE case}

\begin{figure}
    \centering
    \includegraphics[width=\columnwidth]{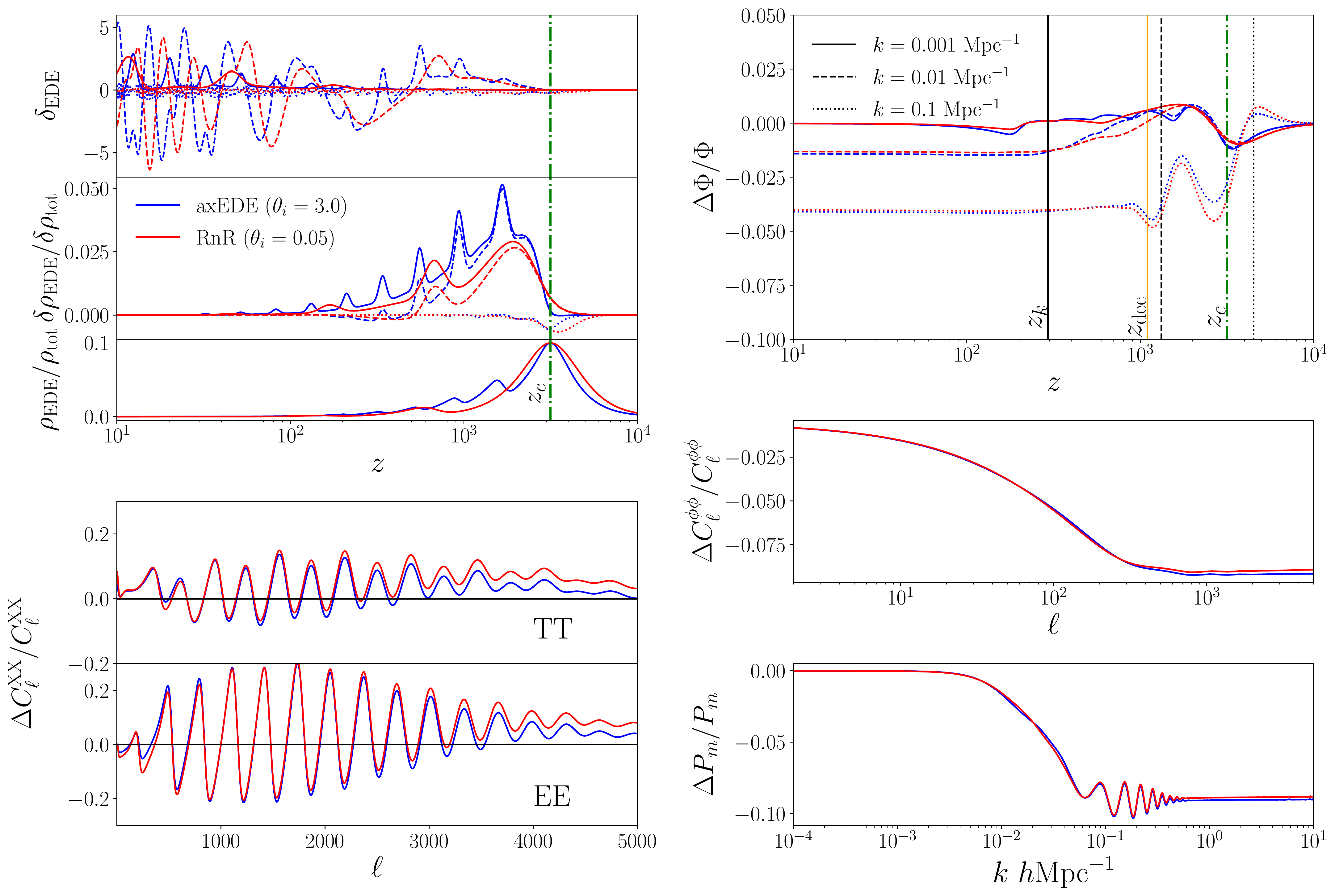}
    \caption{The same as Figs.~\ref{fig:pheno_EDE_comp} and \ref{fig:pheno_EDE_spectra} but for axEDE and RnR EDE. 
    For these, we fix $f_{\rm EDE} = 0.1$, $\log_{10}z_c = 3.5$ and the \lcdm\ parameters at the \textit{Planck} best-fit values. 
    }
    \label{fig:EDE_vs_RnR_comp}
\end{figure} 

We now apply these insights to a more realistic case and study the results of data analyses. 
In Fig.~\ref{fig:EDE_vs_RnR_comp}, we show the same quantities as in Figs.~\ref{fig:pheno_EDE_comp} and \ref{fig:pheno_EDE_spectra} but for axEDE and RnR EDE. 
The only difference between these models is the value of the initial field displacement, $\theta_i$: in axEDE the field has a relatively large displacement so that it starts in a flatter region of the potential, whereas for RnR EDE the field has a relatively small displacement such that the potential can always be approximated as a power law (in this case $V \propto \phi^6$). 

As discussed in detail in Ref.~\cite{Smith:2019ihp}, and from Eq.~\ref{eq:cs2_axEDE}, the choice of $\theta_i$ impacts the effective sound speed of the field, with a larger $\theta_i$ leading to a smaller effective sound-speed, and subsequently a decrease in pressure support. 
One can see this in the upper left panel of Fig.~\ref{fig:EDE_vs_RnR_comp}, with the fractional density perturbation for $\theta_i = 3$ reaching a higher maximum value than for $\theta_i = 0.05$. 
Note that the `spikes' in the axEDE case are caused by the `driving force' due to the background oscillations in the field. 
A similar feature is present in RnR, but with smoother spikes appearing at a lower frequency. 
The linear Klein-Gordon equation is sourced by the term $\phi' h_s'$, leading to a driving force that is 90$^\circ$ out of phase with respect to the oscillations of the background field, as can be seen in a comparison between the middle and lowest panels in the top left of Fig.~\ref{fig:EDE_vs_RnR_comp}. 

The difference in the pressure support for modes entering the horizon around $a_c$ has a similar impact on the CMB as previously described for the phenomenological case. 
However, unlike the comparisons shown in Fig.~\ref{fig:pheno_EDE_comp}, the change in $\theta_i$ not only affects the perturbations but also impacts the background evolution of the EDE. 
The fractional contribution of the EDE to the background energy density is larger in RnR EDE at high redshift than axEDE, leading to a larger increase in the damping scale, $r_d^{\rm RnR} = 44.50$ Mpc vs.~$r_d^{\rm axEDE} = 44.28$ Mpc, and to RnR EDE having more power compared to axEDE at small angular scales. Both  differences in the perturbative and background evolution between the two models lead to axEDE being preferred over RnR EDE when fit to data that includes the SH0ES value of $H_0$.

\section{Early Dark Energy in light of \Planck{} data}
\label{sec:EDE_Planck}

We now turn to the results of analyses of the EDE models presented in Sec.~\ref{sec:EDE_models} in light of up-to-date cosmological data, and in particular the \textit{Planck} CMB power spectra. 
Most EDE models, when fit to \textit{Planck}, produce posterior distributions that show similar degeneracies in the various parameters. 
In Sec.~\ref{sec:EDE_pheno_analysis},  we first discuss the phenomenological EDE model, also called the ADE model \cite{Lin:2019qug}, and draw generic conclusions about the background and perturbation dynamics preferred by the data. 
In Sec.~\ref{sec:axEDE_analysis}, we turn to the the axion-like EDE model, since it has been studied in depth in the recent literature and is a representative toy-model for EDE scalar fields. 
We discuss and compare results for other models in sec.~\ref{sec:other_EDE}. 

The baseline analysis includes the full \textit{Planck} 2018 TT,TE,EE and lensing power spectra \cite{Planck:2018vyg}, BAO measurements from BOSS DR12 at ${z = 0.38, 0.51, 0.61}$~\cite{Alam:2016hwk}, SDSS DR7 at $z = 0.15$~\cite{Ross:2014qpa} and 6dFGS at $z = 0.106$~\cite{Beutler:2011hx}, and a compilation of uncalibrated luminosity distances to  SN1a from Pantheon \cite{Scolnic:2017caz}. The SH0ES measurement\footnote{Recently, the SH0ES and Pantheon+ teams have provided updated data with a new likelihood that takes into account the co-variance between the different measurements. Dedicated EDE analyses in light of these data are still lacking for most models, but we anticipate the impact of the more refined likelihood to be fairly minor since the EDE models studied here do not predict large deviations in the shape of $H(z)$ at late-times. For a first application of this data to the specific ``axion EDE'' model, see Ref.~\cite{Simon:2022adh}. } is included through a Gaussian prior on the Hubble parameter $H_0$. The exact value of the prior may vary depending on the analysis considered, as the measurement has been updated regularly over the last three years. 

These analyses use the \textit{Planck} conventions for the treatment of neutrinos: two massless and one massive species are included with $m_{\nu} = 0.06$ eV \cite{Planck:2018vyg}. 
In addition, large flat priors are imposed on the dimensionless baryon energy density $\omega_b$, the dimensionless cold dark matter energy density $\omega_{\rm cdm}$, the Hubble parameter today $H_0$, the logarithm of the variance $\ln(10^{10}\mathcal{A}_s)$ of curvature perturbations centered around the pivot scale $k_p = 0.05$ Mpc$^{-1}$ (according to the \Planck{} convention), the scalar spectral index $n_s$, and the re-ionization optical depth $\tau_{\rm reio}$.

\subsection{What does it mean to resolve a tension?}

Before presenting results, let us briefly clarify what we mean by an extension of $\Lambda$CDM `resolves' the Hubble tension (here EDE), as there has been some debate on this topic in the recent literature. 
On the one hand, one may consider that  resolving the tension involves analysing a given model in light of a full compilation of datasets that {\it do not} include local measurements of $H_0$, and finding that the model predicts a higher value of $H_0$, in statistical agreement with the SH0ES (and other direct) determination. 
On the other hand, a more modest approach seeks to establish whether a model can provide a good fit to {\it all} the data (i.e. including SH0ES) and be favored over $\Lambda$CDM (by some measure of preference). 
As we discuss in the following, EDE models typically manage to achieve the latter definition of ``success'', but fail according the former. 
The current situation with EDE may appear unsatisfactory, leading to the conclusion that, as of yet, no proper resolution has been put forth.  

It has been argued that EDE's inability to predict a larger $H_0$ without including constraints from direct measurements is tied to the fact that EDE models introduce several new parameters, only one of which ($f_{\rm EDE}$) strongly correlated with $H_0$, while the others are undefined when $f_{\rm EDE}\to 0$.  
If the data (other than SH0ES) do not statistically significantly favor non-zero $f_{\rm EDE}$ by themselves, a fit to these data may be affected by `prior volume' effects: the auxiliary parameters will have no effect on the data, thus artificially increasing the $\Lambda$CDM-like volume, and leading to marginalized posteriors that ostensibly place strong upper limits on the presence of EDE, without being tied to a true degradation of the fit to the data (i.e. a decrease in the likelihood, or conversely a increase in the effective $\chi^2)$. 
On the other hand, once a direct determination of $H_0$ is included in the analysis, it can unveil regions of parameter space which both increases the indirect value of $H_0$ and provides a good fit to all data. 
One way to mitigate (or at least test) the impact of prior volume effects on a model is to perform ``profile likelihood'' analyses \cite{Lewis:2002ah,Audren:2012wb,Henrot-Versille:2016htt,Herold:2021ksg}.

In this review, we present results of analyses that both include and exclude direct measurements of $H_0$, in order to compare and contrast the conclusions that can be drawn from both approaches. 
In addition, we present recent results from likelihood profile analyses that show very different results from the standard Bayesian analyses in the absence of information from SH0ES. Let us note that it is generally recognized that this debate will become moot once near-future CMB and large-scale structure data, whose sensitivity to EDE models will be exquisite, become available.

\subsection{Preliminary study: results for a phenomenological EDE model}
\label{sec:EDE_pheno_analysis}

To gain some insight into the results for specific models of EDE, we begin by discussing results for the phenomenological EDE model (also dubbed "Acoustic Early Dark Energy (ADE)", see Sec.~\ref{sec:EDE_models}), whose dynamics were presented in the Sec.~\ref{sec:EDE_pheno}. 
We recall that this model is specified by the fraction $f_{\rm EDE}(z_c)$ of EDE, the critical redshift $z_c$ after which the field dilutes, the equation of state specified in Eq.~\ref{eq:w_EDE_pheno}, or more precisely by the exponent $n$ entering in the definition of $w_f$, and the effective sound speed $c_s^2$ of the EDE fluid.
The first questions we address with this preliminary study are: $i)$ what background dynamics can resolve the tension, i.e., what value of $f_{\rm EDE}(z_c)$, $z_c$ and $n$ (or equivalently $w_f$) are favored by {\it all} the data (including SH0ES), and  $ii)$ what are the constraints on the field perturbations and how do they correlate with the background dynamics. 
A similar analysis was first presented in Ref.~\cite{Lin:2019qug}.

\begin{figure}
    \centering
    \includegraphics[width=0.8\columnwidth]{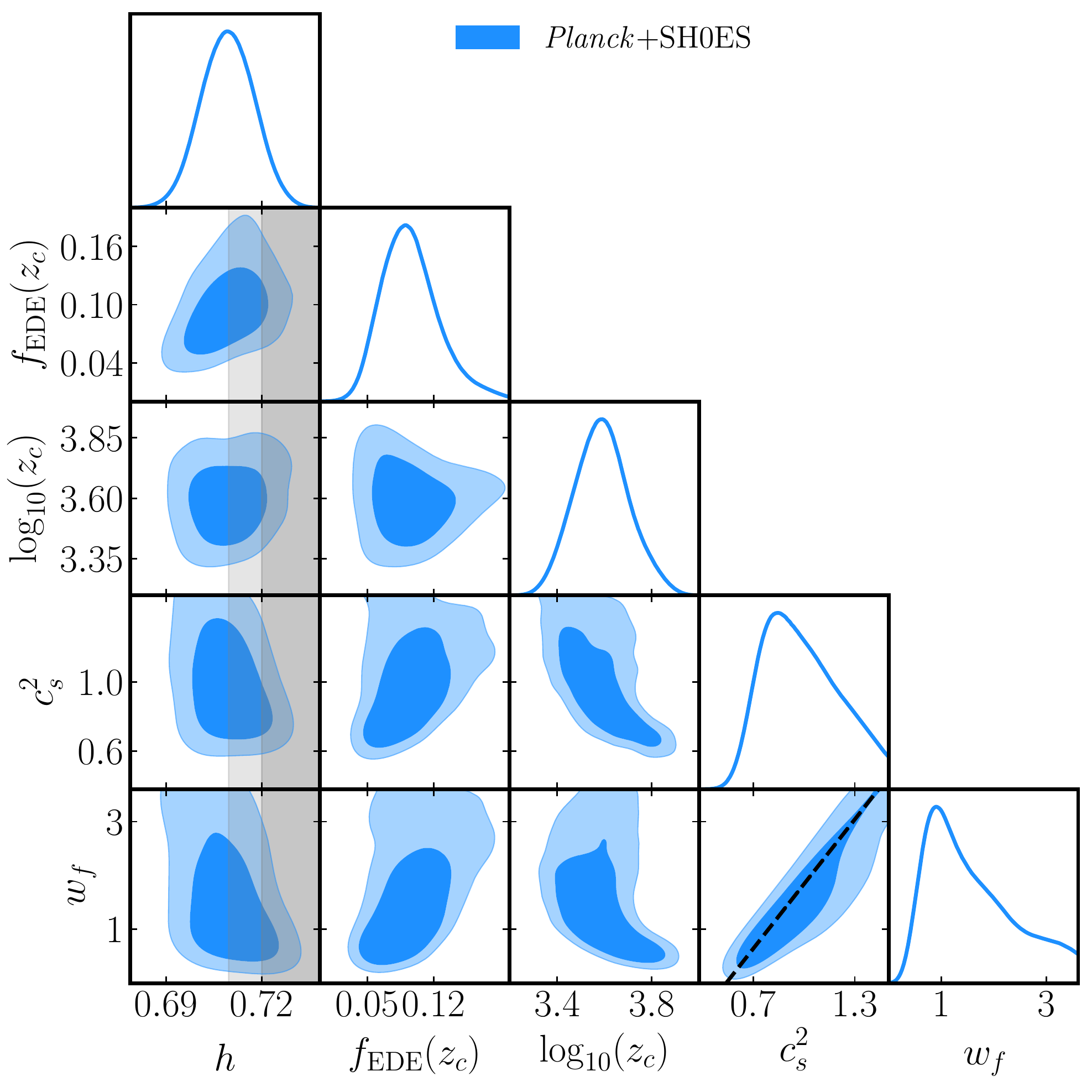}
    \caption{2D posterior distributions of $\{h,f_{\rm EDE}(z_c), z_c, w_f, c_s^2\}$ reconstructed from analyzing \Planck+BAO+Pantheon+SH0ES data. The black dashed line shows the degeneracy direction $c_s^2=0.54+0.25\times w$.
    }
    \label{fig:EDE_pheno}
\end{figure}

We show in Fig.~\ref{fig:EDE_pheno} the 2D posterior distributions of $H_0$ along with the four EDE parameters of the model $\{f_{\rm EDE}(z_c), z_c, w_f, c_s^2\}$, reconstructed from analyzing the combination \Planck{}+BAO+Pantheon+SH0ES data. We follow Ref.~\cite{Lin:2019qug} and impose the priors $c_s^2 \in [0,1.5]$ and $w_f\in [0.3.6]$.
Values of $w_f$ and $c_s^2 >1$  can be achieved with a non-canonical kinetic term \cite{Lin:2019qug}.

One can see that data favors $f_{\rm EDE}(z_c)=0.094_{-0.035}^{+0.023}$, with ${\rm log}_{10}(z_c)=-3.59_{-0.11}^{+0.13}$ and $w_f=1.46_{-0.99}^{+0.33}$.
Consequently, models where the EDE fluid dilutes like matter ($n=1$, $w_f = 0$) such as the standard axion are excluded at more than $2\sigma$. 
The rate of dilution correlates with ${\rm log}_{10}(z_c)$ and $f_{\rm EDE}(z_c)$, as later dilution (i.e. lower ${\rm log}_{10}(z_c)$) requires faster dilution (higher $w_f$) and a larger EDE contribution at the peak (larger $f_{\rm EDE}(z_c)$). 
Hence data requires a fine balance in EDE parameters for the EDE phase to last long enough to impact the value of the sound horizon, without strongly affecting modes that enter the horizon around $z_c$. 
In addition, perturbations are tightly constrained, with $c_s^2 = 0.97_{-0.28}^{+0.15}$, and correlate with the background equation of state along the degeneracy line $c_s^2=0.54+0.25\times w$ (shown with a black dashed line), with a mild preference for $c_s^2\lesssim 1$.
As we have discussed in Sec.~\ref{sec:EDE_pheno}, this is because the effect on the Weyl potential (and consequently on the acoustic driving of CMB perturbation) of faster rates of dilution can be compensated for by greater pressure support (larger $c_s^2$). These results update those presented originally in  Ref.~\cite{Lin:2019qug}.

The features exhibited by this phenomenological analysis are also present (broadly speaking) in more involved EDE modeling, as we discuss below. 

\subsection{Axion-like EDE in light of \Planck{} and SH0ES data}
\label{sec:axEDE_analysis}

\begin{figure}
    \centering
    \includegraphics[width=1\columnwidth]{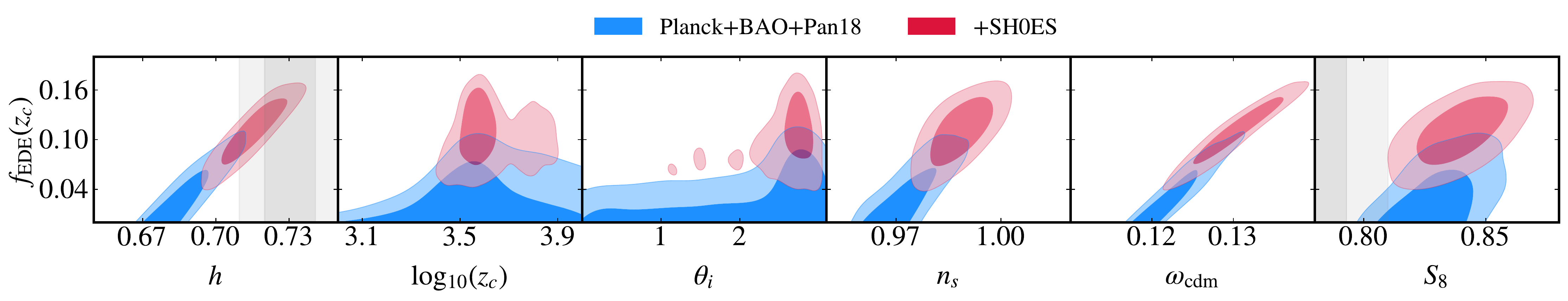}
    \caption{2D posteriors for $f_{\rm EDE}(z_c)$ vs $\{h,\log_{10}(z_c),\theta_i,n_s,\omega_{\rm cdm},S_8\}$ when analyzing {\it Planck}TTTEEE+BAO+Pantheon with and without the SH0ES prior ($H_0=73.04\pm1.04$ \cite{Riess:2021jrx}), as analyzed in Ref.~\cite{Simon:2022adh}.}
    \label{fig:EDE_Planck+SH0ES}
\end{figure}

We now turn to the well-studied axion-like EDE model, for which (in addition to the standard $\Lambda$CDM parameters) a logarithmic prior on $z_c$, and flat priors for $f_{\rm EDE}(z_c)$ and $\theta_i$ are considered as:
\begin{eqnarray}
       3  & \le ~~  \log_{10}(z_c) ~~  \le & 4, \nonumber \\
    0 & \le ~~ f_{\rm EDE}(z_c) ~~ \le & 0.5,  \nonumber \\
    0 & \le~~~~~~~\theta_i ~~~~~~~ \le & \pi\,.\nonumber
\end{eqnarray}

Note that, in principle, the exponent $n$ (or equivalently the equation of state once the field rolls $w_f=(n-1)/(n+1)$) could be considered as an extra free parameter of the model. 
We find that the combination of \Planck+BAO+Pantheon+SH0ES data yields $n=3.37_{-0.99}^{+0.41}$, updating the original result from Ref.~\cite{Smith:2019ihp}, and simply fixing the value of $n$ to a number chosen within this interval has very little impact on the model's performance at resolving the tension (see also Ref.~\cite{Agrawal:2019lmo} for a similar analysis in the RnR model). 
In the following, we present constraints setting $n=3$, as is the conventional choice in the literature. 
For results with the explicit choice\footnote{
It is worth stressing that the $n=2$ case leads to similar posteriors for $\{f_{\rm EDE}$, $H_0$, $z_c\}$, with only slightly worse $\chi^2$. The main difference lies in the reconstructed initial field value $\theta_i=1.53^{+0.84}_{-0.37}$, which as we have argued before, controls the effective sound speed $c_s^2$ and the shape of the energy injection.} 
$n=2$, we refer to the Appendix of  Ref.~\cite{Smith:2019ihp}.

We show in Fig.~\ref{fig:EDE_Planck+SH0ES} the 2D reconstructed posteriors for the parameters $$f_{\rm EDE}(z_c)~~{\rm  vs}~~ \{h,\log_{10}(z_c),\theta_i,n_s,\omega_{\rm cdm},S_8\}$$ when analyzing {\it Planck}TTTEEE+BAO+Pantheon with and without the SH0ES prior ($H_0=73.04\pm1.04$ \cite{Riess:2021jrx}), taken from Ref.~\cite{Simon:2022adh}. 
See also Refs.~\cite{Smith:2019ihp,Hill:2020osr,Murgia:2020ryi} for similar analyses with older datasets.
In the absence of a prior on $H_0$, the EDE model is not favored by the data in the Bayesian framework, and the combination {\it Planck}TTTEEE+BAO+Pantheon simply provides an upper limit on the EDE contribution $f_{\rm EDE}(z_c)< 0.091$, with $h=0.688^{+0.006}_{-0.011}$.  

However, as pointed out in various articles \cite{Murgia:2020ryi,Smith:2020rxx,Schoneberg:2021qvd,Herold:2021ksg}, posteriors are highly non-Gaussian with long tails towards high-$H_0$. 
These constraints should hence be interpreted with some degree of caution, and not naively scaled using an intuitive Gaussian law. 
This is further supported by the fact that the best-fit point lies at the $2\sigma$ limit of the reported constraints ($f^{\rm b.f.}_{\rm EDE}(z_c)= 0.088$ and $h^{\rm b.f.}=0.706$). 

Once the SH0ES prior is included in the analysis, one reconstructs $f_{\rm EDE}(z_c)=0.109^{+0.030}_{-0.024}$, at $\log_{10}(z_c) =3.599(3.568)^{+0.029}_{-0.081}$  with $h = 0.715\pm0.009$, which at face value is in tension with the results that do not include SH0ES. 
This is tied to the effect discussed in previous sections and the non-Gaussianity of the posterior (without SH0ES). 
To measure the level of tension with SH0ES, the following tension metric is suggested \cite{Raveri:2018wln,Schoneberg:2021qvd} 
\begin{equation}
    Q_{\rm DMAP}\equiv \sqrt{\chi^2({\rm w/~SH0ES})-\chi^2({\rm w/o~SH0ES})}\,
\end{equation}
(in units of Gaussian $\sigma$), which agrees with the usual Gaussian-metric tension for Gaussian posteriors, but can better capture any non-Gaussianity in the posterior.
This metric can easily be interpreted from the rule-of-thumb stating that for a single additional data point (SH0ES here), the $\chi^2$ value of a ``good'' model should not degrade by more than $\sim 1$ (for a rigorous discussion, see Ref.~\cite{Raveri:2018wln}). 
Applying this to EDE, one finds the tension metric $Q_{\rm DMAP}=1.9\sigma$, while the metric gives 4.8$\sigma$ in $\Lambda$CDM.
Additionally, in the combined analysis, one finds $\Delta\chi^2\equiv\chi^2_{\rm EDE}-\chi^2_{\Lambda{\rm CDM}}=-23.7$ with SH0ES and therefore a strong preference in favor of the EDE model (a simple estimate gives $\sim 4\sigma$ preference, assuming $\Delta\chi^2$ is $\chi^2-$distributed with 3 degrees of freedom).
Therefore, the EDE model is able to accommodate large values of $H_0$, while providing a good fit to {\it Planck}+BAO+SN1a. 
It is in this sense that {\it EDE can resolve the Hubble tension}. 

Although {\it Planck}+BAO+SN1a data alone do not favor EDE and predict large $H_0$ to claim of the discovery of new physics, 
it turns out that this is not unexpected. 
Mock analyses of {\it Planck} data alone that include an EDE signal have shown that they {\it cannot} detect even a $10\%$ contribution of EDE at $z_c\sim 3500$ and only lead to upper limits. 
Fortunately, an experiment like CMB-S4 would unambiguously detect such a signal \cite{Smith:2019ihp}.

Let us also highlight that data provide a strong constraint on the initial field value $\theta_i$ (in units of $f$, the axion decay constant). 
As shown in Sec.~\ref{sec:EDE_pheno}, $\theta_i$ has two main impacts on the EDE model - its dominant impact is setting the effective sound speed at the perturbation level, and a secondary effect is modifying the shape of the energy injection at the background level. 
We recall that for an oscillating EDE, the sound speed can be both time- and scale-dependent (see Eq.~\ref{eq:cs2_axEDE}). 
In fact, the data prefer an EDE for which modes inside the horizon around $z_c$ have effective sound-speed $\lesssim 0.9$ \cite{Smith:2019ihp}, consistent with phenomenological ADE results in the previous section which favor $c_s^2<1$ for $w_f\sim 1/2$. This can be achieved if the potential is flat around the initial field value $\theta_i$ such that the term $\theta_i^{n-1}/\sqrt{E_{n,\theta\theta}(\theta_i)}\gg 1$ in Eq.~\ref{eq:cs2_axEDE}. 
Since the range of $k-$modes within the horizon is a sharp function of the initial field value $\theta_i$, data provide a fairly strong  constraint on $\theta_i \sim \pi$ (see Fig.~\ref{fig:EDE_Planck+SH0ES}). 
We recall that in addition, $\theta_i$ also contributes to the shape of the energy injection, see Fig.~\ref{fig:EDE_vs_RnR_comp}, that in turns affects the sound horizon and damping scales.
These constraints on the dynamics, and as a result on the shape of the potential, explain why simpler power-law potentials ( as in the Rock'n'Roll model) fair less favorably in resolving the tension. 

Finally, translating the phenomenological parameters back into the theory parameters, the axEDE model suggests the existence of an axion-like field with mass $m \simeq 10^6 H_0$, decay constant $f\simeq 0.15 M_{\rm pl}$, and initial field value $\phi_i/f \sim \pi $. 
This prompted studies to understand whether the near-Planckian values of the decay constant and field excursion can lead to additional phenomenology (as in the EDS model \cite{McDonough:2021pdg, Lin:2022phm}) and constraints, as it might violate the axion weak gravity conjecture \cite{Kaloper:2019lpl,Rudelius:2022gyu}.

\subsection{Profile likelihood analysis of axEDE }
\label{sec:prof}

A question often arises about the inclusion of SH0ES data for EDE constraints and for comparison to \lcdm: as discussed earlier, ideally, the proposed model when analyzed with CMB data alone should predict a higher $H_0$, improve the $\chi^2$ more than the additional degrees of freedom and be preferred over \lcdm. 
While the best-fit EDE with CMB data alone does exhibit the first two features, the MCMC posteriors are broad and show no $H_0$ solution, nor a preference for EDE. In fact, they ostensibly exclude the amount of EDE favored by the analysis that includes SH0ES. 
This signals the influence of `prior-volume effects'.

Indeed, a common feature of all EDE models is that, as the amount of EDE $f_{\rm EDE} \to 0$, the other EDE parameters (e.g., $z_c$, $\theta_i$, etc) become unconstrained. 
In the literature, it has been argued that this leads to a large prior volume at $f_{\rm EDE} \simeq 0$, impacting results from MCMCs and skewing them towards $f_{\rm EDE} \to 0$ \cite{Smith:2020rxx,Murgia:2020ryi,Herold:2021ksg}. 
These posteriors however do not trace the true data likelihood at each point in the extended \lcdm+EDE parameter space\footnote{For further discussion about the mitigation of projection and prior volume effect, see Refs.~\cite{Gomez-Valent:2022hkb,Hadzhiyska:2023wae}.}. 
While the literature largely concentrates on employing Bayesian techniques like MCMCs to constrain EDEs, alternate methods like likelihood profiles provide additional insight into the data response to these new models. 

Constraints derived with a profile likelihood approach (e.g. \cite{Herold:2021ksg,Herold:2022iib,Reeves:2022aoi,Holm:2022kkd}) can lead to confidence intervals that are very different from the Bayesian credible intervals\footnote{Strictly speaking, there are several ways to define credible intervals from a posterior distribution. 
For instance, one may choose to have the same fraction of samples at
both ends of the distribution, or require that the value of the marginalized probability be the same at each limit.  }. 
Profile likelihoods, rooted in the ``frequentist'' framework, trace the 1D data likelihood along a parameter direction - they optimise the $\chi^2$ over all other parameters while holding the parameter of interest fixed. 
Constraints for that parameter can then be extracted directly from this $\chi^2$ curve, with $\Delta \chi^2 \simeq 1$ giving the $1\sigma$ region in the case of a Gaussian profile when far from the prior boundaries on the parameter \cite{Neyman:1937uhy}\footnote{In a 
more generic (non-Gaussian) case, one can use the Feldman-Cousins prescription to define confidence intervals \cite{Feldman:1997qc}. For EDE, these match the simple Gaussian prescription \cite{Herold:2021ksg}.}. 
For parameter spaces such as \lcdm\ analysed using CMB data, where the profile is Gaussian and overwhelmingly dominates the prior contribution, constraints from profile likelihoods match those from MCMCs \cite{Planck:2013nga}. 
However for EDE, affected by prior-volume effects, these deviate \cite{Herold:2021ksg,Herold:2022iib}. 

\begin{figure}
    \centering
    \includegraphics[width=0.8\textwidth]{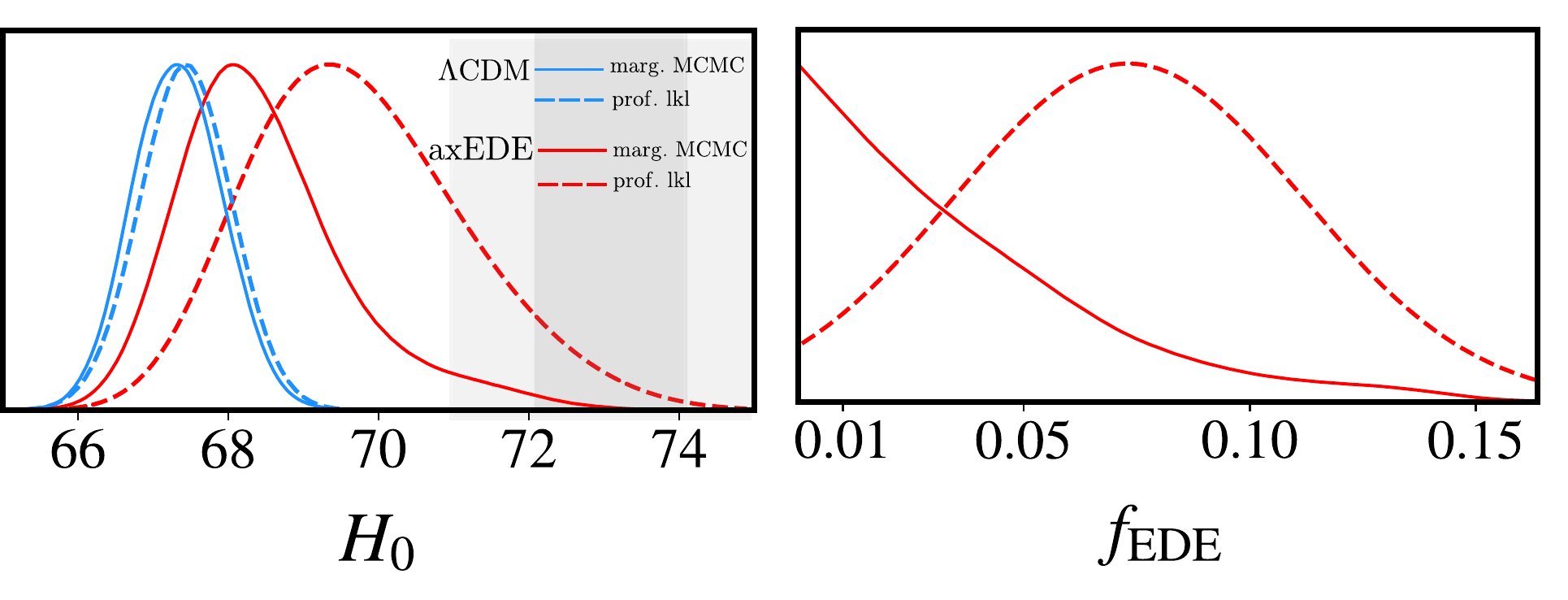}
    \caption{
    Posteriors obtained from a profile likelihood (dashed) and MCMC methods (solid) are shown for \lcdm\ (blue) and axEDE (red) for $H_0$ and $f_{\rm EDE}$, for \textit{Planck} 2018 CMB TTTEEE data.
    For the profile likelihoods, we plot the normalised curves of $e^{-\chi^2/2}$. 
    While the posterior and profile likelihood match under \lcdm, the profile likelihood indicates both a higher $H_0$ and $f_{\rm EDE}$ under the axEDE model. 
    Note that although the location and width of the dashed curves will differ for different EDE models, this upward trend remains as they all are affected by prior-volume effects. 
    Most notably, while the MCMC 1D posterior for $f_{\rm EDE}$ shows consistency with $f_{\rm EDE} = 0$, the profile likelihood shows preference for EDE at $\gtrsim 2\sigma$. 
    }
    \label{fig:prof_lkl_h_fede}
\end{figure}

While profile likelihoods were already suggested as a check of potential prior-volume effects \cite{Lewis:2002ah,Audren:2012wb}\footnote{These references suggest using the mean likelihood within a given bin to estimate the prior volume effect, a quantity that is simple to estimate directly from the MCMC sample, but that does not necessarily carry a well-defined statistical meaning.}, they have received much attention recently in the context of EDE constraints. 
In Fig. \ref{fig:prof_lkl_h_fede}, we compare the MCMC 1D posteriors for $H_0$ and $f_{\rm EDE}$ in \lcdm\ and axEDE to the profile likelihoods of the same parameters for \textit{Planck} 2018 CMB TTTEEE data. 
There is little difference between posteriors and likelihood profiles of \lcdm\ from the two approaches \cite{Planck:2013nga}. 
On the axEDE front however, there is stark contrast - tracing the true data likelihood leads to wider constraints on both $H_0$ and $f_{\rm EDE}$. 
Moreover, data show preference for higher $H_0$ and $f_{\rm EDE}$. As a result the (frequentist) confidence and (Bayesian) credible intervals extracted from the two strongly differ. 
While an MCMC analysis concludes that $f_{\rm EDE}$ is consistent with 0, a profile likelihood shows preference for $f_{\rm EDE} > 0$ at over $2\sigma$ without a SH0ES $H_0$ prior. 

This preference is already hinted at in the vast EDE literature that does not utilise profile likelihoods, by simply quoting the best-fit point (which by definition corresponds to the peak of the profile likelihood).
Consistently, the best-fit point in EDEs does not coincide with the maximum of the MCMC posterior when an $H_0$ prior is excluded - that is, the greatest density of posterior-sampled points is not co-located with the best-fit point or the maximum of the likelihood. 
In contrast, for \lcdm, the best fit for any parameter is very close to the maximum of the 1D MCMC posterior. 
This raises the question of which is the most appropriate tool to use to constrain EDEs, or at least, suggests that both analyses methods should be performed to test the robustness of a constraint: when they diverge, care should be taken in strongly interpreting either result\footnote{We stress that we do not claim profile likelihoods are definitively better than the Bayesian approaches, as the latter analyses also carry an estimate of ``Occam's razor'' by disfavoring models with unnecessary extra parameters, an aspect also known as the ``look-elsewhere effect'' in Frequentist terms and that is complicated to estimate in general. 
However, it can be problematic to exclude models based purely on their complexity when they, in fact, provide a quality fit to the data (and can, in this specific context, accommodate direct $H_0$ measurements).}.

As attention turns to profile likelihoods for EDE constraints, we note that for model-comparison purposes, one might still rely on Bayesian tools that account for the increase in parameter dimensions, using Bayesian evidence criteria, bearing in mind their dependence on the prior. 
Alternative methods have been suggested to overcome this limitation (e.g. Refs.~\cite{Kass:1995loi,Raveri:2018wln,Raveri:2021wfz,schwarz1978estimating,2013PDU.....2..166V,Handley:2019wlz,Hergt:2021qlh}). 
In principle, the Frequentist framework can also test whether the preference for a deviation from $\Lambda$CDM is statistically significant, given $N$ number of additional degrees of freedom (taking into account the ``look-elsewhere effect''). 
This can be done assuming the $\Delta\chi^2$ follows a $\chi^2$ distribution with $N$ degrees of freedom. 
In the case of EDE, because the parameters $\{z_c,\theta_i\}$ are unconstrained (i.e. have no effects on the likelihood) as $f_{\rm EDE}\rightarrow 0$, this test statistic does not fully encapsulate the true significance, as required by Wilks' theorem \cite{Wilks:1938dza}. 
Still, it yields results more conservative than {\it local significance} tests which compute the preference at fixed $\{z_c,\theta_i\}$ with a single degree of freedom.  More detailed analyses estimating the true significance, for instance following Refs.~\cite{Gross:2010qma,Ranucci:2012ed,Bayer:2020pva} or dedicated mock data analyses are still lacking.

\subsection{Role of \Planck{} polarization }

\begin{figure}
    \centering
    \includegraphics[width=0.49\columnwidth]{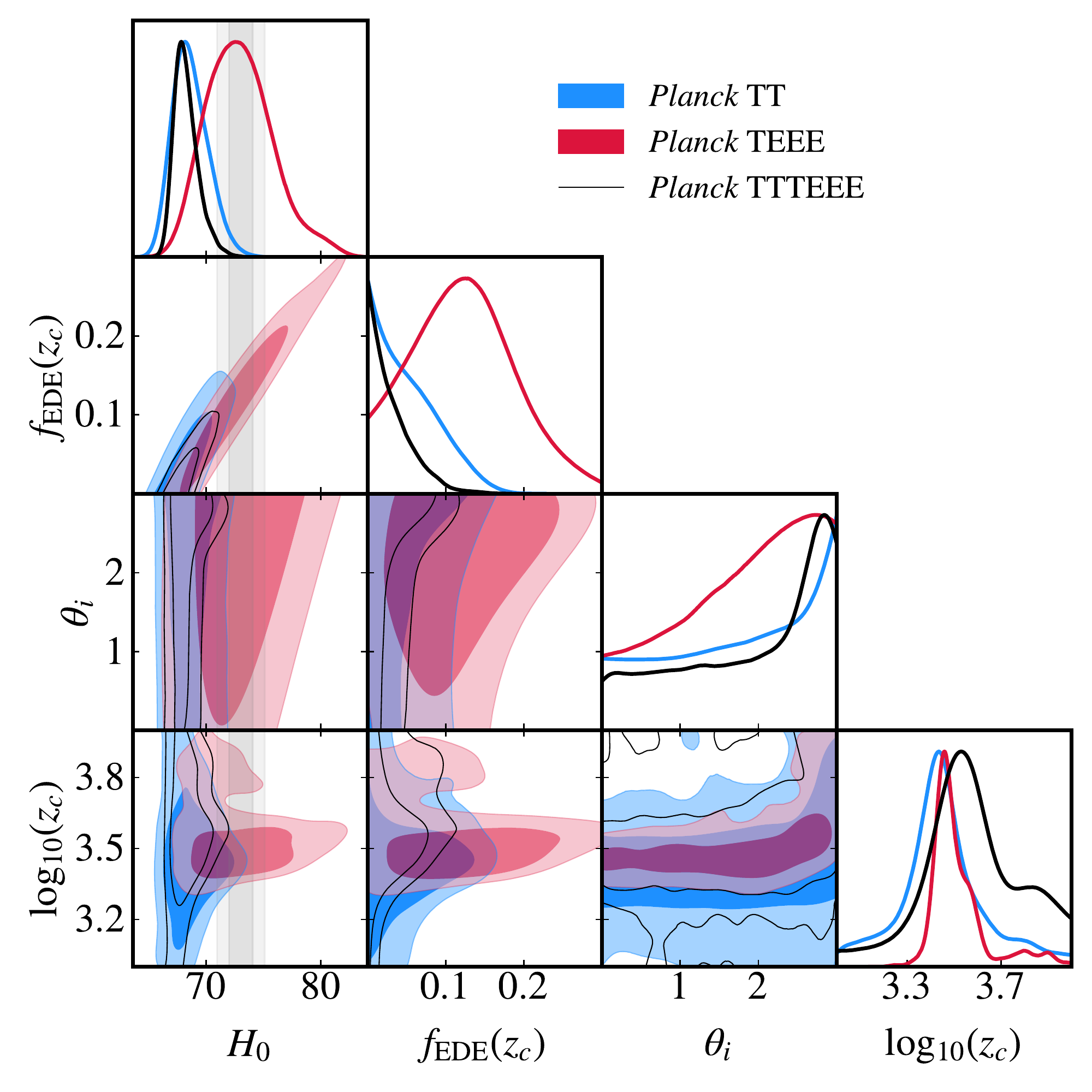}
        \includegraphics[width=0.49\columnwidth]{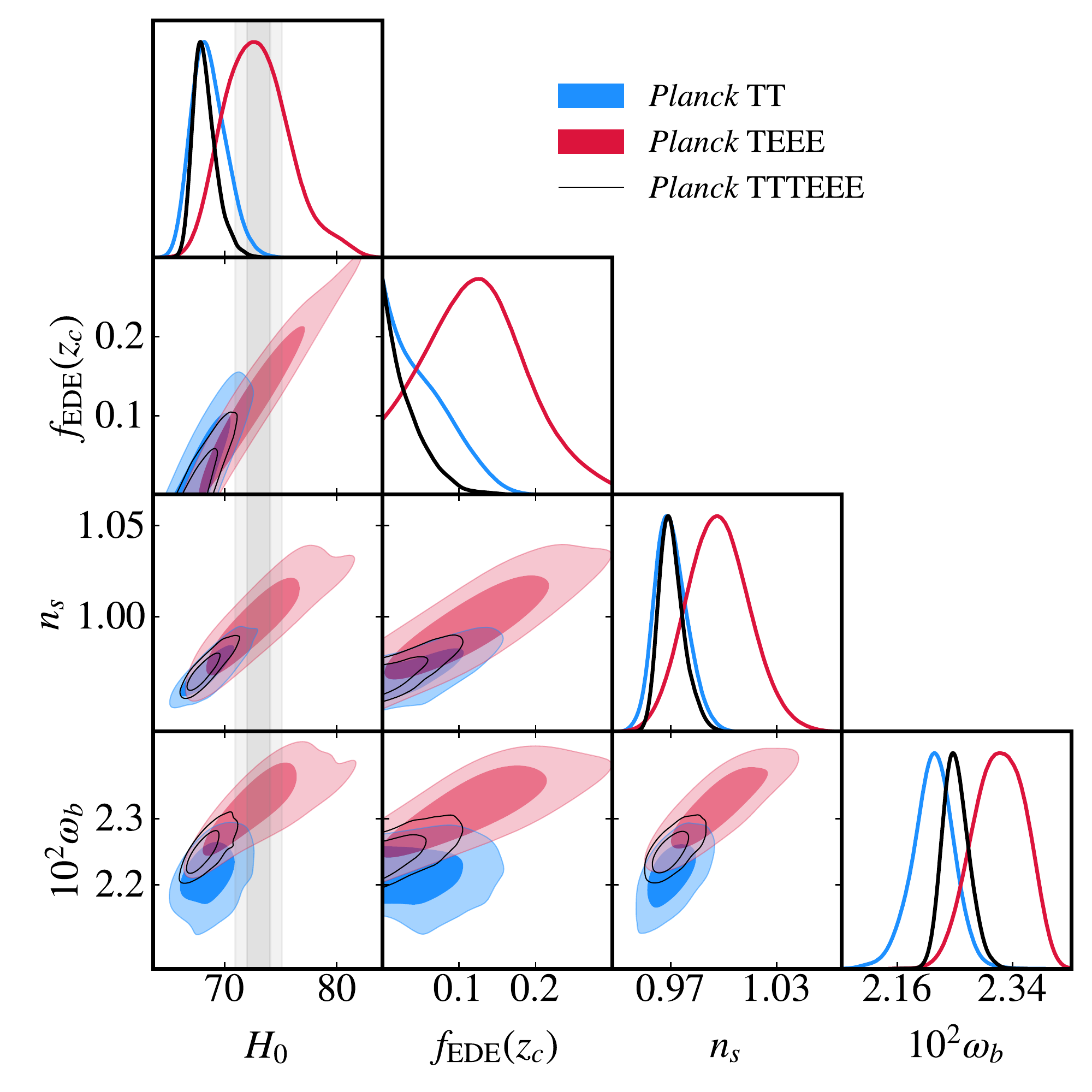}
    \caption{ Posterior distributions of $\{H_0,f_{\rm EDE}(z_c),\theta_i,\log_{10}(z_c)\}$ (left panel) in the axEDE cosmology from analyzing \Planck TT and \Planck TEEE. 
    The right panel replaces $\theta_i$ and $\log_{10}(z_c)$ with the posterior distributions of $n_s$ and $\omega_b$. 
    }
    \label{fig:PlanckTT_TEEE_EDE}
\end{figure}

To understand the origin of the preference for EDE from \Planck\ alone, it is instructive to compare results from TT and TEEE separately. 
The reconstructed posterior distributions of $\{H_0,f_{\rm EDE}(z_c),\theta_i,\log_{10}(z_c)\}$ are shown in the left panel of Fig.~\ref{fig:PlanckTT_TEEE_EDE}. 
Interestingly, one can see that {\it Planck} TEEE favors non-zero $f_{\rm EDE}(z_c)$, at the $2\sigma$ level, with $z_c$ and $H_0$ values in good agreement with the results of analyses that include SH0ES.
{\it Planck}TT data on the other hand do not show preference for EDE, although when explored alone, the constraints on EDE are weak with $f_{\rm EDE}\lesssim 0.15$ at 95\% C.L..
It is intriguing that the combination of {\it Planck}TT and TEEE data lead to significantly stronger constraints than what one may naively expect from the individual constraints.  
To better understand this, in the right panel of Fig.~\ref{fig:PlanckTT_TEEE_EDE}, we show the $\Lambda$CDM parameters that are most discrepant between TT and TEEE. 
The origin of these EDE constraints seems to be tied to the mild disagreement in the $n_s$ and $\omega_b$ posteriors.  

\begin{figure}
    \centering
    \includegraphics[width=0.8\columnwidth]{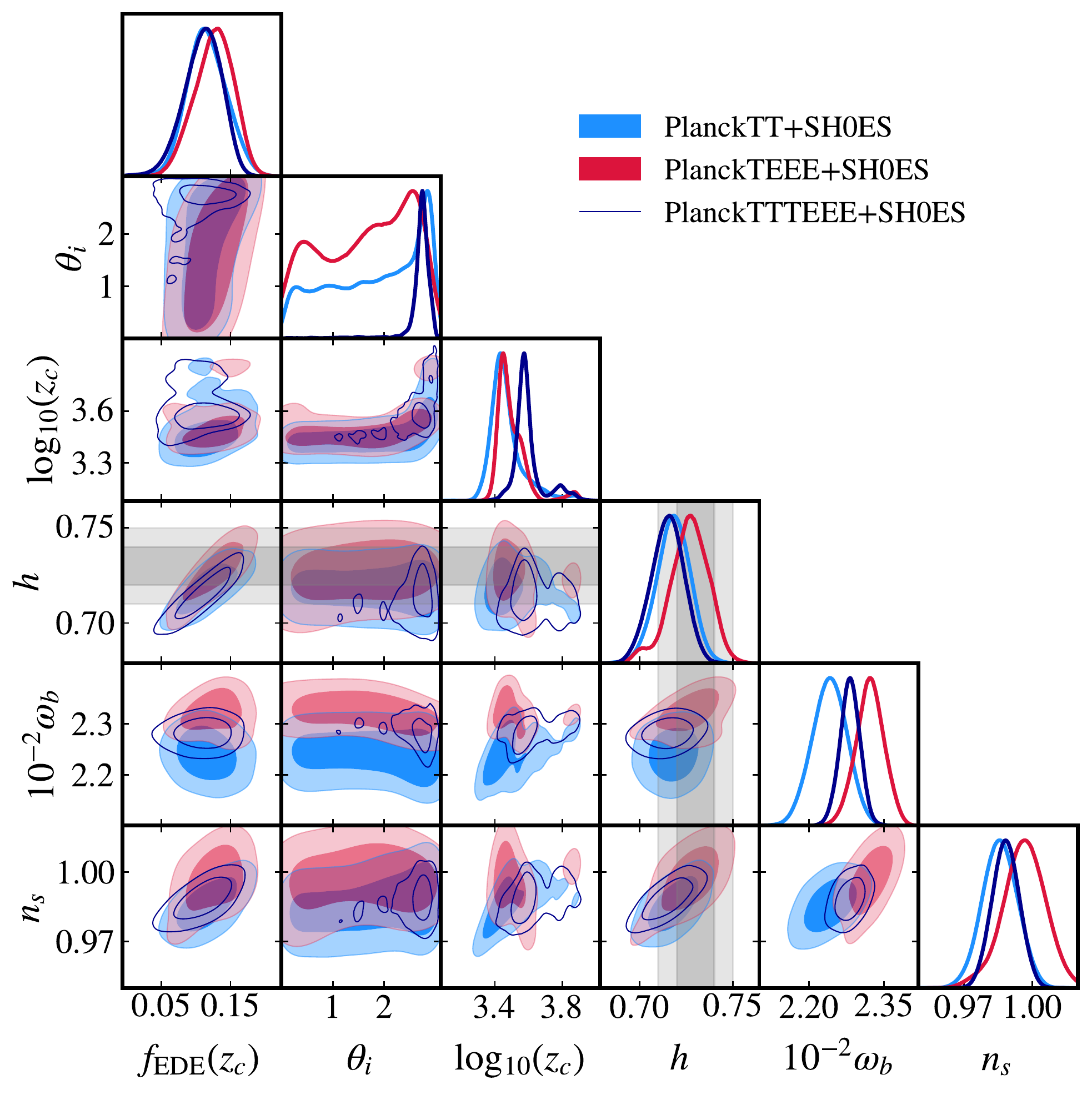}
    \caption{2D posteriors of $\{f_{\rm EDE}(z_c),\log_{10}(z_c),\theta_i, h, \omega_b, n_s\}$ reconstructed from analyzing either \Planck TT or TEEE or the combination TTTEEE data, along with the SH0ES prior. 
    All analyses also include BAO and Pantheon data. }
    \label{fig:EDE_TT-vs-TEEE_rec}
\end{figure}

To contrast the role of TT and TEEE data in constraining EDE, and the importance of the mild $\omega_b$ mismatch between these data, it is also instructive to perform an analysis of TT and TEEE data separately, including the prior on $H_0$. 
We show in Fig.~\ref{fig:EDE_TT-vs-TEEE_rec} the reconstructed 2D posterior of $\{f_{\rm EDE}(z_c),\log_{10}(z_c),\theta_i, h, \omega_b, n_s\}$ in these analyses. 
Evidently, both datasets individually allow for essentially any value of $\theta_i$, and slightly favor lower values\footnote{ 
The combined analysis favors $\log_{10}(z_c)\sim 3.5$, while both separate analyses favor $\log_{10}(z_c)\sim 3.4$. 
Yet, values are compatible at $1\sigma$ and always roughly around $z_{\rm eq}$. 
This seems to result from its correlation with $\theta_i$ visible in Fig.~\ref{fig:EDE_TT-vs-TEEE_rec}.
} of $z_c$. 
It is only when they are combined that a very narrow range of $\theta_i$ values is favored, at the intersection of the posteriors at $\sim 1\sigma$ in the $\omega_b-\theta_i$ plane. 
Therefore, it is possible that high field values being favored by the data is driven by either small residual systematic errors in either TT or TEEE (or both), or by a statistical fluctuation that also drives the mismatch in $\omega_b$. Since the preference for large $\theta_i$ indicates that pure power-law potentials are disfavored,
this can be important from the model-building perspective, as pure power-law potentials may be more easily motivated than the axion-like cosine potential considered in this study. 

Ref.~\cite{Smith:2022hwi} demonstrated that the preference in {\it Planck}TEEE data is sensitive to assumptions in modeling the {\it Planck}TE polarization efficiency (PE) calibration.
There are two techniques for setting the TE PE parameters that should give equivalent results, but in practice, estimates in {\it Planck} are slightly discrepant at the $\sim 2\sigma$ level (see Eqs. (45) -  used as a baseline, and (47) of Ref.~\cite{Planck:2018vyg}). 
Ref.~\cite{Smith:2022hwi} found that the {\it Planck} preference for EDE decreases when the TE PE parameters are fixed to the non-standard values. 
Interestingly, the shift in this nuisance parameter goes in the same direction as results from the ACT collaboration, which found that a potential systematic error in their TE spectra can slightly reduce the preference for EDE within ACT DR4 data \cite{Hill:2021yec}.

It has also been noted that the galactic-dust contamination amplitude is strongly correlated with the primordial tilt $n_s$. 
The fiducial analysis does not treat the galactic-dust contamination amplitude as a free parameter, but given the mild discrepancy seen in Fig.~\ref{fig:PlanckTT_TEEE_EDE} in the determination of $n_s$, it is interesting to test how freeing these parameters can affect the preference for EDE. 
Ref.~\cite{Smith:2022hwi} showed that this only has a marginal effect on the preference for EDE (slightly increasing it).

\subsection{Understanding the EDE-$\Lambda$CDM degeneracy in the CMB}
\label{sec:EDE_LCDM_deg}

\begin{figure}
    \centering
    \includegraphics[width=1\columnwidth]{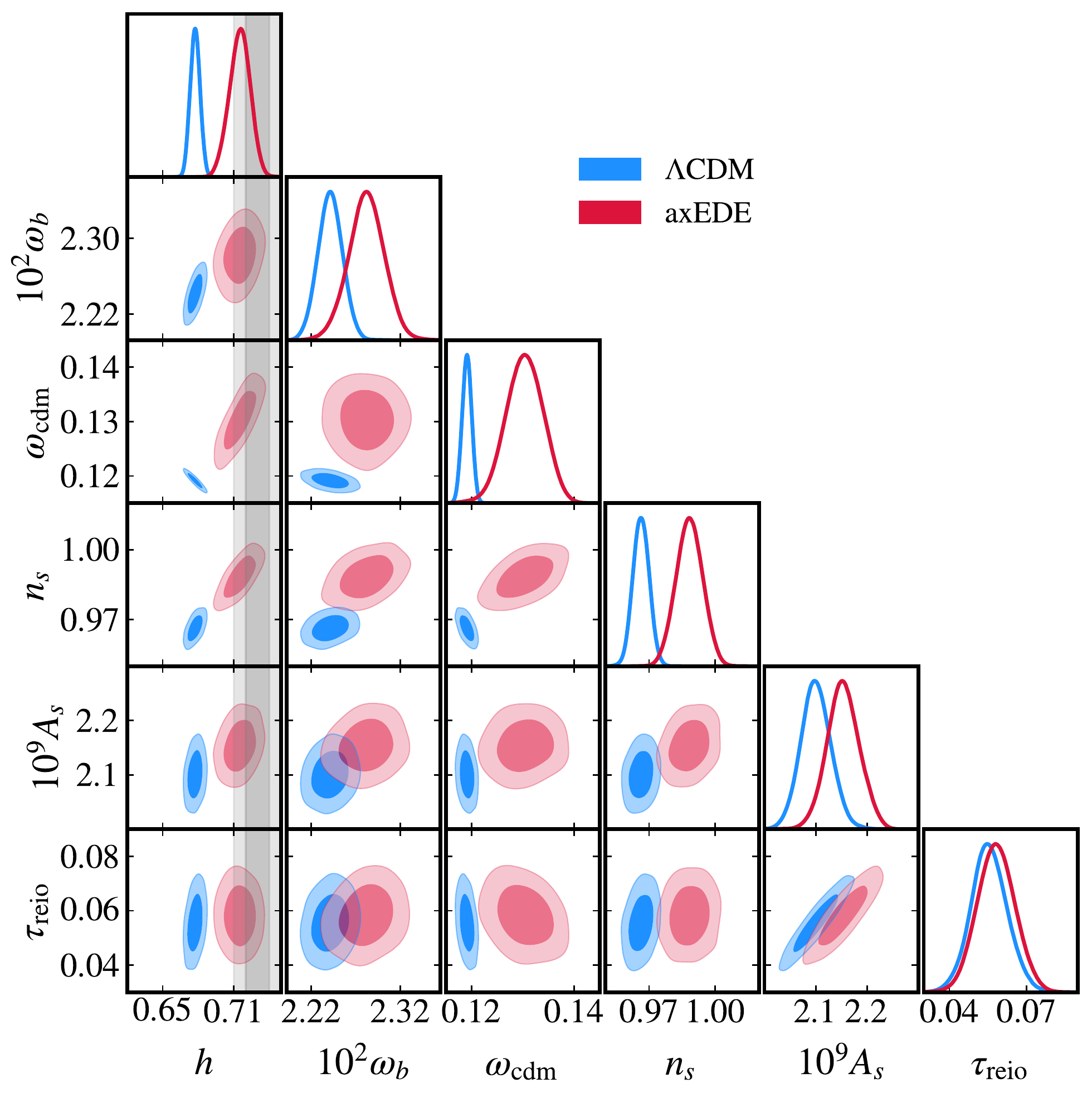}
    \caption{2D posteriors for the six $\Lambda$CDM parameters in the concordance $\Lambda$CDM and axEDE models fit to \Planck+BAO+Pantheon data, illustrating how parameters reshuffle to accommodate the EDE contribution and a higher $h$.}
    \label{fig:LCDM-vs-EDE}
\end{figure}

\begin{figure}
    \centering
    \includegraphics[width=0.8\columnwidth]{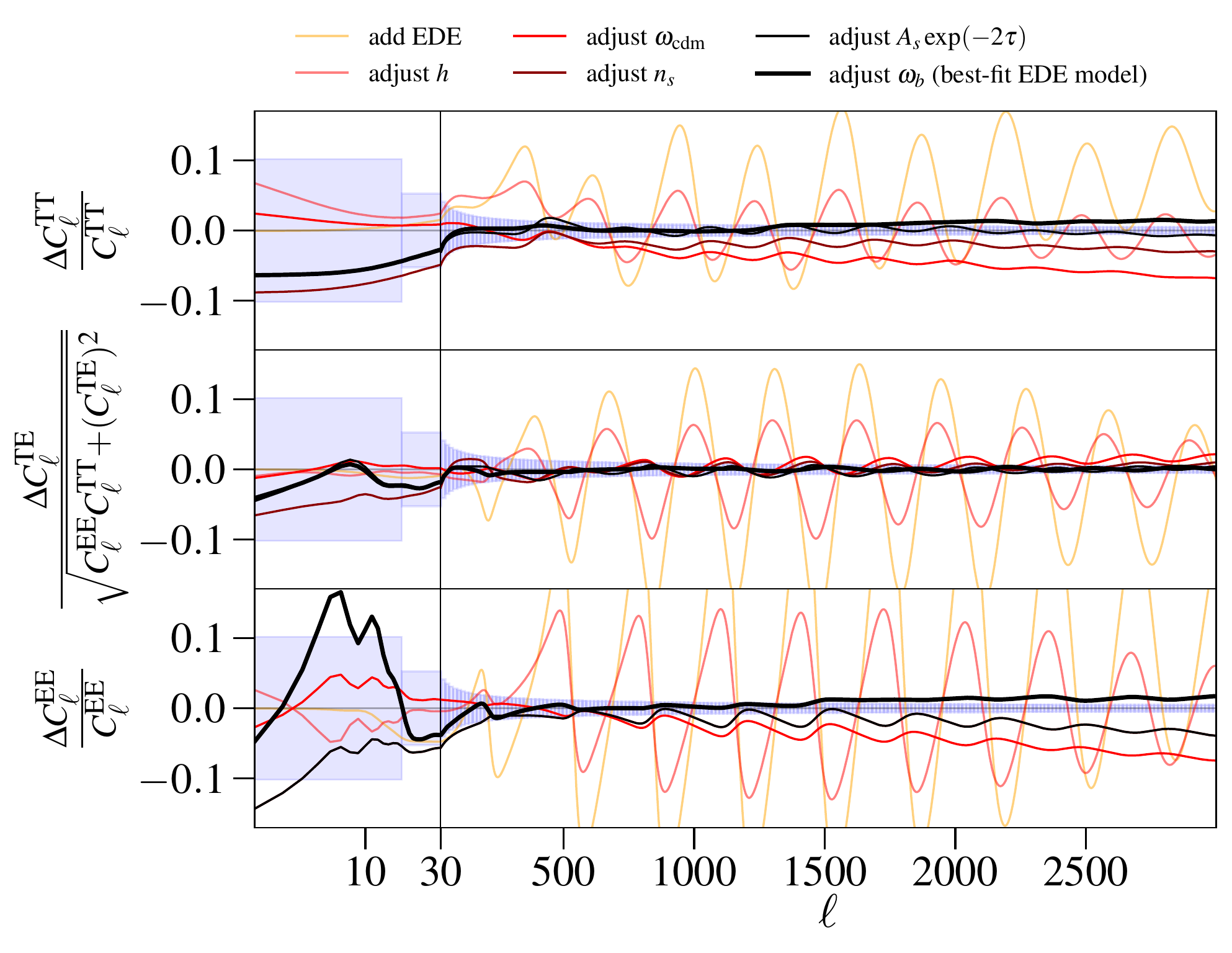}
    \includegraphics[width=0.8\columnwidth]{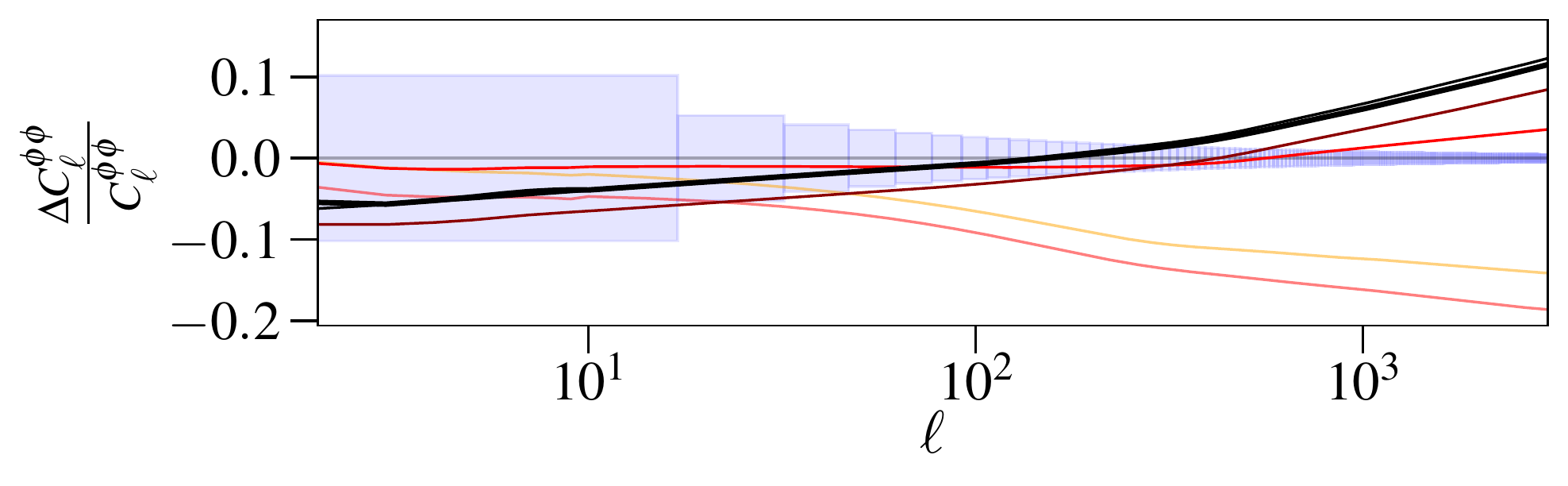}
    \includegraphics[width=0.8\columnwidth]{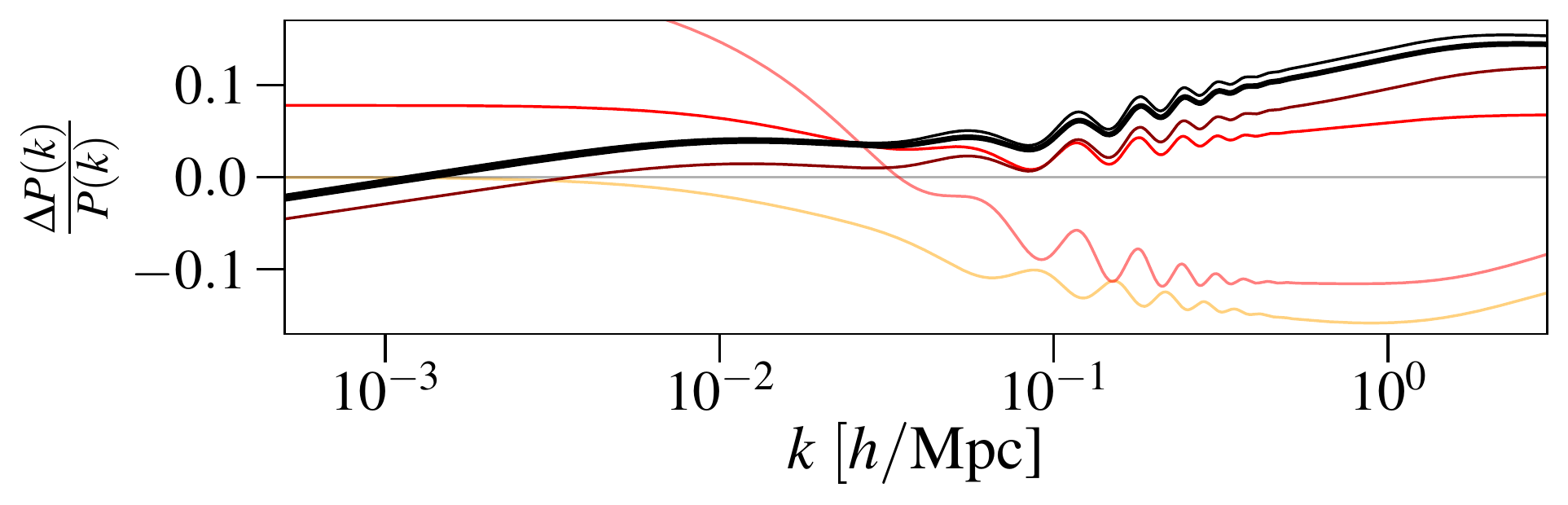}
    \caption{From top to bottom: Evolution of the residual CMB TT, TE, EE, lensing $\phi\phi$ and linear matter power spectra as cosmological parameters are adjusted to the best-fit EDE cosmology one-by-one. 
    The best-fit $\Lambda$CDM model to \Planck{} data is taken as reference. 
    Note that for TT, TE and EE we follow {\it Planck}'s convention and plot using a logarithmic scaling from $\ell = 2-30$ and a linear scaling from $\ell = 30-3000$. 
    The fit to {\it Planck} TTTEEE data provided by the axEDE model is smaller by $\Delta\chi^2=-6.4$ compared to that of $\Lambda$CDM.}
    \label{fig:CMBresiduals}
\end{figure}

Fig.~\ref{fig:EDE_Planck+SH0ES} shows that incorporating EDE and a higher $h$ also require readjusting other cosmological parameters to fit CMB data. We compare all six $\Lambda$CDM parameters in the concordance $\Lambda$CDM model when fit to \Planck+BAO+Pantheon data to the axEDE model fit to \Planck+BAO+Pantheon+SH0ES in Fig.~\ref{fig:LCDM-vs-EDE}.

One can see that (apart from $h$) $\omega_{\rm cdm}$ and $n_s$ are shifted the most between the $\Lambda$CDM and EDE cosmologies. 
To illustrate the degeneracy between EDE and these cosmological parameters, we show in Fig.~\ref{fig:CMBresiduals} the evolution of the residual CMB TT and EE power spectra as cosmological parameters are adjusted to the best-fit EDE cosmology one by one. 
The best-fit $\Lambda$CDM model to \Planck{} data is taken as reference.

When EDE is added to the Universe (orange curve), the most striking effect is a shift in the position of the peaks, as expected from the reduction of the size of the sound horizon induced by EDE and discussed in detail in Sec.~\ref{sec:EDE_pheno}.
This effect is more strongly visible in the polarization spectra.
The increase in $h$, from $\sim0.67$ to $\sim0.72$ partly compensates this shift, reducing the amplitude of oscillations (light red curve). 
In principle, $h$ could be further increased to compensate the offset.
However, the power at $\ell\sim500$ is significantly larger than in $\Lambda$CDM. 
As discussed in Sec.~\ref{sec:EDE_pheno}, this corresponds to the multipole range where the effect of EDE perturbations on the acoustic driving of oscillations in the photon-baryon fluid and the time-dependence of the gravitational potential close to recombination (affecting the eISW term) is the most prominent. 
In particular, the gravitational potential tends to decay faster than in $\Lambda$CDM, leading to a notable increase of the eISW power at intermediate multipoles. 
In the EDE cosmology, this can be compensated for by increasing the CDM density, which stabilizes the potential decay and reduces the amplitude of the eISW effect (dark red curve) \cite{Vagnozzi:2021gjh}. 
The effect of CDM on the angular diameter distance also compensates the reduced sound horizon, and the residual oscillation pattern further decreases. 
Note that, while the degeneracy with $\omega_{\rm cdm}$ can balance the eISW effect introduced by EDE, it also limits the ability of the model to reach very high-$H_0$ due to the impact on the angular diameter distance.

Once $h$ and $\omega_{\rm cdm}$ have been adjusted, one can see that the spectrum shows a strong tilt, due to a different diffusion angular scale $\theta_d$: the effect of EDE on the angular sound horizon is stronger than on the damping scale, and therefore the increase in $h$ and $\omega_{\rm cdm}$ cannot simultaneously keep $\theta_d$ and $\theta_s$ fixed. 
This effect is similar to that of an extra radiation species $\Delta N_{\rm eff}$, and can be compensated for by increasing $n_s$ (brown curve). 
This is a remarkable consequence of the EDE cosmology: $n_s\sim 1$ is allowed at 2$\sigma$, impacting the viability of models of inflation and the evidence for a slow-roll phase of inflation in EDE cosmologies \cite{Takahashi:2021bti,Cruz:2022oqk}. 
In fact, such large values of $n_s$ can be probed by future measurements of CMB spectral distortions, offering a way to test the EDE scenario \cite{Lucca:2020fgp}.
The resulting residual shows an overall offset in amplitude, that can be compensated for by adjusting the combination $A_s\exp(-2\tau_{\rm reio})$ (thin black curve). 
Finally, the remaining oscillations (particularly visible in polarization) are due to an offset in the ratio $\omega_b/\omega_{\rm cdm}$ from increasing $\omega_{\rm cdm}$ earlier. 
A slight increase in $\omega_b$ compensates for these. 

\begin{figure}
    \centering
    \includegraphics[width=\columnwidth]{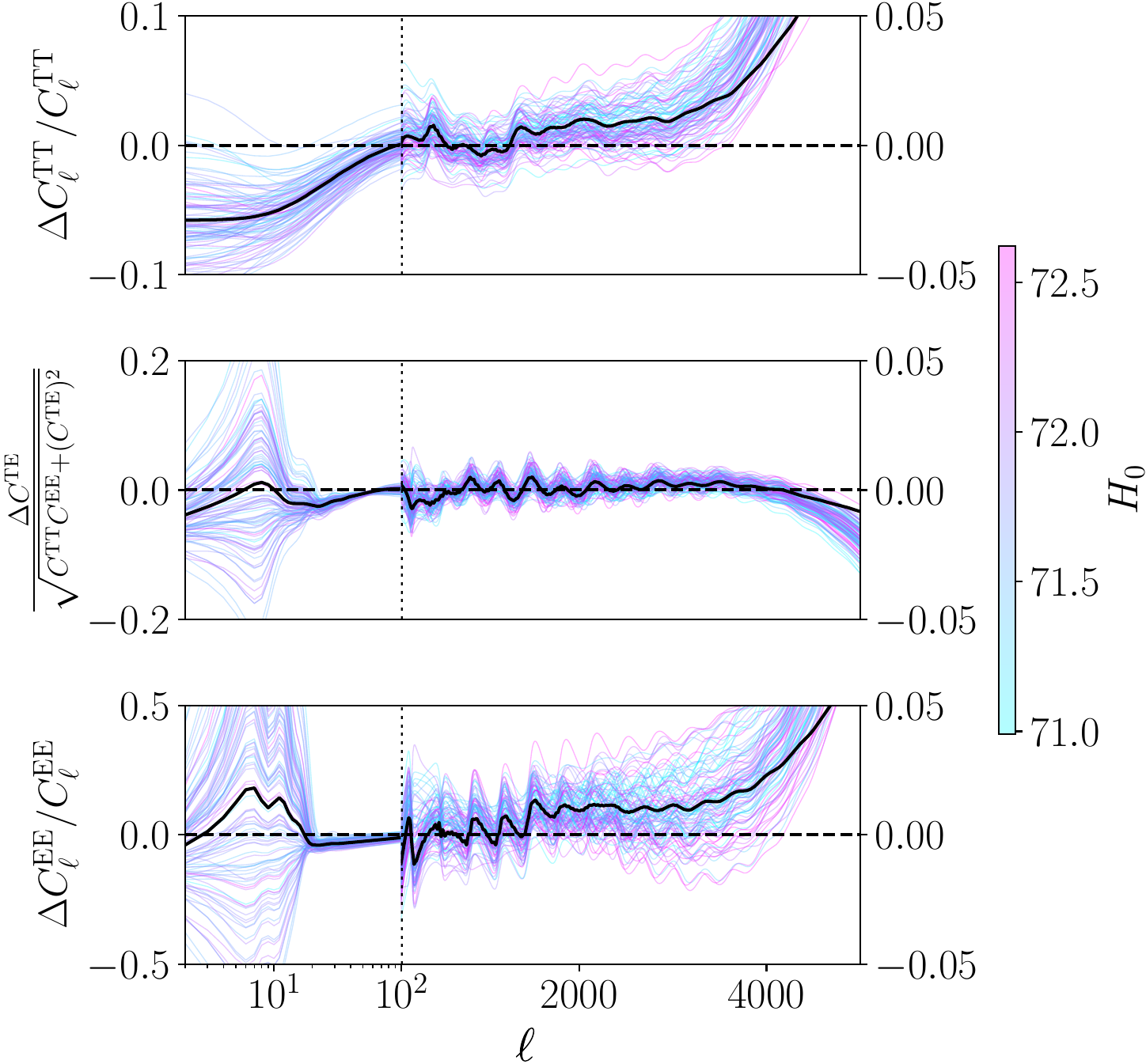}
    \caption{Residuals of 100 samples randomly drawn from the MCMC chains in the axEDE model (fit to \Planck+BAO+Pantheon+SH0ES), normalized to the $\Lambda$CDM best-fit to \Planck. The color code indicates the value of $H_0$ in each sample.}
    \label{fig:DH_axEDE}
\end{figure}

The predictions for the residuals of the CMB TT and EE power spectra with respect to $\Lambda$CDM in the best-fit EDE cosmology are shown with the solid black line.
To emphasize the irreducible effects imprinted by EDE in data (i.e. those that cannot be absorbed by a re-shuffling of $\Lambda$CDM parameters) that future surveys can target, we show in Fig.~\ref{fig:DH_axEDE} the residuals of 100 samples randomly drawn from the MCMC chains, normalized to the $\Lambda$CDM best-fit to \Planck. The color code indicates the value of $H_0$ in each sample. Note that we zoom in on the differences in the high-$\ell$ part of the plot. 

 First, the spectra show a bump-like feature around $\ell\sim 1500$
with increased power toward large $\ell$ in TT and EE that is present in essentially all the samples, and is due to larger $n_s$ and $\omega_b$ (which affects the diffusion-damping scale besides the oscillations discussed earlier). 
This feature can be further probed with high-resolution ground-based measurements, such as ACT and SPT, and in the future, the Simons Observatory and CMB-S4.
Multipoles $\ell\sim 100-500$ are particularly sensitive to details of the EDE dynamics as extensively discussed in Sec.~\ref{sec:EDE_pheno}, as these are the multipoles entering the horizon concurrently with the largest fractional contribution of the EDE energy density. 
Yet, the data require no strong departure from the $\Lambda$CDM prediction in that multipole range, and the samples oscillate around zero. 
Finally, one can also see large residual patterns at large angular scales ($\ell < 30$) in all three power spectra. The increased depth of the Sachs-Wolfe plateau in TT is (mostly) a consequence of slightly different values of $A_s$ and $n_s$, and is common to all samples, albeit at different amplitudes. On the other hand, the large bump/dip in TE and EE, which is due to small differences in $\tau$ and $\omega_b$, should not be over-interpreted, as the amplitudes of these vary greatly within the random samples of 100 points. It is therefore possible to easily adjust any forthcoming large-scale EE measurements with a slight shift in parameters and this signal is not a constraining feature for EDE.
Note also that there is no strong correlation between the value of $H_0$ in each sample and the size of the residuals: it is possible to find samples of points with large $H_0$ values that show only small deviations from $\Lambda$CDM (and vice versa).

Finally, the modifications to the matter power spectrum and CMB lensing potentials once EDE is included are primarily an overall power suppression, due to the fact that we added a non-clustering contribution to the energy density of the Universe. 
This clarifies that the increase in $\sigma_8$ is not due to the EDE {\it per se}, but in fact due to the reshuffling of $\Lambda$CDM parameters (in particular $\omega_{\rm cdm}$). 
Adjusting $h$ has the effect of shifting the overall spectrum horizontally, and to increase the amplitude at large scales. 
This latter effect comes from keeping $\omega_{\rm cdm}$ fixed at this stage, and the small-$k$ branch is mostly controlled by $(g(a_0,\Omega_m)/\Omega_m)^2$ where $\Omega_m=\omega_{\rm cdm}/h^2$ is more strongly decreased by the $h$ increase than the effect on the growth rate $g(a_0,\Omega_m)$ due to the longer dark energy domination \cite{lesgourgues_mangano_miele_pastor_2013}.
Increasing $\omega_{\rm cdm}$ partly compensates this effect on large-scales, while also leading to earlier growth ($z_{\rm eq}$ is shifted to higher redshift) and therefore more power on small scales. Finally, adjusting $A_s$ and $n_s$ shifts up and blue-tilts the spectrum, while increasing $\omega_b$ slightly reduces the overall amplitude of the spectrum and leads to slightly more contrasted BAO oscillations.  Very similar effects are visible in the lensing potential power spectrum (up to the BAO oscillations which are smoothed out).

\subsection{A summary of \Planck{} constraints on the EDE dynamics}
\label{sec:other_EDE}

\begin{table}
    \centering
    \begin{tabular}{c|c|c|c}
    \hline         

\multicolumn{4}{c}{{\it Planck}+BAO+SN1a}\\
\hline         
\hline         

Model &Parameters & w/o SH0ES & w/ SH0ES  \\
\hline 
           & $h$ & $ 0.688(0.706)^{+0.006}_{-0.011}$ & $0.715(0.719)\pm 0.009$ \\
axEDE \cite{Poulin:2018cxd,Smith:2019ihp,Simon:2022adh}  &  $f_{\rm EDE}$ &  $< 0.091 (0.088)$ &$0.109(0.122)^{+0.030}_{-0.024}$  \\  
$n=3$   & $\log_{10}(z_c)$ &  unconstrained  (3.55)&$3.599(3.568)^{+0.029}_{-0.081}$ \\  
     & $\theta_i$  & unconstrained (2.8) & $2.65(2.73)^{+0.22}_{-0.025}$  \\
      & &\multicolumn{2}{c}{1.9$\sigma$}
\\
   \hline
      RnR   \cite{Agrawal:2019lmo}    &  $h$ & 
 $0.6847(0.6788)_{-0.0083}^{+0.0057}$ &$0.7033(0.7076)_{-0.0096}^{+0.0092}$   
 \\
      $n=2$   &  $f_{\rm EDE}$ & $<0.055(0.004)$ & $0.067(0.0794)\pm0.025$ 
      \\
        
     &  $z_c$& unconstrained (9950) & $3180(2907)^{+495}_{-604}$ 
     \\
           & &\multicolumn{2}{c}{3.0$\sigma$}\\

         \hline
          NEDE  \cite{Niedermann:2019olb,Niedermann:2020dwg}& $h$  &$0.688(0.692)_{-0.013}^{+0.0078}$ & $0.713(0.714)_{-0.01}^{+0.011}$  \\
       &  $f_{\rm EDE}$ &   $<0.117(0.055)$& $0.125(0.130)_{-0.03}^{+0.038}$ \\
                & $\log_{10}(m_{\rm NEDE})$   &  unconstrained (2.4)& $2.5(2.6)_{-0.1}^{+0.2}$ \\
       &   $3w_{\rm NEDE}$ &  unconstrained (2.06)&$2.11(2.06)_{-0.21}^{+0.17}$ \\
 & &\multicolumn{2}{c}{1.9$\sigma$}
\\
\hline
  EMG \cite{Braglia:2020auw,Schoneberg:2021qvd} & $h$&  $0.6837(0.6853)_{-0.0055}^{+0.0049}$&$0.707(0.715)_{-0.01}^{+0.0097}$ \\
    & $f_{\rm EDE}(z_c)$ & $<0.0346 (0.024)$& $0.079(0.108)_{-0.023}^{+0.032}$ \\
    & ${\rm Log}_{10}(z_c)$ & unconstrained (3.59) & $3.612(3.626)_{-0.058}^{+0.061}$ \\
    & $\xi$& $<0.86(0.017)$ &$0.167(0.172)_{-0.091}^{+0.065}$\\

     & &\multicolumn{2}{c}{1.9$\sigma$} \\

\hline
      ADE   \cite{Lin:2019qug,Lin:2020jcb}  &  $h$ &  &  $0.706 (0.7057) \pm0.0085$ \\
       $w=c_s^2=1$ & $f_{\rm EDE}$ &   N.A.   &$0.082(0.082)\pm0.025$  \\
       & $\log_{10}(z_c)$ & &  $-3.46 (-3.45) \pm0.06$ \\
      
             \hline
        DA EDE  \cite{Berghaus:2019cls,Berghaus:2022cwf} 
       & $h$  & & $0.7085(0.7143)^{+0.0093}_{-0.0080}$   \\ 
       & $f_{\rm EDE}$ & N.A. & $0.050(0.063)^{+0.018}_{-0.015}$  \\
       & $\log_{10}(z_c)$ & & $4.79(4.96)^{+0.30}_{-0.20}$  \\
       & $\log_{10}\Upsilon$ [Mpc]$^{-1}$ & & $8.01(7.47)\pm 0.76$  \\
          \hline
      EDS  \cite{McDonough:2021pdg} &  $h$ & & $0.711 (0.7252) \pm 0.012$ \\
      & $f_{\rm EDE}$ & N.A. & $0.099 (0.142) ^{+0.056}_{-0.041}$ \\
      & $\log_{10}(z_c)$ & & $3.602 (3.58) ^{+0.071} _{-0.19 }$ \\
      & $\theta_i$ & & $< 3.14 (2.72 )$ \\
      & $c_{\theta}$ & & $-0.0024 (-0.0010)^{+0.0091}_{-0.015}$\\
                         \hline

      $\alpha$-EDE (B)  \cite{Braglia:2020bym}    & $h$ &  &  $0.709\pm0.011$ \\
      & $f_{\rm EDE}(z_c)$ &  N.A.  &  $ 0.082 \pm 0.02$ \\
      & $\log_{10}(1+z_c)$& &$ 3.510^{+0.044}_{-0.05}$ \\
      & $\Theta_i$ & & $<0.184$\\
\hline 

dULS  \cite{Gonzalez:2020fdy}    & $h$ &  &  $0.699^{+0.0084}_{-0.0086}$ \\
      & $f_{\rm EDE}(z_p)$ &  N.A.  &  $ 0.063^{+0.023}_{-0.025}$ \\
      & $\log_{10}(1+z_p)$& &$ 3.880^{+0.033}_{-0.455}$ \\
\hline 

\hline 
    \end{tabular}
    \caption{A summary of constraints on EDE parameters and the reconstructed Hubble parameter from analyzing {\it Planck}+BAO+SN1a+SH0ES with a Bayesian approach across various models suggested in the literature. 
    Where possible, we quantify the residual tension with SH0ES using the tension metric $Q_{\rm DMAP}\equiv\sqrt{\chi^2_{\rm min}({\rm EDE})-\chi^2_{\rm min}(\Lambda{\rm CDM})}$ introduced in Refs.~\cite{Raveri:2018wln,Schoneberg:2021qvd}. 
    In the dULS model, there is a significant offset between what Ref.~\cite{Gonzalez:2020fdy} calls `$z_c$' and the redshift at the peak contribution of the EDE energy density. 
    Here we report both $f_{\rm EDE}$ and $z_c$ at the peak $z_p$. 
    Since Ref.~\cite{Gonzalez:2020fdy} does not report a value for $f_{\rm EDE}$, we have estimated one using their reported constraint to $\Omega_{\rm dULS}$. 
    Note that we do not report numbers for the $\nu$EDE \cite{Sakstein:2019fmf} and chain EDE models \cite{Freese:2021rjq} as dedicated MCMC analyses are still lacking.}
    \label{tab:MCMC}
\end{table}

\begin{figure*}
    \centering
        \centering

    \includegraphics[width=1\columnwidth]{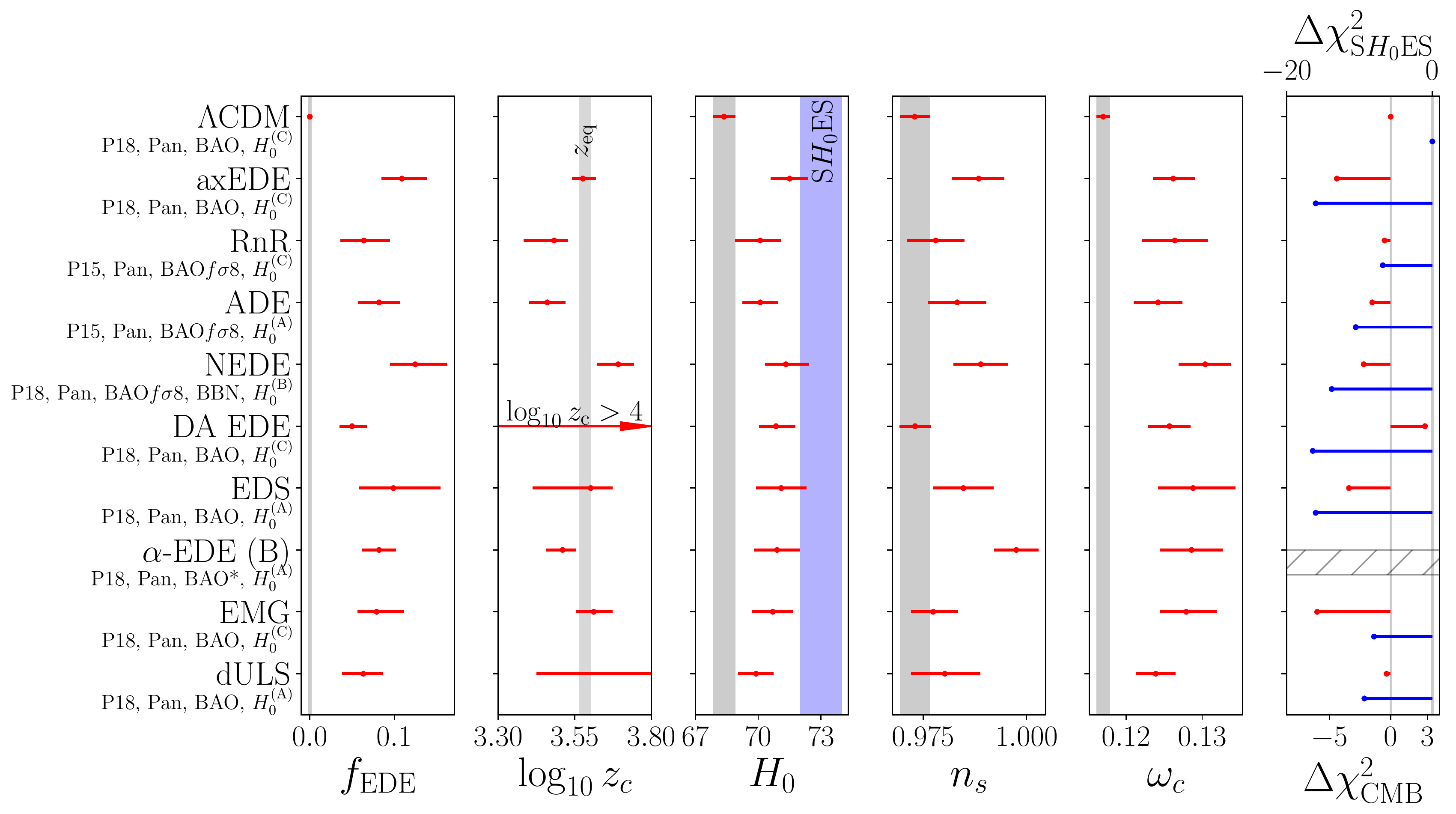}
    
    \caption{Comparison of the reconstructed 1D posteriors of the most relevant parameters in a representative sample of EDE models suggested in the literature, compared with $\Lambda$CDM.  
    Note that all models were not analyzed using the exact same datasets (see legends for details), but all analyses include (at least) {\it Planck} data (2015 or 2018), BOSS BAO DR12, Pantheon and SH0ES. 
    The label $H_0^{(A)}$ refers to the value from Ref.~\cite{Riess:2019cxk},  $H_0^{(B)}$ from Ref.~\cite{Riess:2020fzl} and  $H_0^{(C)}$ from Ref.~\cite{Riess:2021jrx}. 
    In the rightmost plot, we show $\Delta \chi^2_{\rm CMB}$ in red and $\Delta \chi^2_{\rm SH_0ES}$ in blue. 
    Note that for the dULS model, $z_c$ is significantly larger than the redshift $z_p$ at which the fractional EDE energy density peaks. 
    Since $z_p$ more closely corresponds to $z_c$ in the other EDE models, here we report constraints on $z_p$. 
    As Ref.~\cite{Gonzalez:2020fdy} does not report a value for $f_{\rm EDE}$, we have computed an approximate value using their constraint on $\Omega_{\rm dULS}$. 
    We do not have access to the  $\Delta \chi^2$ for the $\alpha$-EDE model.
    }
    \label{fig:EDE_whis}
\end{figure*}

While our discussion was mostly focused on the axEDE model as the first  
(and a representative) candidate for EDE, other models have been subsequently studied in the literature in light of similar cosmological data. 
By comparing constraints from the various models, we can draw some generic conclusions about the dynamics required by the data to resolve the Hubble tension.
We list in Tab.~\ref{tab:MCMC} and show in Fig.~\ref{fig:EDE_whis} the reconstructed 1D posteriors of the most relevant parameters in a representative sample of EDE models taken from the literature. 
We include results of analyses that all include (at least) {\it Planck} data (2015 or 2018), BOSS BAO DR12, Pantheon and SH0ES.  
Although this direct comparison should be taken with a grain of salt given that all models were not analyzed using the exact same data-sets, this simple exercise allows to draw some generic conclusions about the required EDE dynamics:
\begin{itemize}
    \item all models indicate a maximum contribution $f_{\rm EDE}(z_c)\simeq 5-12\%$ of EDE in the early universe;
    \item the maximum contribution is consistently found to be reached at $\log_{10}(z_c) \simeq 3.3-3.8$, i.e. around matter radiation equality;
    \item all models lead to $H_0\simeq 70-72$ km/s/Mpc (in fact no model can lead to values above $H_0\sim 72.5$ km/s/Mpc at $1\sigma$);
    \item all models have an increased primordial tilt of perturbations $n_s \simeq 0.98-1$ and an increased DM density $\omega_{\rm cdm}\simeq  0.12-0.14$\,.
\end{itemize}
This demonstrates that the degeneracies discussed in the previous sections are generic, irrespective of the details of the EDE model. 
If EDE is responsible for the current Hubble tension, one generically expects a new $\sim 10\%$ DE-like contribution around matter-radiation equality accompanied with a higher DM density and primordial tilt. 
This shift in parameters has additional consequences on the growth of structure at late-times, and leads to a very different matter power spectrum than predicted in $\Lambda$CDM, that can further probe the model (see Secs.~\ref{sec:S8} and \ref{sec:BOSS}). 
Furthermore, it may have important consequences for inflation, as the evidence for slow-roll behavior is currently tied to the detection of a small deviation from $n_s=1$, and we expand on this in Sec.~\ref{sec:inflation}.

To summarize, besides specifics of background dynamics, Sec.~\ref{sec:EDE_pheno_analysis} showed that data (in particular {\it Planck} polarization data) favor particular dynamics for the perturbations of the EDE field (as first shown by Ref.~\cite{Lin:2019qug}), with the approximate relation $c_s^2=0.54+0.25\times w$ where $w_f=(n-1)/(n+1)$. 
For an oscillating EDE with $n=3$, the data enforce dynamics in which modes inside the horizon around $z_c$ have effective sound-speed of less than $\simeq 0.9$ \cite{Smith:2019ihp}. 
This can be achieved if the potential is flat around the initial field value $\theta_i$ such that  $\theta_i^{n-1}/V''_{n}\gg 1$, which is realized in the axEDE model for $\theta_i \sim \pi$ (see Fig.~\ref{fig:EDE_Planck+SH0ES}).
This explains why the RnR model, which matches the axEDE model in the limit $\theta \ll 1$, cannot attain as high an EDE contribution and therefore as good a resolution of the tension. 

It is likewise interesting to compare the results of the RnR model with that of the EMG, given that their only difference is the addition of a non-minimal coupling to the Ricci scalar in the EMG model of the form $\xi\sigma^2 R/2$. 
As shown in Ref.~ \cite{Braglia:2020auw}, EMG models perform significantly better than the RnR model for $\xi > 0$, with data favoring $\xi\simeq 0.167_{-0.091}^{+0.065}$ once the SH0ES prior is included. 
The non-minimal coupling enhances the effect of the energy injection, while reducing the growth of CDM perturbations once the field starts to roll. 
In fact, this at least partly compensates for the effect of a larger $\omega_{\rm cdm}$ on the linear matter power spectrum, and reduces the imprints of the EMG at late-times relative to the minimally-coupled EDE model,  with potentially weaker constraints from galaxy surveys.

Similar results are obtained in the NEDE model which has very different micro-physical dynamics, as the NEDE fluid decays away through a phase-transition rather than coherent oscillations of the scalar field, showing that current data cannot yet characterize the transition from the DE phase to the dilution phase of EDEs. 
Note that in the results for the NEDE model presented here, we impose $c_s^2=w$ when the fluid dilutes. 
However, even when $c_s^2$ is free to vary independently of $w$, the data enforce $c_s^2\simeq 2/3$, i.e. to be very close to $w$ \cite{Niedermann:2020dwg}. 
Finally, in the $\alpha-$EDE model, the balance between the shape of the energy injection at the background level and the dynamics of the perturbations leads instead to an upper-limit on $\theta_i$ \cite{Braglia:2020bym}, that is fairly tight given the choice of $(p,n)$ and $\beta$ (which dictates the shape of the potential, see Eq.~\ref{eq:alphaEDE}). 

This discussion highlights that while there are broad qualitative similarities between all the models, there are some quantitative differences that can help disentangle the exact EDE dynamics with CMB and LSS data. 

In the following, we explore searches including additional data that can further distinguish the models. 

\section{Early Dark Energy in light of ACT and SPT data}
\label{sec:EDE_ACT_SPT}

Until now, we have limited our discussion to {\it Planck} CMB data (along with BAO and SN1a). 
Interestingly, recent literature has shown that the latest ACT data hints at a $2-3\sigma$ preference for EDE over $\Lambda$CDM, with $H_0$ values that are in good agreement with the SH0ES determination. 
We review these hints of EDE in light of the latest ACT and SPT data below. 

\begin{figure}
    \centering
    \includegraphics[width=0.49\columnwidth]{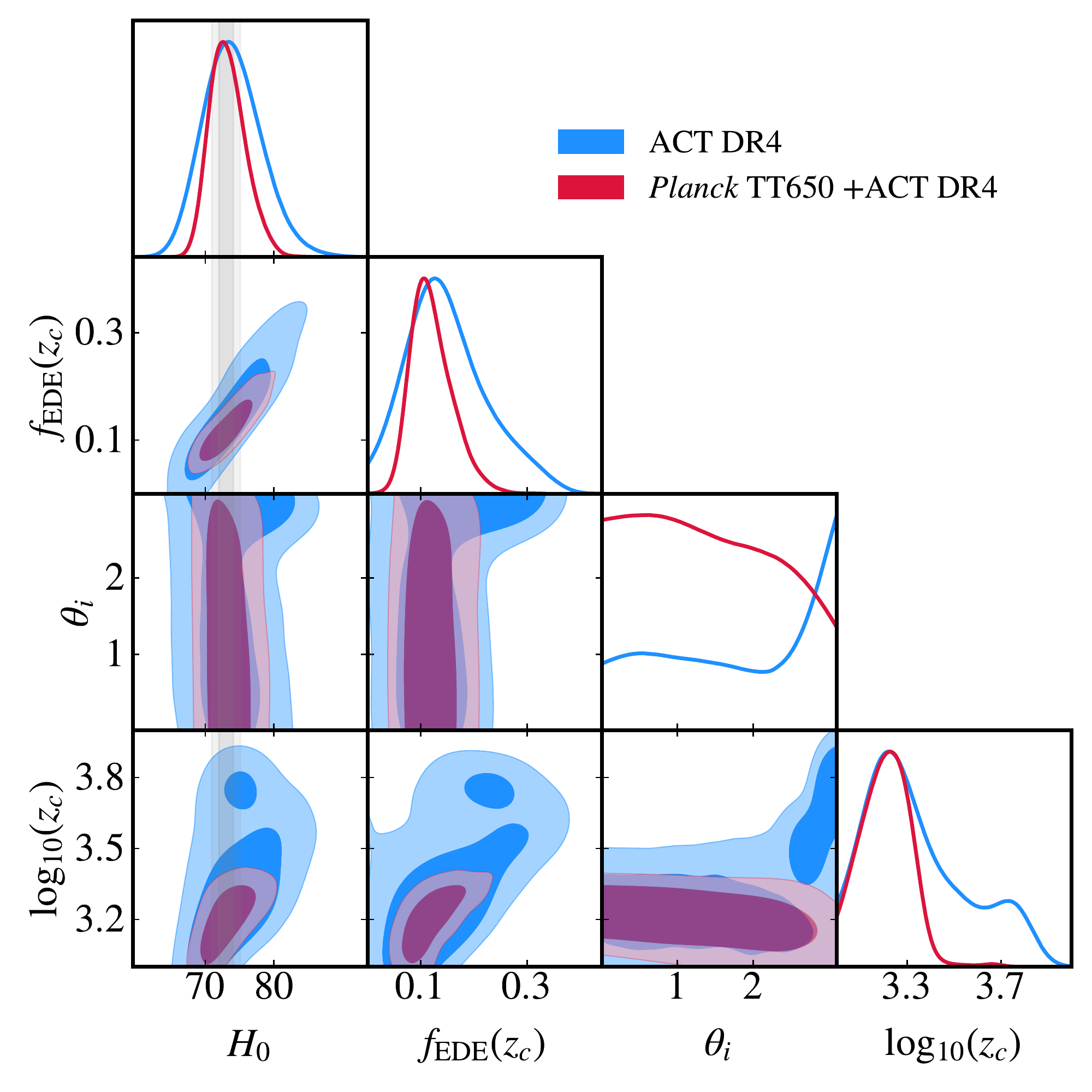}
        \includegraphics[width=0.49\columnwidth]{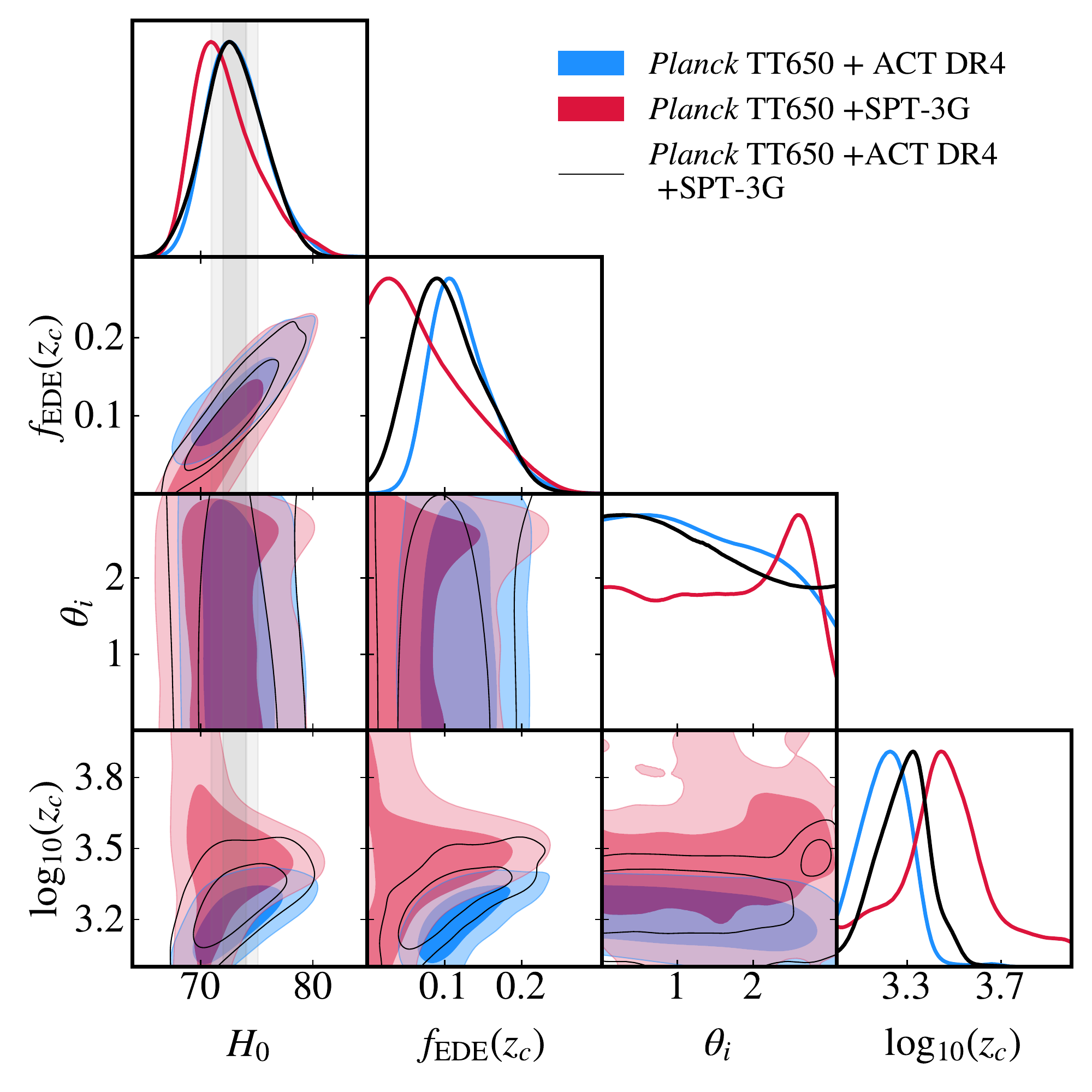}
    \caption{ Posterior distributions of $\{H_0,f_{\rm EDE}(z_c),\theta_i,\log_{10}(z_c)\}$ reconstructed from various analyses. 
    The left panel shows ACT data (with and without \Planck TT data upto $\ell \leq 650$). 
    The right panel shows a comparison between ACT and SPT, and their combination.}
    \label{fig:ACT_EDE}
\end{figure}

\subsection{Hints of EDE in ACT}

When analyzing CMB data, one can test the robustness of results obtained with \Planck{} by trading \Planck{} data for a combination of WMAP, sensitive to $\ell \lesssim 800$ and a high-resolution ground based CMB telescopes, such as ACT \cite{Aiola:2020azj} or SPT \cite{Dutcher:2021vtw,SPT-3G:2022hvq}. 
For instance, ACT data cover multipoles $\sim 500-5000$ in TT and $\sim 350-5000$ in EE and TE, with exquisite precision such that the combination of WMAP+ACT was shown to constrain $\Lambda$CDM parameters at a precision similar (within a factor of 2) to that of \Planck{}, and in agreement at $\sim 2\sigma$ \cite{Aiola:2020azj}. Similar precision is attained by SPT \cite{SPT-3G:2022hvq}, which covers the multipole range $\ell \sim 500-3000$.

We begin by presenting results with a similar baseline, where instead of WMAP, we make use of restricted \Planck{} TT data at $\ell\leq 650$ which yield results in excellent agreement with WMAP and are therefore commonly used as a proxy for WMAP data \cite{Hill:2021yec}.
The posterior distributions of $\{H_0,f_{\rm EDE}(z_c),\theta_i,\log_{10}(z_c)\}$ are shown in Fig.~\ref{fig:ACT_EDE} for various data combinations under an axEDE cosmology. 
One can see that ACT data favor a non-zero contribution of EDE at $\gtrsim 2\sigma$, with $f_{\rm EDE}(z_c)=0.152_{-0.092}^{+0.055}$ and $h=0.742_{-0.049}^{+0.036}$, improving the $\Delta\chi^2=-9.3$ with respect to $\Lambda$CDM. 
Very similar results were originally reported in Ref.~\cite{Hill:2021yec} by the ACT collaboration with a different analysis pipeline. 
Once \Planck{}TT650 is included, the preference for EDE increases to $\sim 3\sigma$, with $\Delta\chi^2=-15.4$. 
Note that ACT+\Planck{}TT650 favors a region of $z_c$ below $z_{\rm eq}$, and $\theta_i$ is essentially unconstrained,  while \Planck{}TTTEEE+SH0ES favored  $z_c\sim z_{\rm eq}$ with $\theta_i \sim \pi$. 
Ref.~\cite{Hill:2021yec} established that the preference for EDE in ACT originates from a residual pattern at $\ell \sim 500$ in EE. 
This is visible in Fig.~\ref{fig:ACT_Planck_EDE}, left panel, where we show the residuals between EDE and the $\Lambda$CDM best-fit model to ACT DR4 data, with and without the inclusion of {\it Planck}TT650 data. 
The $\sim 6$ ACT DR4 data points in EE around $\ell \sim 500$ are significantly better fit by EDE than by $\Lambda$CDM. 
However, one can clearly see that there is also a large power excess at similar multipoles in TT in this model, which is not observed in {\it Planck} data.
As a result, once large-scale {\it Planck}TT650 data are included, the residual features around $\ell \sim 500$ in EE and TT become less prominent. 
Instead, one can see a residual pattern of oscillations in all spectra that still yields a significantly better fit to ACT data, although the improvement now more equally arises from TT, TE and EE spectra.

\subsection{Impact of SPT}

\begin{figure}
    \centering
        \includegraphics[width=0.55\columnwidth]{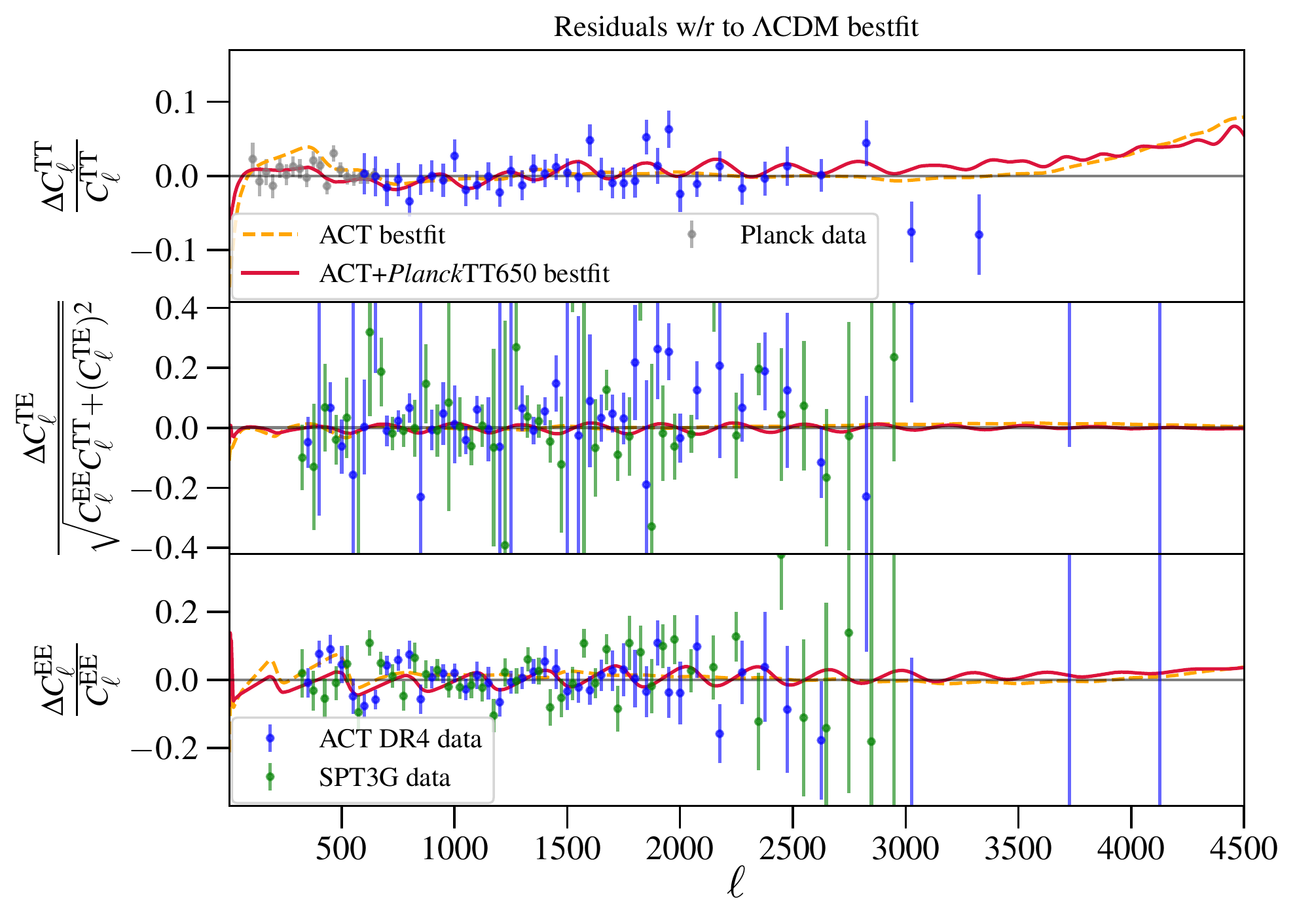}
        \includegraphics[width=0.44\columnwidth]{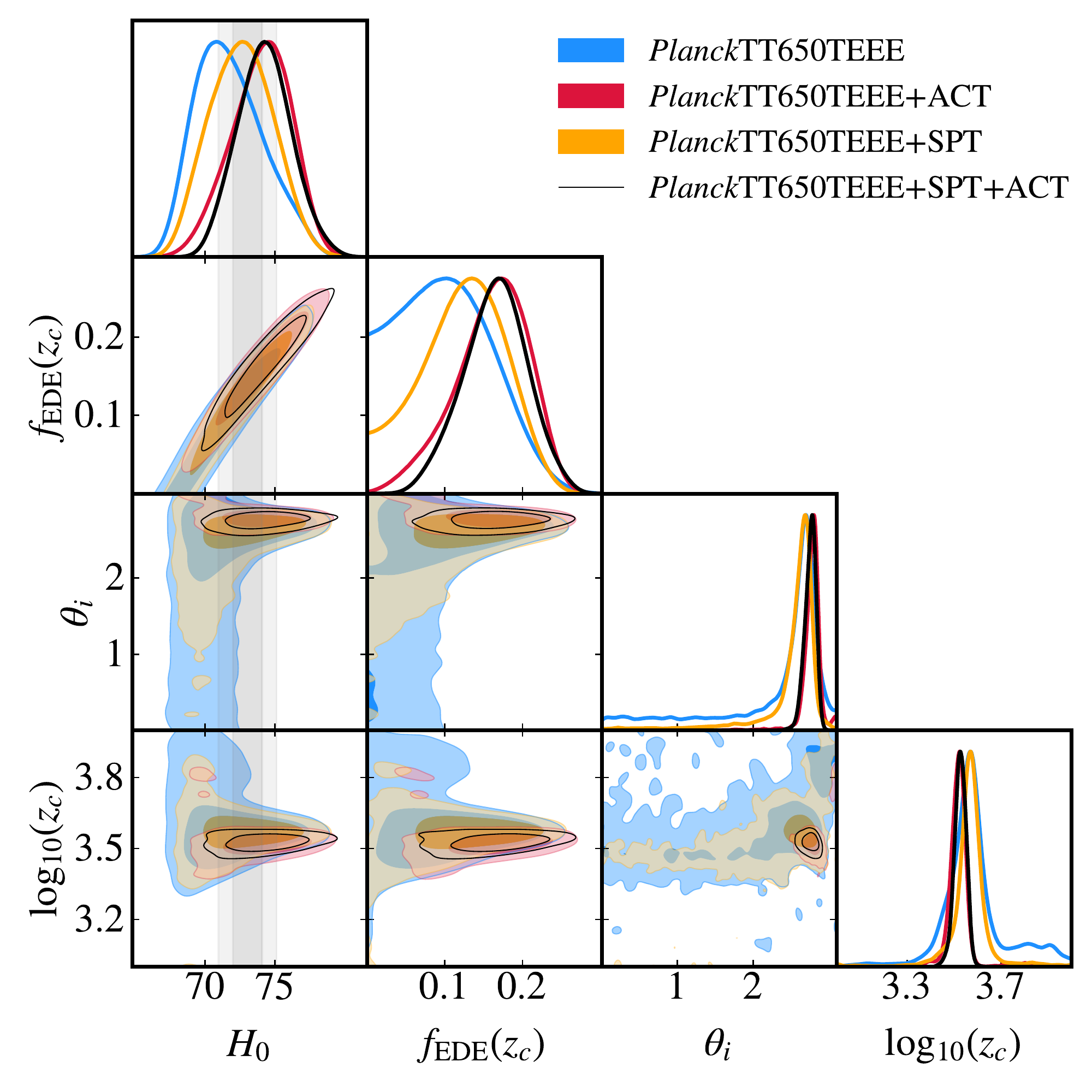}

    \caption{{\it Left panel:} Residuals between EDE and $\Lambda$CDM when fit to ACT DR4 data alone and in combination with {\it Planck}TT650. We also show the residuals of ACT DR4, SPT3G TEEE and {\it Planck}TT650 in the $\Lambda$CDM model.  {\it Right panel:}  Comparison between the posterior distributions of $\{H_0,f_{\rm EDE}(z_c),\theta_i,\log_{10}(z_c)\}$ reconstructed in the axEDE cosmology when analyzing ACT DR4 and SPT3G TEEE (and their combination) with \Planck TT650TEEE. From Ref.~\cite{Smith:2022hwi}.}
    \label{fig:ACT_Planck_EDE}
\end{figure}

Analyses with SPT 3G TEEE data\footnote{As this review was nearing completion, new results by the SPT collaboration have been released \cite{SPT-3G:2022hvq}, specifically new TT data.
The first dedicated analysis of the axion-like EDE model in light of these data were presented in Ref.}~\cite{Smith:2023oop}. have been presented in Refs.~\cite{LaPosta:2021pgm,Smith:2022hwi} (for earlier works considering SPT Pol, see Refs.~\cite{Chudaykin:2020acu,Chudaykin:2020igl}). 
On the right-hand side of Fig.~\ref{fig:ACT_EDE}, we show the results of analyzing {\it Planck}TT650+SPT 3G TEEE data compared to {\it Planck}TT650+ACT DR4, as well as the combination {\it Planck}TT650+ACT+SPT. 
One can see that SPT 3G TEEE data do not favor EDE, although error bars are large and not in particular tension with ACT DR4. 
In combination with ACT DR4, the preference for EDE is only slightly reduced compared to results with ACT DR4 alone. 
The most striking difference appears in $\log_{10}(z_c)$, with SPT 3G showing the usual degeneracy between $f_{\rm EDE}(z_c)$ and $H_0$ for values of $\log_{10}(z_c)$,  similar to the \Planck{}TTTEEE+SH0ES analysis, while ACT DR4 favors lower values as mentioned previously. 
The difference between ACT DR4 and SPT 3G TEEE data can be roughly understood from Fig.~\ref{fig:ACT_Planck_EDE}, left panel: around $\ell \sim 500$, where ACT DR4 data pull for an excess of power with respect to $\Lambda$CDM, SPT 3G TEEE data prefer a small deficit.

Because of the small apparent differences between ACT DR4 and SPT 3G TEEE, it is instructive to add {\it Planck} TEEE data to the analysis, which have a much larger signal to noise ratio at $\ell \sim 500$. 
The right panel of Fig.~\ref{fig:ACT_Planck_EDE} shows that all analyses with {\it Planck}TT650TEEE data consistently detect non-zero $f_{\rm EDE}$.
When \Planck{} high$-\ell$ TT data are replaced by ACT DR4 data, the preference for EDE exceeds the $3\sigma$ level.
The inclusion of SPT 3G TEEE data further increases the preference, and Ref.~\cite{Smith:2022iax} reports $f_{\rm EDE}(z_c)=0.163(0.179)_{-0.04}^{+0.047}$ with 
$\log_{10}(z_c) = 3.526(3.528)_{-0.024}^{+0.028}$ and $\theta_i =2.784(2.806)_{-0.093}^{+0.098}$, in good agreement with the EDE parameters reconstructed in the analysis of \Planck+SH0ES presented earlier, with a $\Delta\chi^2= -16.2$ (3.3$\sigma$ preference) in favor of EDE over $\Lambda$CDM. In addition, the model predicts the value of $H_0=74.2_{-2.1}^{+1.9}$ km/s/Mpc, in good agreement with direct determinations.
We now turn to discussing the impact of including \Planck{} high-$\ell$ TT data on the analyses.

\subsection{A new tension between ACT and \Planck{} temperature data?}

\begin{figure}
    \centering
    \includegraphics[width=0.50\columnwidth]{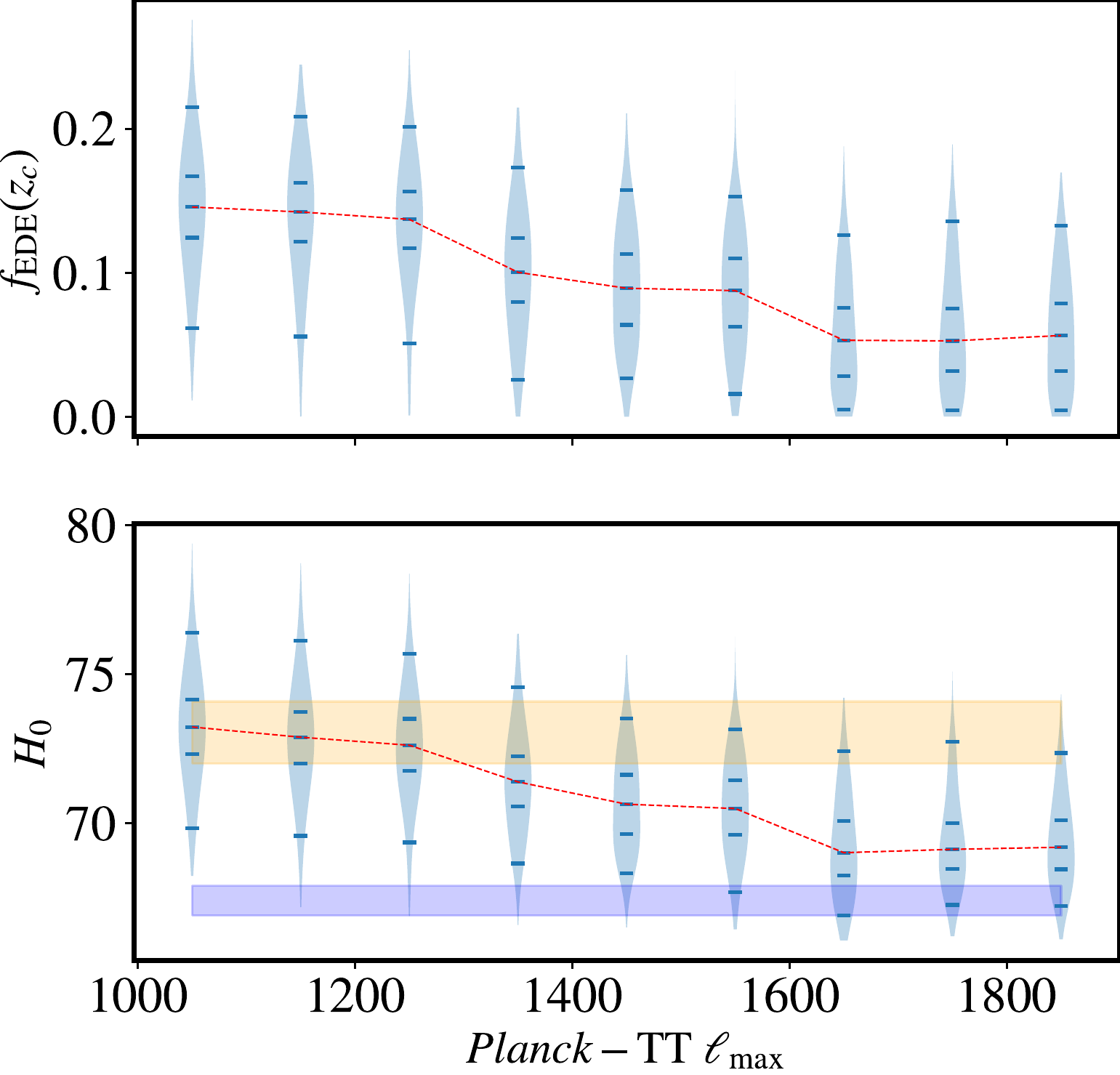}
        \includegraphics[width=0.49\columnwidth]{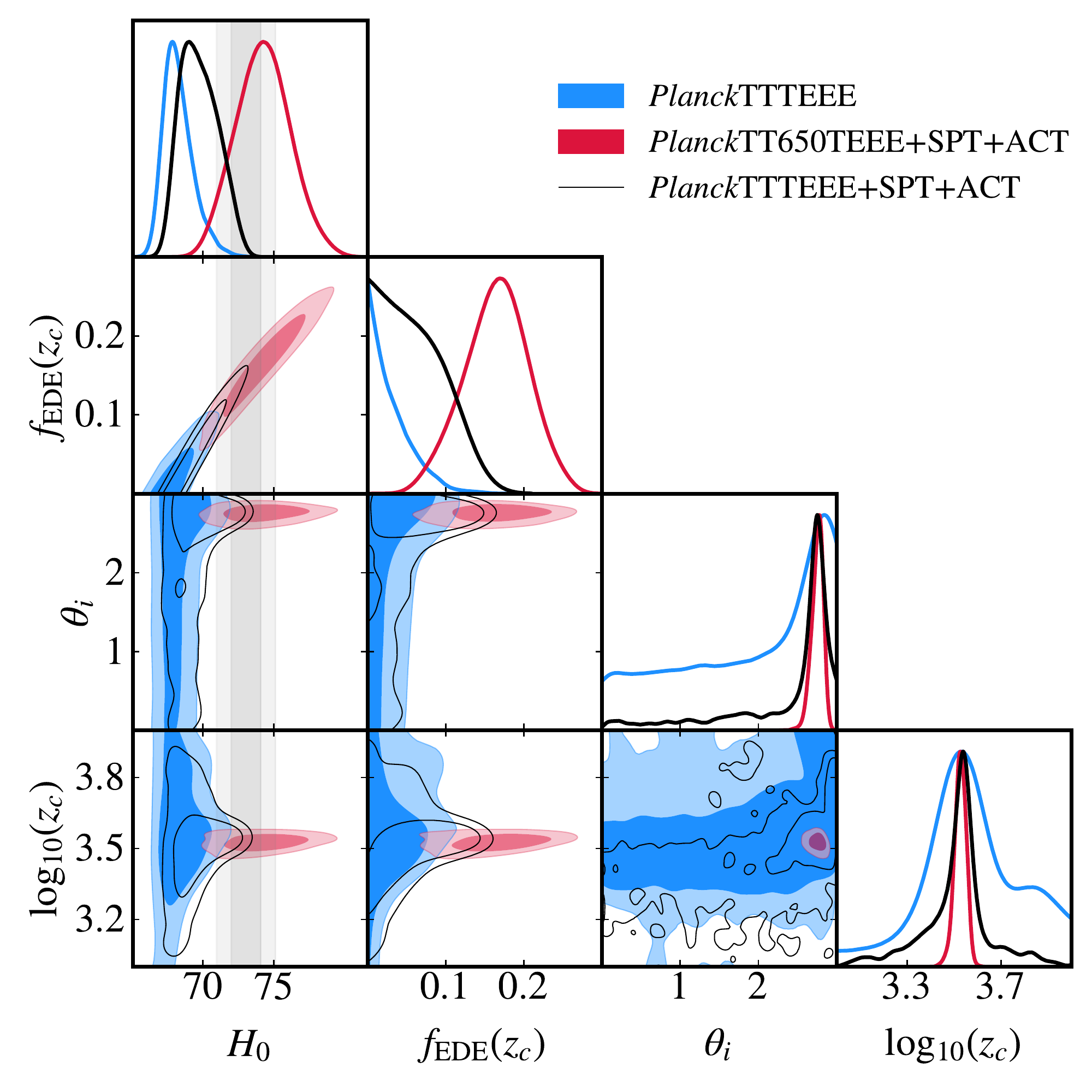}
    \caption{The left panel shows the 1D posterior distribution of $f_{\rm EDE}(z_c)$ and $H_0$ as the range of multipoles included in the \Planck{} TT data is increased to $\ell_{\rm max}$. The right panel shows $\{H_0,f_{\rm EDE}(z_c),\theta_i,\log_{10}(z_c)\}$ reconstructed from \Planck TTTEEE compared to \Planck TT650TEEE+ACT+SPT, and the combination \Planck TTTEEE+ACT+SPT. 
    These figures are taken from Refs.~\cite{Smith:2022hwi}.}
    \label{fig:EDE_FullPlanck_ACT}
\end{figure}

We show in Fig.~\ref{fig:EDE_FullPlanck_ACT} the reconstructed 1D posterior of $f_{\rm EDE}(z_c)$ and $H_0$ as the range of multipoles (upto some $\ell_{\rm max}$) included in the \Planck{} TT data is increased \cite{Smith:2022hwi}. 
The results are unchanged until $\ell_{\rm max} \simeq 1300$. 
Beyond that, the inclusion of small angular-scale data from the \textit{Planck} temperature power spectrum decreases the preference of EDE to $2.3 \sigma$ (in the absence of a $H_0$ prior). 
This is consistent with the fact that \textit{Planck} high-$\ell$ TT data have most of their constraining power at those scales, and drive parameters very close to their $\Lambda$CDM values, limiting the ability to exploit degeneracies between $\Lambda$CDM and EDE parameters. 

We show the results of analyses of \Planck{}TTTEEE, \Planck{}TT650TEEE+ACT+SPT and \Planck{}TTTEEE+ACT+SPT in the right panel of Fig.~\ref{fig:EDE_FullPlanck_ACT}. 
Notably, the reconstructions with \Planck{}TTTEEE and \Planck{}TT650TEEE+ACT+SPT appear to be in mild tension, and as a consequence, the combination \Planck{}TTTEEE+ACT+SPT leads to the constraint $f_{\rm EDE} < 0.128$, which is weaker than that obtained when analyzing \Planck{}TTTEEE alone, $f_{\rm EDE} < 0.091$ . 

It remains to be understood whether this mild tension between high-$\ell$ \Planck{}TT data and ACT DR4 data is due to a statistical fluctuation or a sign of yet unknown systematic errors in either experiment\footnote{It is interesting to note that the first analysis of the recent SPT TT data leads to constraints on EDE in agreement with \Planck, putting some pressure on the EDE model favored by ACT DR4 \cite{Smith:2023oop}.}. 
For further examination of the statistical agreement between these three CMB data sets, we refer the reader to Refs.~\cite{Handley:2020hdp,DiValentino:2022oon}.
There have also been several studies looking into the consistency between the `low' ($\ell \lesssim 1000$) and `high' TT multipoles in {\it Planck} (see e.g. Refs.~\cite{Addison:2015wyg,Planck:2018vyg,Planck:2016tof}). 
The high-$\ell$ TT power spectrum has a slight ($\sim 2 \sigma$) preference for higher $\omega_{\rm cdm}$, higher amplitude ($A_s e^{-2\tau_{\rm reio}}$), and lower $H_0$.
However, an exhaustive exploration of these shifts indicated that these are all consistent with expected statistical fluctuations \cite{Planck:2016tof}. 
Although there may be localized features in the high-$\ell$ TT power spectrum which are due to improperly-modeled foregrounds (see Sec.~6.1 in Ref.~\cite{Planck:2018vyg}), under the assumption of $\Lambda$CDM, there is no evidence that these data are broadly biased. 

\subsection{ACT and SPT constraints on other EDE models}

Until now, our discussion has focused on results of analyses of the axion-like EDE model in light of ACT and SPT data, as a representative example. 
There have been, however, several notable analyses of other EDE models with these data, showing similar hints of a preference over $\Lambda$CDM. 

First, the NEDE model was analyzed in Ref.~\cite{Poulin:2021bjr} in light of ACT+WMAP data, finding that ACT supports the existence of two different ``modes'' corresponding to different trigger field masses (i.e. different transition redshifts). 
The high-mass mode has $\log_{10}(m_{\rm NEDE})=2.916_{-0.079}^{+0.13}$ 
(corresponding to $z_* = 7870_{-900}^{+1200}$), with associated $f_{\rm NEDE}(z_*) = 0.071_{-0.024}^{+0.02}$, and $H_0 = 70.3_{-0.95}^{+0.89}$ km/s/Mpc, while the low-mass mode has $\log_{10}(m_{\rm NEDE}) = 1.687_{-0.25}^{+0.22}$ with $f_{\rm NEDE}(z_*) = 0.12^{+0.03}_{-0.055}$ and $H_0 = 70.51_{-2.2}^{+1.1}$ km/s/Mpc. 
The high-mass mode represents an improvement with respect to $\Lambda$CDM of $\Delta\chi^2_{\rm min}({\rm NEDE})=-17.6$, while the low-mass mode has $\Delta\chi^2_{\rm min}({\rm NEDE})=-8.4$. 
Note that the high-mass mode -- with a slightly lower $H_0$ -- has a significantly lower $\chi^2$ than the low-mass mode, and is thus favored over the mode that would fully resolve the Hubble tension. 

When combining full {\it Planck} (instead of WMAP) with ACT, it was found that the NEDE model still improves the fit over $\Lambda$CDM by a small amount, $\Delta\chi^2_{\rm min}({\rm NEDE})\simeq -5.7$, but $f_{\rm NEDE}(z_*)$ is compatible with zero at $1\sigma$. 
In addition, the $2\sigma$ constraint on the NEDE contribution significantly strengthens, from  $f_{\rm NEDE}(z_*) < 0.116$ (without ACT) to $f_{\rm NEDE}(z_*)<0.082$ (with ACT). 
This is in contrast with the results for axEDE (for which the combined constraint is weaker than {\it Planck} alone, see previous section), and indicates that the combination of {\it Planck} and ACT has the potential to distinguish between different EDE cosmologies.  

\begin{figure}
    \centering
    \includegraphics[width=0.49\columnwidth]{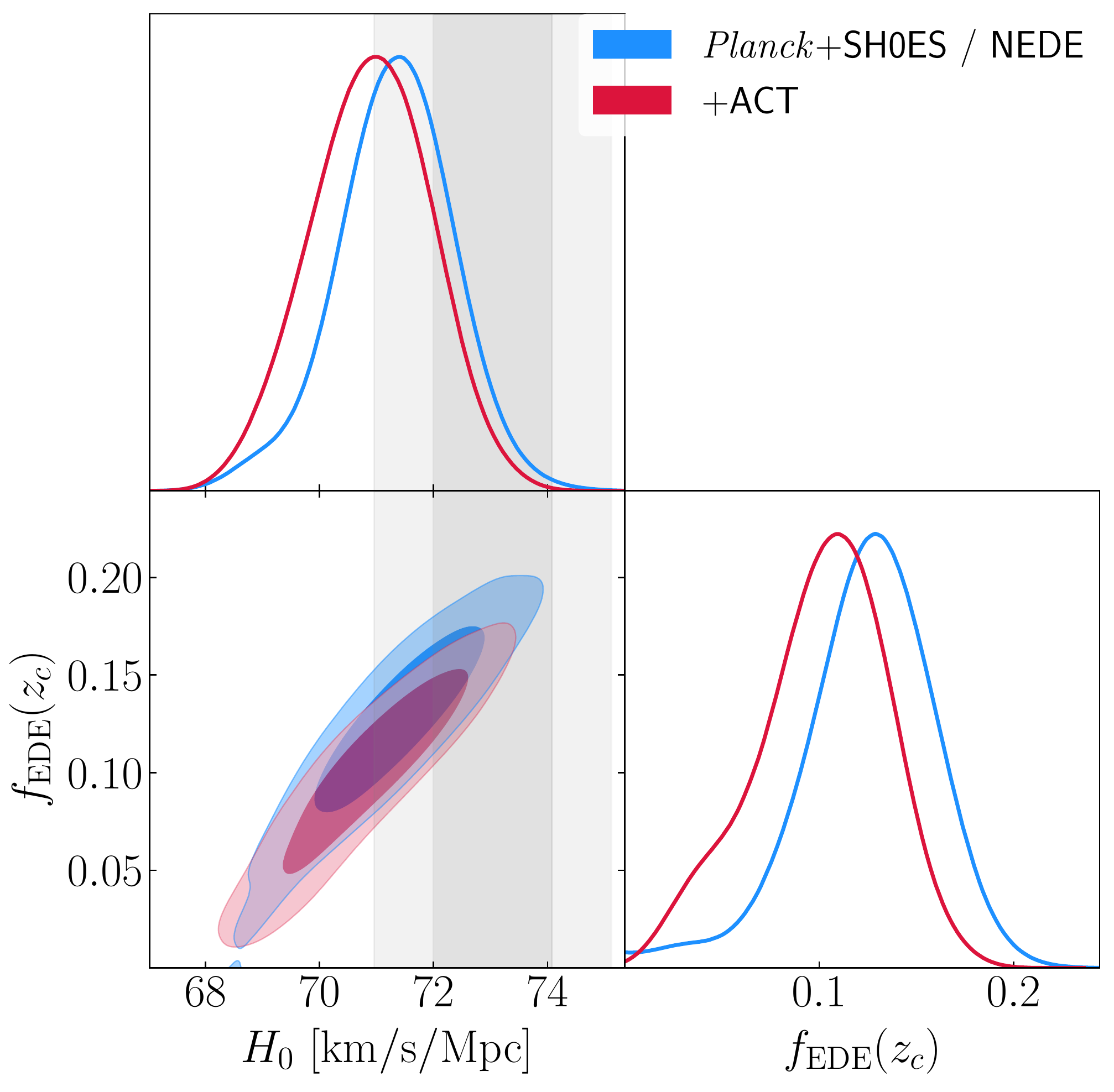}
    \includegraphics[width=0.49\columnwidth]{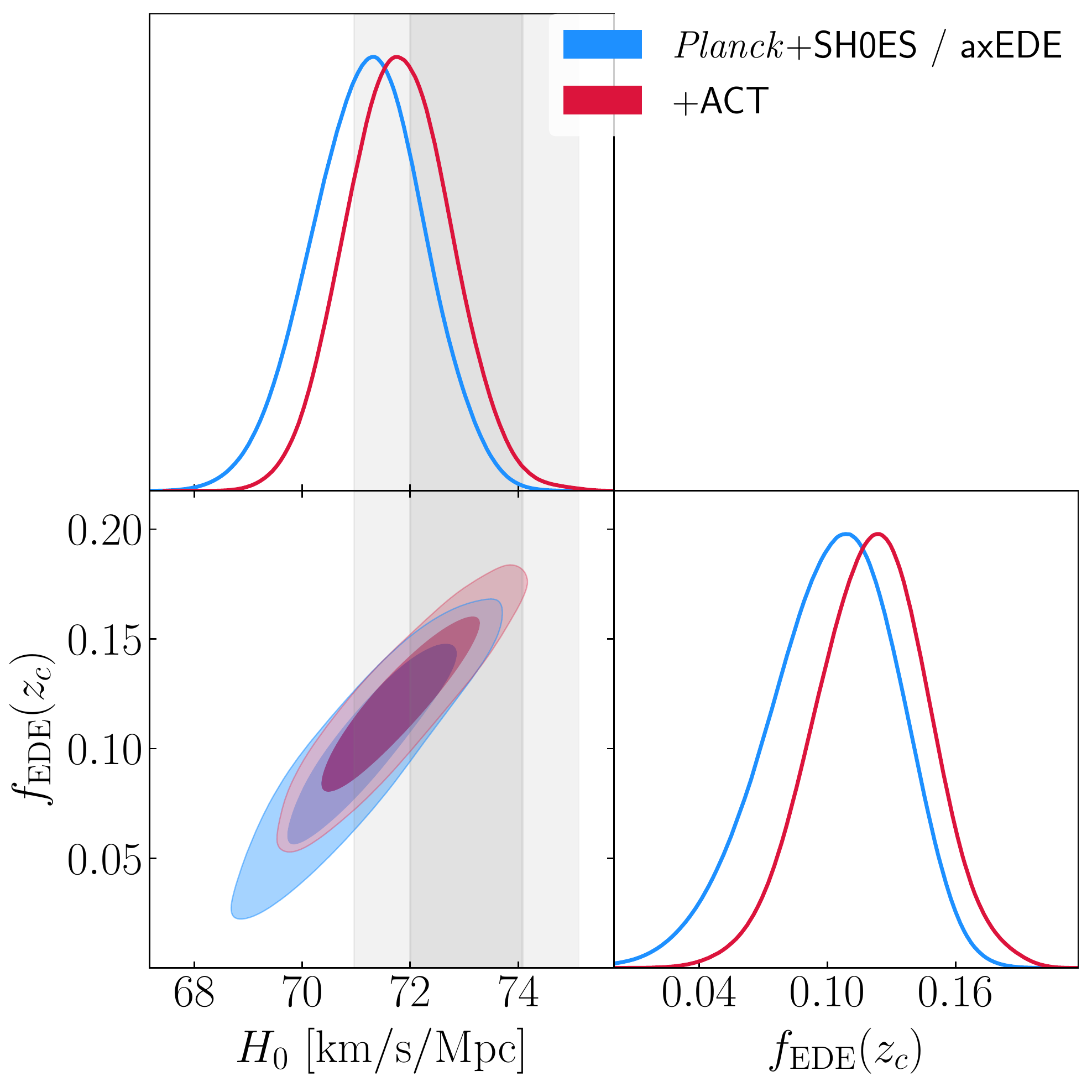}
    \caption{The reconstructed $H_0-f_{\rm EDE}(z_c)$ posteriors in the NEDE and axEDE models, when considering either  \Planck+SH0ES data and the addition of ACT (always including BAO and SNeIa data).}
    \label{fig:EDE_NEDE_ACT_H0}
\end{figure}

\begin{figure}
  \centering
   \includegraphics[width=0.7\columnwidth]{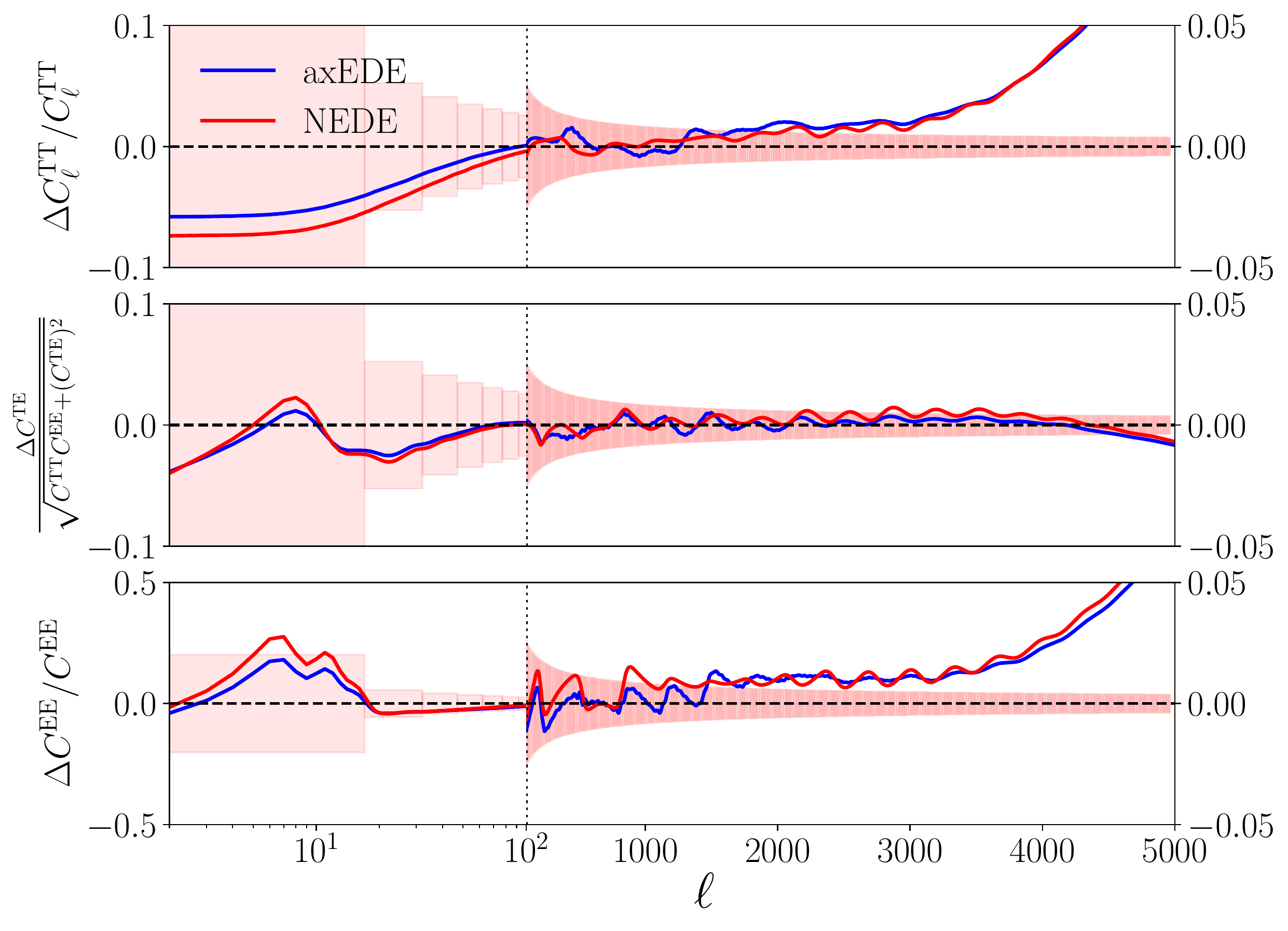}
   \caption{The differences in residuals between the best-fit $\Lambda$CDM model and axEDE/NEDE. The models were fit to \textit{Planck} CMB (TTTEEE+lensing) power spectra, BAO, SNeIa and SH0ES.
   Although they produce qualitatively similar signals, more accurate data at small angular scales are sensitive to the small quantitative differences between the models. The pink boxes give the cosmic variance uncertainty binned with a width of 15. Note that the linear part of these panels have a smaller $y$ range in order to better compare the axEDE and NEDE models. 
   }
   \label{fig:best_comp}
\end{figure}

We illustrate the difference in the constraints on the NEDE and axEDE models from the combination of \Planck\ and SH0ES, with and without ACT data (always including BAO and SNeIa data) in Fig.~\ref{fig:EDE_NEDE_ACT_H0}.  
One can see that once ACT data are added to the NEDE model, the preference for non-zero $f_{\rm EDE}$ and large $H_0$ decreases, while it increases in the axEDE model. 
To investigate the source of the difference, in Fig.~\ref{fig:best_comp}, we show the difference in CMB TTTEEE, lensing, and matter power spectra between the best-fit $\Lambda$CDM model and axEDE/NEDE when fit to \Planck\ and SH0ES. 
Although both models provide qualitatively similar residuals in the CMB data, displaying the characteristic increase in power at small scales, and quantitatively similar fits to \Planck{}, small differences between the models in the transition mechanism from the DE phase to the dilution phase can have important consequences with more accurate data.

This analysis was superseded by that performed in Ref.~\cite{Cruz:2022oqk} which, in addition to ACT-DR4 data, also considered the impact of SPT-3G data on the NEDE model.
They showed that prior-volume effects played a part in these constraints, as the constraint on $f_{\rm NEDE}(z_*)$  significantly weakens when fixing the mass of the trigger field. 
Once the SH0ES prior is added, a non-zero value of $f_{\rm NEDE}(z_*)$ is still strongly favored at $4.8 \sigma$ ($5.2\sigma$ without ACT), but the tension level with SH0ES remains $\sim 2.9\sigma$ ($1.6\sigma$ without ACT). 
SPT data on the other hand have a relatively weak impact on the model, and although they do not favor a non-zero $f_{\rm NEDE}(z_*)$ on their own, the preference over $\Lambda$CDM becomes $4.7\sigma$ when SH0ES is included in the analysis, with a tension level of $1.8\sigma$.

The ADE model was studied in light of updated \textit{Planck} and ACT data in Ref.~\cite{Lin:2020jcb}, which highlighted the important role played by \textit{Planck} TEEE data in the constraints: these restrict the ability of the ADE model to reach values larger than $H_0 \sim 71$ km/s/Mpc at $1\sigma$, although the combination of \textit{Planck}+ACT+BAO+Pantheon+SH0ES still favors ADE at $\sim2.8\sigma$. 
On the other hand, ACT data do not play a significant role in favoring or disfavoring the ADE model, with ADE constraints barely affected by the inclusion of ACT data. 
These result are again in contrast with the axion EDE and NEDE models, and stress the considerable power of precision CMB data in distinguishing between different EDE dynamics.

\section{Challenges to the EDE models}
\label{sec:EDE_challenges}

\subsection{A second cosmic coincidence problem}

An obvious question raised by EDE models is a coincidence or `why then' problem
- why does the EDE contribution become significant precisely close to matter-radiation equality, when it is most impactful on the CMB? 
This echoes the open question in cosmology regarding dark energy - the 'why now' or 'cosmic coincidence' problem that asks why the densities of dark energy and matter are similar today. 

Models of `tracking dark energy' \cite{Griest:2002cu,Dodelson:2001fq}, in which the density of a scalar field is always a fixed fraction of the energy density of the dominant component may provide an interesting starting point for model-building, and connect various eras of accelerated expansion in the universe. 
A similar idea - assisted quintessence \cite{Kim:2005ne} was recently applied in the EDE context, but the authors of Ref.~\cite{Sabla:2021nfy} found that it could not alleviate the tension.  
Alternatively, chameleon EDE \cite{Karwal:2021vpk} and trigger EDS \cite{Lin:2022phm} connect the dynamics of the EDE model to the onset of DM domination by coupling the EDE scalar field to DM. 
These models trigger the dilution of EDE through their interaction with DM, tying $z_c$ to matter-radiation equality and the DM mass to the EDE scalar field, with potential implications for the $S_8$ tension. 

Yet, the answer to this coincidence question may be simple probability. 
As suggested in the ``Axiverse'' scenario \cite{Arvanitaki:2009fg}, it is possible that there exist additional EDE-like fields in the Universe that may have been relevant (or not) at different eras, with a non-negligible probability that one such field becomes dynamical close to matter-radiation equality \footnote{A similar `probabilistic' scenario was introduced in Ref.~\cite{Kamionkowski:2014zda}, in order to solve the `why-now' problem of late-time DE and leave observables unaffected.}. 
In fact, it that context, it is plausible that the EDE field responsible for the Hubble tension is the first  member of a family of such fields, that have yet to be detected (or second, if one counts late-DE) (see also Refs.~\cite{Griest:2002cu,Linder:2010wp,Samsing:2012qx,Hojjati:2013oya,Freese:2021rjq}). 
It is also plausible that an EDE field acting around matter-radiation equality was detected first {\it precisely because} our cosmological data gains most of their sensitivity right around recombination. 
Fields acting at much earlier times would leave the CMB mostly unaffected \cite{Karwal:2016vyq,Poulin:2018dzj}, while a field acting at much later times could play the role of dark energy.
In any case, if the presence of an EDE-like field is confirmed with future data, the question of what if any connection exists between DE and EDE will become particularly relevant. 
Indeed, this new `why then' problem may provide insight into the long-standing `why now' problem.
 
\subsection{Fine-tuning problems of some scalar-field EDE models}

The ``axion-like'' potential given by Eq.~\ref{eq:potential}  suffers from fine-tuning issues that makes it challenging to embed in a UV-complete theory. 
While one may generate such a potential from an instanton expansion, where the leading terms with order $n<3$ (and higher order ones) have vanished due to a delicate balance in the controlled expansion, 
this is generically hard to achieve, as was pointed out in Refs.~\cite{Gonzalez:2020fdy,Rudelius:2022gyu,McDonough:2022pku}. 
Attempts at realizing this potential in string theories have been discussed\footnote{Let us stress that the recent work presented in Ref.~\cite{Cicoli:2023qri} concludes that EDE with a modified axion-like potential can be a viable cosmological model in a string-theory context, with model-building challenges similar to that of inflation, late DE and fuzzy dark matter.} in Ref.~\cite{McDonough:2022pku,Cicoli:2023qri}. 
In addition, as mentioned earlier, the near-Planckian field excursions favored in the analysis may run into problems with the weak-gravity conjecture \cite{Kaloper:2019lpl,Rudelius:2022gyu}.
Similarly, for models not protected by approximate shift-symmetries like the axion-like model, one expects that radiative corrections (arising from self-interactions or couplings to other fields) to the scalar-field mass will lead to additional lower-order contributions to the potential that scale as $\phi^2$. 
In both cases, these additional terms appear at energy scales that are connected to those of the higher-order terms, such that it may not possible to ignore lower-order terms in the analyses. 
When included, they may dominate the dynamics of the model, slowing the dilution of the field and leaving imprints on observables that are ruled out. 
For instance, if $m^2\phi^2$ terms dominate, the field dilutes like matter, spoiling the success of the EDE model \cite{Poulin:2018dzj}.\footnote{If we include a quadratic term in the axEDE potential $V = m_b^2 \phi^2/2 + m_a^2 f^2(1-\cos \phi/f)^3$, the resolution to the Hubble tension is unspoiled as long as $m_b/m_a \lesssim 10^{-4}$.}

Nonetheless, the axEDE model can offer insight into the background and perturbative phenomenological properties of EDE preferred by data that replacement, well-motivated EDE models can target to resolve the tension. 
For example, while DA EDE can emulate these properties at the background level, due to vastly different perturbation dynamics, it is unable to replicate the success of axEDE \cite{Berghaus:2022cwf}. NEDE, $\alpha-$attractors, and the dULS model, on the other hand, are theoretically better motivated and achieve results similar to axEDE, as they can reproduce both the required background and perturbative phenomenology to increase $H_0$ while also providing a good fit to CMB data.

Hence, two avenues of further research may be particularly interesting: exploring alternate, well-motivated scalar fields for EDE and directly reconstructing the EDE potential from data. 
Lower-order terms in the EDE scalar potential that address these fine-tuning issues, or the presence of multiple fields as in the NEDE model may have similar effects as additional ultra-light axions like in the `dark sector' model, and may simultaneously alleviate the $S_8$ tension \cite{McDonough:2022pku}. 
Alternatively, successful non-EDE scalar-field models have also been suggested \cite{Aloni:2021eaq,Joseph:2022jsf,Buen-Abad:2022kgf} as Hubble tension solutions. 
On the other hand, abandoning parametric potentials altogether, one may directly reconstruct the EDE potential from data, as routinely done for the inflaton potential \cite{Planck:2018jri} and dark energy \cite{Park:2021jmi,Goldstein:2022okd}. 
Ref.~\cite{Moss:2021obd} attempts to reconstruct EDE dynamics non-parametrically, but at the level of the fluid energy density, rather than the scalar-field potential. 
In any case, future high-accuracy CMB and clustering data will be crucial in distinguishing the various EDE models suggested so-far, and may shed light on the dynamics required to resolve the Hubble tension beyond a toy-model description.

\subsection{The trouble with $S_8$}
\label{sec:S8}

Besides the statistically-significant Hubble tension, another interesting tension is emerging in cosmology relating to the amplitude of density fluctuations on large scales, captured by estimations of the parameter $S_8\equiv \sigma_8(\Omega_m/0.3)^{0.5}$. 
The amplitude $\sigma_8$ is the root-mean-squared of matter fluctuations on a $8 h^{-1}$Mpc scale and is defined through
\begin{equation}
    \sigma_8^2 = \int \frac{k^3}{2\pi^2}P_m(k)W_8^2(k)d\ln{k}.
\end{equation} 
where $P_{\rm m}(k,z=0)$ is the linear matter power spectrum today, $W(kR)$ is the top-hat window function in Fourier space, and $R=8$Mpc/$h$ by convention.  
Inferences of $S_8$ from weak lensing surveys such as the CFHTLenS \cite{Heymans:2012gg}, KiDS, as well as from {\it Planck} SZ cluster abundances \cite{Planck:2018vyg,Planck:2015lwi} are about $2-3\sigma$ smaller than that from the CMB. 

In fact, the combination of KiDS data with BOSS and 2dFLenS data leads to a $3\sigma$ lower value than {\it Planck}, and points to a $\sigma_8$-tension rather than discrepancies in $\Omega_m$ \cite{Heymans:2020gsg}. 
However, some surveys such as DES \cite{DES:2021wwk} and HSC \cite{Hikage:2018qbn}, while yielding $S_8$ lower than {\it Planck}, are statistically compatible with it at less than $2\sigma$. 
More data are therefore awaited to firmly confirm this potential breakdown of the $\Lambda$CDM model (see Ref.~\cite{Abdalla:2022yfr} for a more complete review of measurements). 

\begin{figure}
\centering
\includegraphics[width=0.495\columnwidth]{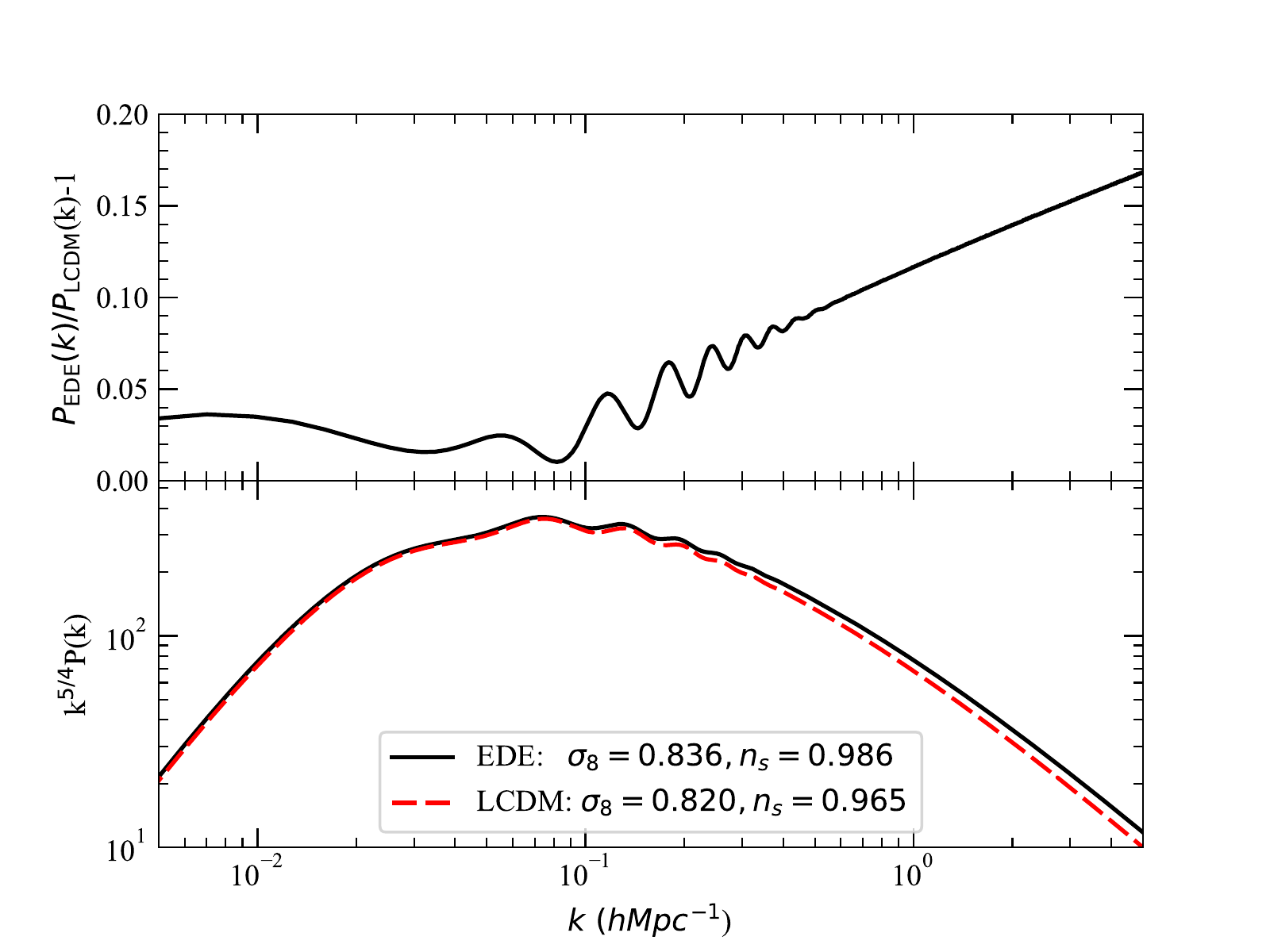}
    \includegraphics[width=0.495\columnwidth]{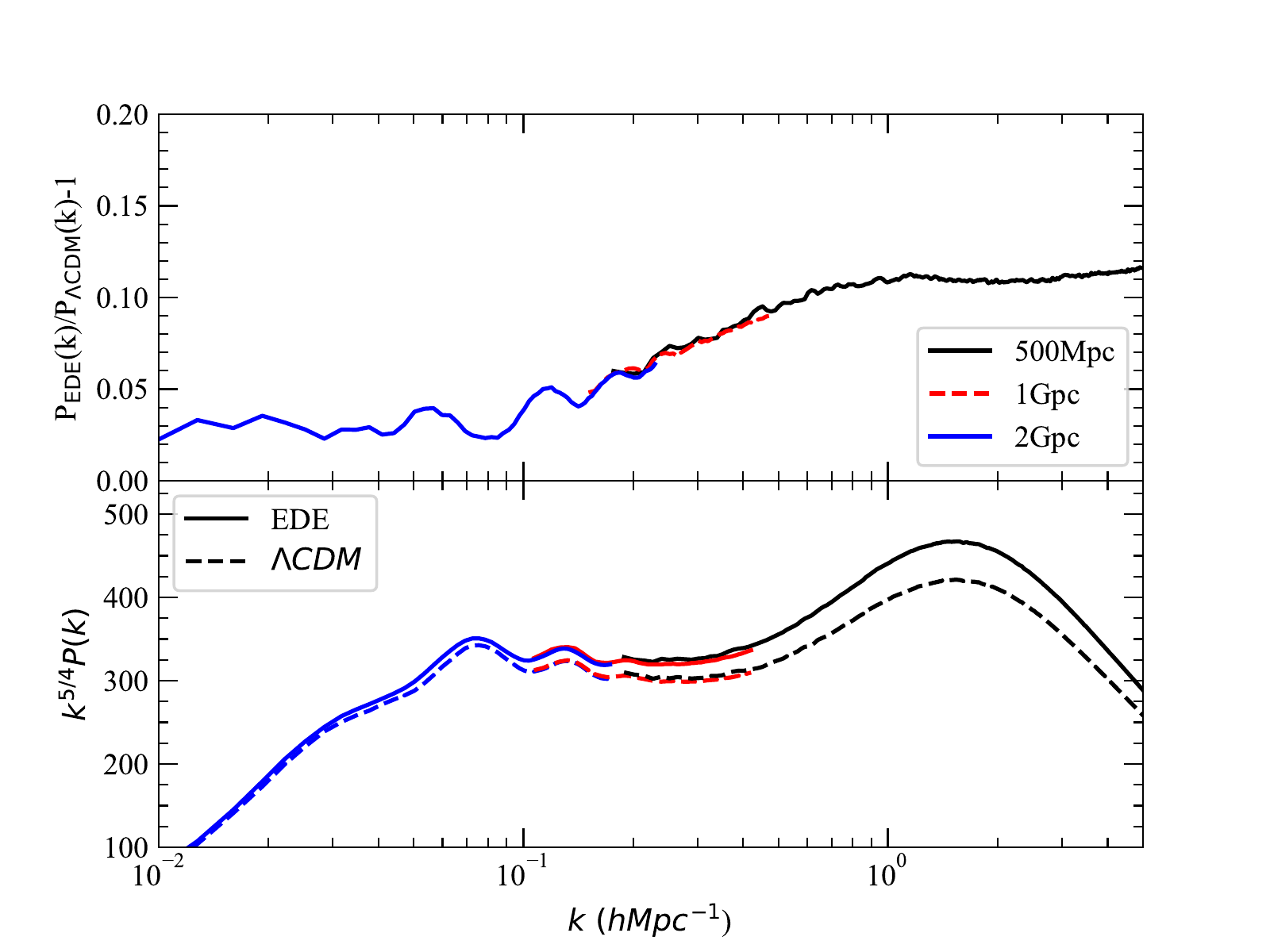}
    \caption{ The linear (left) and non linear (right) matter power spectra in an EDE cosmology and in $\Lambda$CDM. Taken from Ref.~\cite{Klypin:2020tud}. }
    \label{fig:EDE_pk}
\end{figure}

Unfortunately, a generic feature of early-universe models proposed to resolve the Hubble tension is to predict greater $S_8$ and {\it increase} this tension. 
For EDEs, this may appear surprising since, at fixed $\Lambda$CDM parameters, the impact of EDE is to decrease the amplitude of matter fluctuations. 
The subsequent increase in $S_8$ is due to the degeneracy with $\omega_{\rm cdm}$ described in Sec.~\ref{sec:EDE_LCDM_deg}: the impact of the increased expansion rate on the time-evolution of potential wells (in particular the eISW effect), which manifests as an increase in the height of the first acoustic peak (see Sec.~\ref{sec:EDE_pheno} for details), is compensated for by increasing the DM density \cite{Poulin:2018cxd,Vagnozzi:2021gjh}. 
As the DM density increases, the onset of DM domination and the formation of structure begin earlier, leading to larger $\sigma_8$.
In addition, a similar increase in the DM density is required to keep the angular scale of the BAO fixed, with the same consequences for $S_8$. 
It has been suggested that it might simply be impossible to resolve both tensions simultaneously with a single new-physics mechanism \cite{Vagnozzi:2021gjh,Jedamzik:2020zmd,Clark:2021hlo,Allali:2021azp}.
We show in Fig.~\ref{fig:EDE_pk} the linear (left) and non linear (right) matter power spectra in the EDE cosmology and in $\Lambda$CDM as computed in Ref.~\cite{Klypin:2020tud} for a fiducial analysis that includes \textit{Planck}+BAO+Pantheon+SH0ES data. 
One can see a clear increase of power at scales $k \sim 0.1~h$/Mpc and above, although the differences are smaller in the non-linear power spectrum.

Due to the impact on $S_8$, it is possible to use low-$S_8$ measurements to constrain EDE. 
In particular, Ref.~\cite{Hill:2020osr} showed that the combination of DES data, with priors on $S_8$ as measured by KIDS and HSC, leads to the constraint $f_{\rm EDE}(z_c)<0.06$ without SH0ES, and $f_{\rm EDE}(z_c)=0.062^{+0.032}_{-0.033}$ with SH0ES. 
At least part of this constraining power was found to originate from prior-volume effects, since once $\theta_i$ and $z_c$ are fixed to their best-fit values, 
the constraints from a combined analysis with the KiDS/DES/HSC data relax substantially to $f_{\rm EDE}(z_c)<0.092$, and the inclusion of SH0ES leads to $f_{\rm EDE}(z_c)=0.087^{+0.029}_{-0.024}$ \cite{Murgia:2020ryi,Smith:2020rxx}. 
However, it is clear that low-$S_8$ measurements cannot be explained by the presence of EDE \cite{Secco:2022kqg}, and that one can use LSS surveys to constrain the presence of EDE.

\subsection{Is Early Dark Energy excluded by BOSS?}
\label{sec:BOSS}

Beyond the mere value of $S_8$,  measurements of biased tracers of the matter power spectrum can be used to constrain EDE models. 
An example of such an observable is the galaxy power spectrum. 
Until now, we have reported analyses that only make use of compressed information from galaxy surveys done by BOSS, in the form of the BAO angles, and for some analyses, also use the redshift-space distortion information (quantified by $f\sigma_8$) 
The impact of including $f\sigma_8$ information for EDE is minor. For instance, in the latest analysis of the axEDE model which was presented above, the inclusion of $f\sigma_8$ from BOSS DR12 changes the constraints from $f_{\rm EDE}(z_c)< 0.091$ without SH0ES ($f_{\rm EDE}(z_c) = 0.109^{+0.030}_{-0.024}$ with SH0ES) to $f_{\rm EDE}(z_c)<0.088$ without SH0ES ($f_{\rm EDE}(z_c) = 0.102^{+0.030}_{-0.024}$ with SH0ES) and the $Q_{\rm DMAP}$ metric still indicates a decrease of the tension with SH0ES to the $2\sigma$ level \cite{Simon:2022adh}.

Nevertheless, developments of the one-loop prediction of the galaxy power spectrum in redshift space from the Effective Field Theory of Large-Scale Structures\footnote{See also the introduction footnote in e.g.~\cite{DAmico:2022osl} for relevant related works on the EFTofLSS.} (EFTofLSS)~\cite{Baumann:2010tm,Carrasco:2012cv,Senatore:2014via,Senatore:2014eva,Senatore:2014vja,Perko:2016puo}  have made possible the determination of the $\Lambda$CDM parameters from the full-shape analysis of SDSS/BOSS data~\cite{BOSS:2016wmc} at a precision higher than that from conventional analyses (i.e. using BAO/$f\sigma_8$ information), and comparable to that of CMB experiments. 
This provides an important consistency test for the $\Lambda$CDM model, and leads to competitive constraints on models beyond $\Lambda$CDM (see e.g. Ref.~\cite{DAmico:2019fhj,Ivanov:2019pdj,Colas:2019ret,DAmico:2020kxu,DAmico:2020tty,Simon:2022ftd,Chen:2021wdi,Zhang:2021yna,Philcox:2021kcw,Kumar:2022vee,Nunes:2022bhn,Lague:2021frh}). 

\begin{figure}
\centering
\includegraphics[width=0.49\columnwidth]{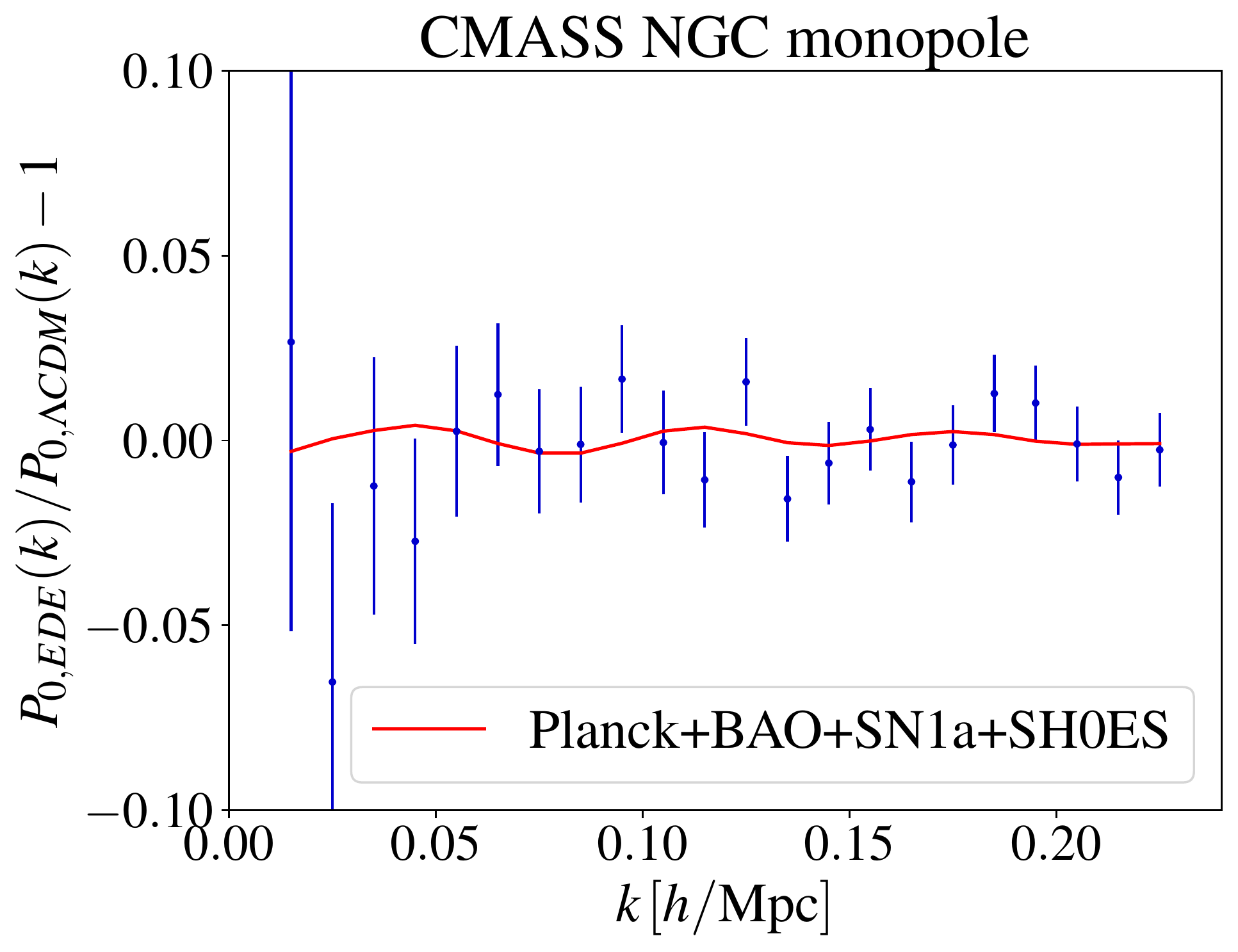}
    \includegraphics[width=0.49\columnwidth]{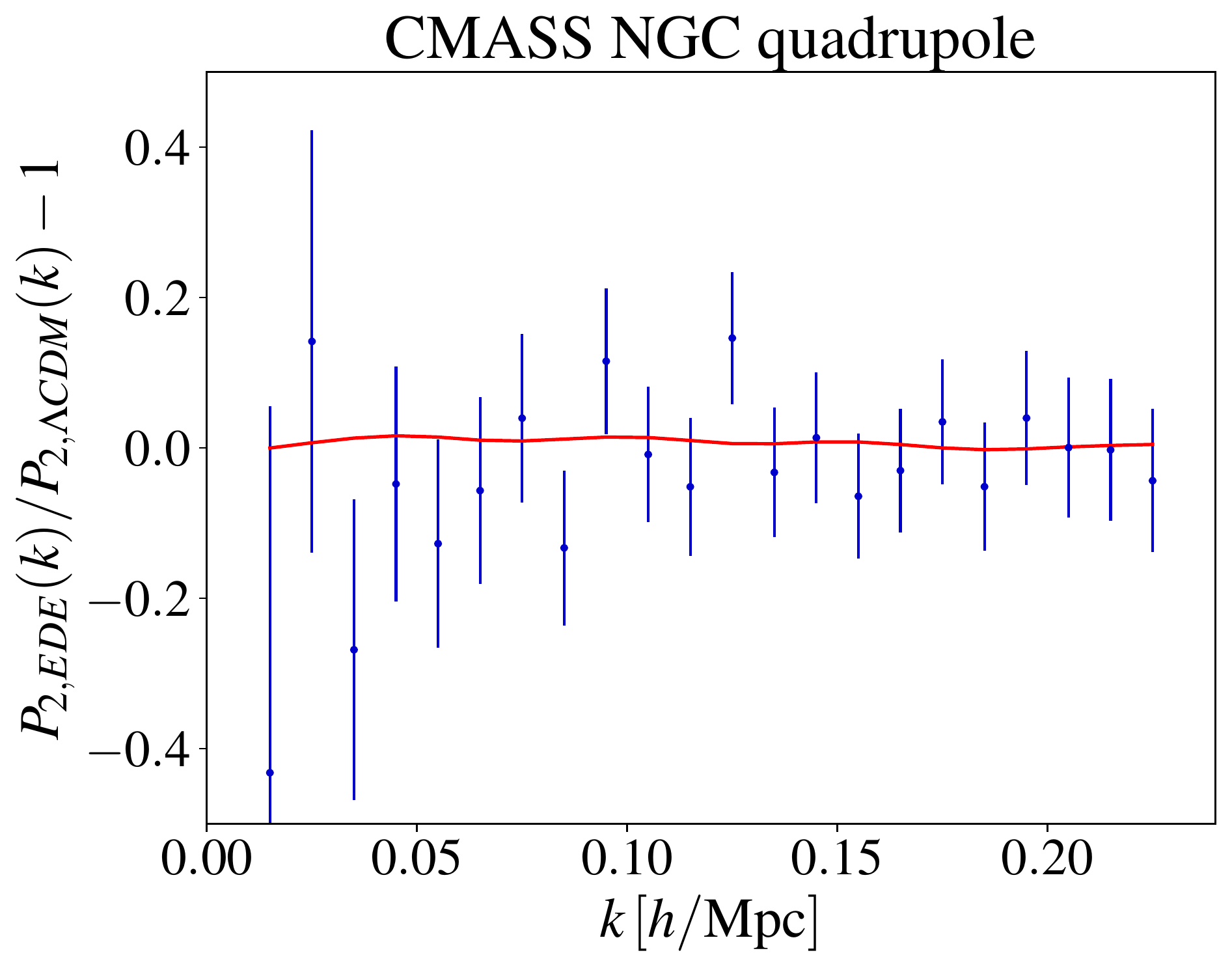}
    \caption{ Residuals of the monopole (left) and quadrupole (right) of the galaxy power spectrum calculated with the EFTofLSS between the  EDE cosmology and $\Lambda$CDM. }
    \label{fig:EDE_pgal}
\end{figure} 
 
It has been argued that the full-shape analysis of the galaxy power spectrum of BOSS using the EFTofLSS (loosely named `EFTBOSS' data)  disfavors the EDE model~\cite{Ivanov:2020ril,DAmico:2020ods}. 
Indeed, fitting the BAO data in 2D and 3D at different comoving distances in a galaxy clustering survey (typically at $z\sim0.1-1$) requires an increase in $\omega_{\rm cdm}$ in the EDE cosmology \cite{Poulin:2018cxd,Jedamzik:2020krr}, which can affect the fit to the full shape~\cite{DAmico:2020ods,Ivanov:2020ril}. 
Thus, galaxy-clustering data offer a way to break the degeneracy introduced by EDE through the constraints they provide on $\omega_{\rm cdm}$ and $\sigma_8$.
To gauge the constraining power of BOSS data, we show in Fig.~\ref{fig:EDE_pgal} the residuals between an EDE cosmology and the $\Lambda$CDM prediction of the monopole and quadrupole of the galaxy power spectrum measured by BOSS in the north galactic cap (NGC) analyzed with the EFTofLSS. 
Here, we fix the cosmological parameters by fitting to Planck+BAO+Pantheon+SH0ES and then simply fit the EFT nuisance parameters to the data, to ``predict'' the signal to be searched for with BOSS data. One can see that the differences between EDE and $\Lambda$CDM are small compared to the data statistical errors, after marginalizing over the theoretical uncertainty, with a change in $\chi^2$ of $+1.1$ in the EDE cosmology. Moreover, the p-value associated with the fit is acceptable ($\sim 16.7\%$) \cite{Simon:2022adh}.
Yet, Refs.~\cite{DAmico:2020ods,Ivanov:2020ril} derived $f_{\rm EDE}(z_c)<0.08$ and $f_{\rm EDE}(z_c)<0.072$ respectively, with two different implementations of the EFT. 
The addition of $S_8$ measurements from KiDS/DES/HSC in Ref.~\cite{Ivanov:2020ril} further tightens the constraint on $f_{\rm EDE}(z_c)<0.053$. 

The original interpretation of the additional constraining power suggested in Refs.~\cite{DAmico:2020ods,Ivanov:2020ril} was disputed in Refs.~\cite{Smith:2020rxx,Murgia:2020ryi,Herold:2021ksg}. 
As mentioned above regarding the constraints from $S_8$ data, the apparent constraining power from the BOSS full-shape analysis may be amplified by the impact of the prior volume artificially favoring $\Lambda$CDM in the Bayesian context. 
This was explicitly verified with a profile-likelihood approach  \cite{Herold:2021ksg,Reeves:2022aoi} which indicated that EDE is favored by the combination of {\it Planck}+BOSS at the $\sim 2\sigma$ level, with $f_{\rm EDE}(z_c)=0.072\pm0.036$ (see discussion in Sec.~\ref{sec:prof}).  
This is similar to the value found in Refs.~\cite{Smith:2020rxx,Murgia:2020ryi} when fixing $\theta_i$ and $z_c$ to their best-fit values. A similar argument was put forward in the case of the NEDE model, for which the inclusion of BOSS data has insignificant impact on the constraints \cite{Niedermann:2020qbw}. 

\begin{figure*}
    \centering
    \includegraphics[width=0.49\columnwidth]{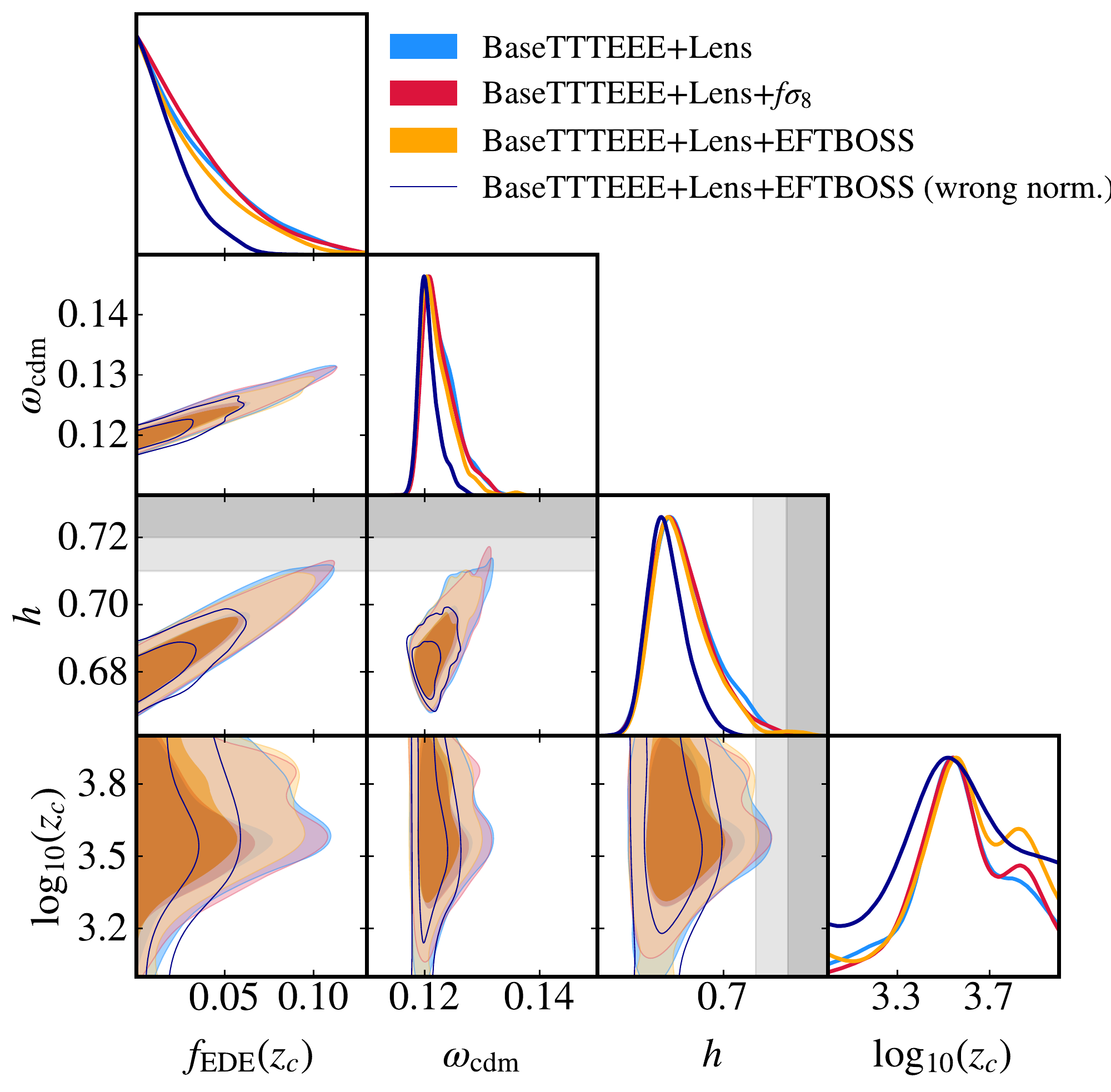}
    \includegraphics[width=0.49\columnwidth]{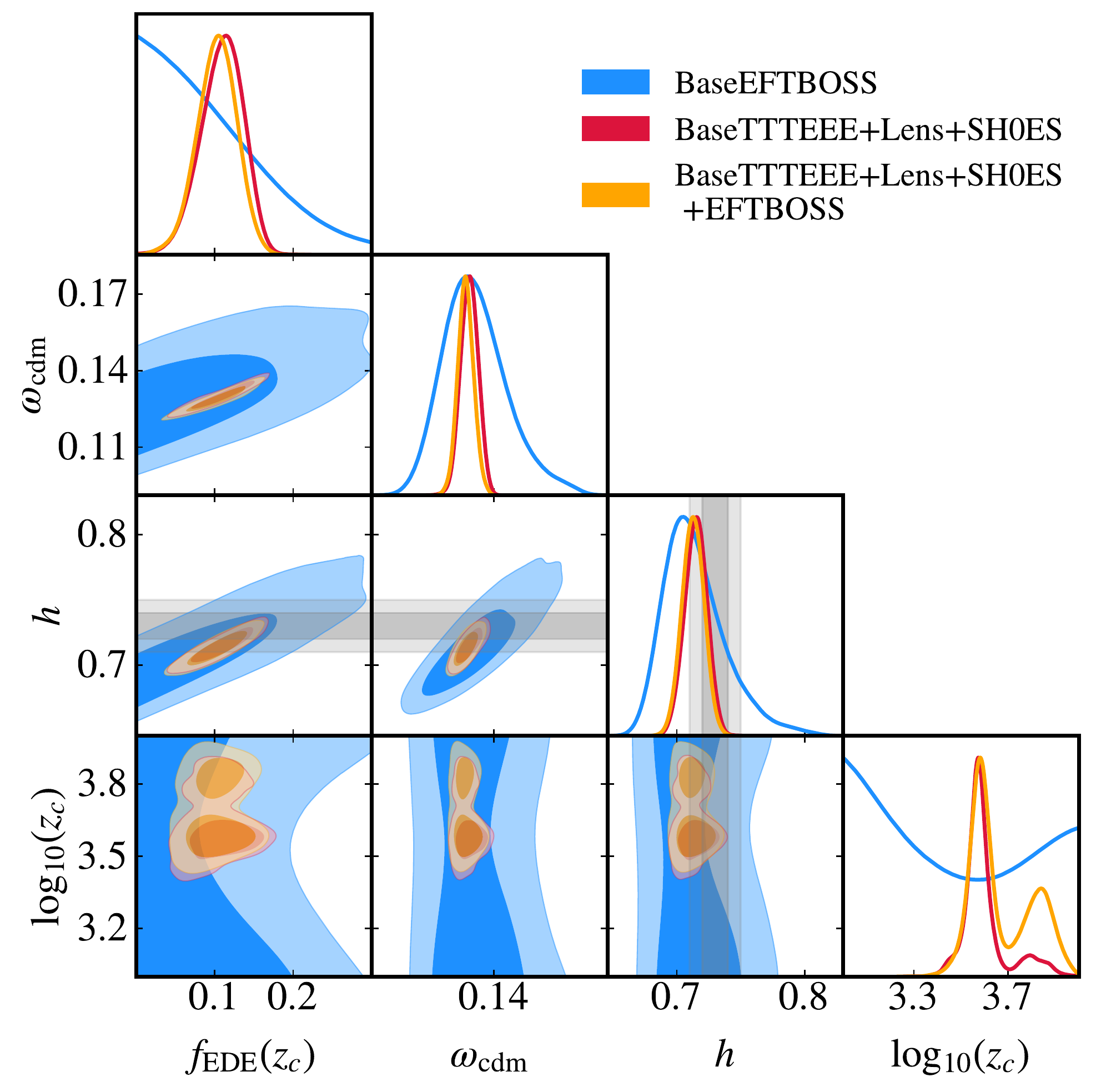}
    \caption{{\it Left panel:} 2D posterior distributions from BaseTTTEEE+Lens, BaseTTTEEE+Lens+$f\sigma_8$ and BaseTTTEEE+Lens+EFTBOSS. 
    We also show the results from the EFTBOSS data with an incorrect normalization for comparison. 
    {\it Right panel:} 2D posterior distributions from BaseEFTBOSS, and BaseTTTTEEE+Lens+SH0ES, with and without EFTBOSS data. BaseTTTTEEE refers to \Planck TTTEE+BAO+Pan18, while BaseEFTBOSS refers to EFTBOSS+BBN+Lens+BAO+Pan18.  }
    \label{fig:EDE_EFT_Planck}
\end{figure*}

Furthermore, Ref.~\cite{Smith:2020rxx} showed that the additional constraints from EFTBOSS come at least in part from a $\sim 20\%$ mismatch in the overall amplitude (typically parameterized by the primordial power spectrum amplitude $A_s$) between BOSS and \Planck, rather than tighter constraints on $\omega_{\rm cdm}$. 
Recently, it was found that the original EFTBOSS data used in these analyses were affected by an inconsistency between the normalization of the survey window function and that of the data measurements, that led to a mismatch in $A_s$.
Constraints obtained when combining \Planck\ and EFTBOSS  are shown in Fig.~\ref{fig:EDE_EFT_Planck}.
After implementing the correction of the normalization of the window function, the combination of \Planck TTTEEE+Lens+BAO+Pan18+EFTBOSS leads to $f_{\rm EDE}(z_c)<0.083$, which is a $\sim10\%$ improvement over the constraints without BOSS data, and a $\sim 5\%$ improvement over the constraints with conventional BAO/$f\sigma_8$ data \cite{Simon:2022adh}. The Hubble tension is reduced to the $2.1\sigma$ level in the EDE cosmology ($1.9\sigma$ without EFTBOSS) compared to $4.8\sigma$ in the $\Lambda$CDM model, and one finds $f_{\rm EDE}(z_c)=0.103^{+0.027}_{-0.023}$ at $z_c=3970^{+255}_{-205}$ when the SH0ES prior is included. 

\begin{figure*}
    \centering
    \includegraphics[width=0.49\columnwidth]{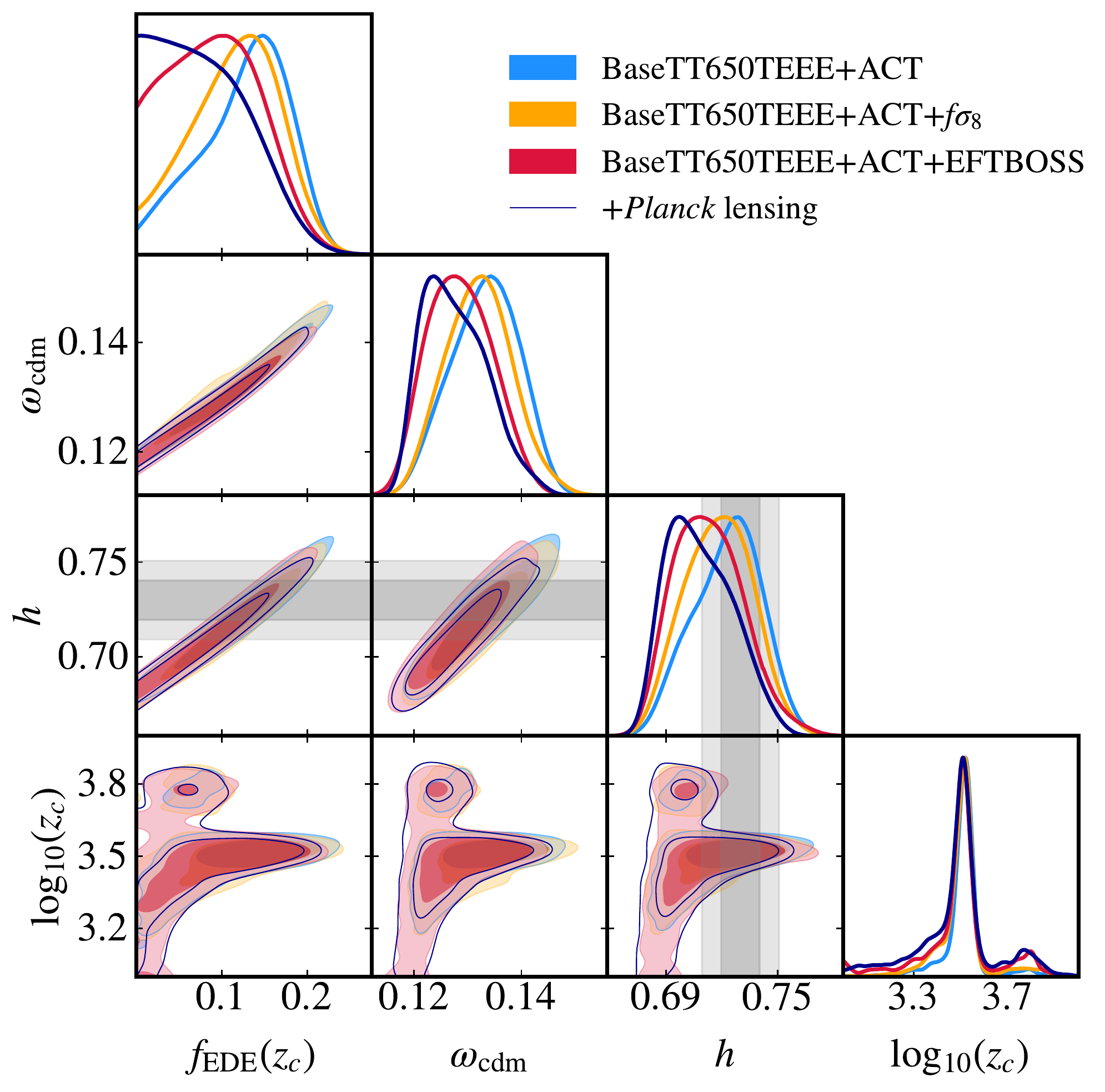}
    \includegraphics[width=0.49\columnwidth]{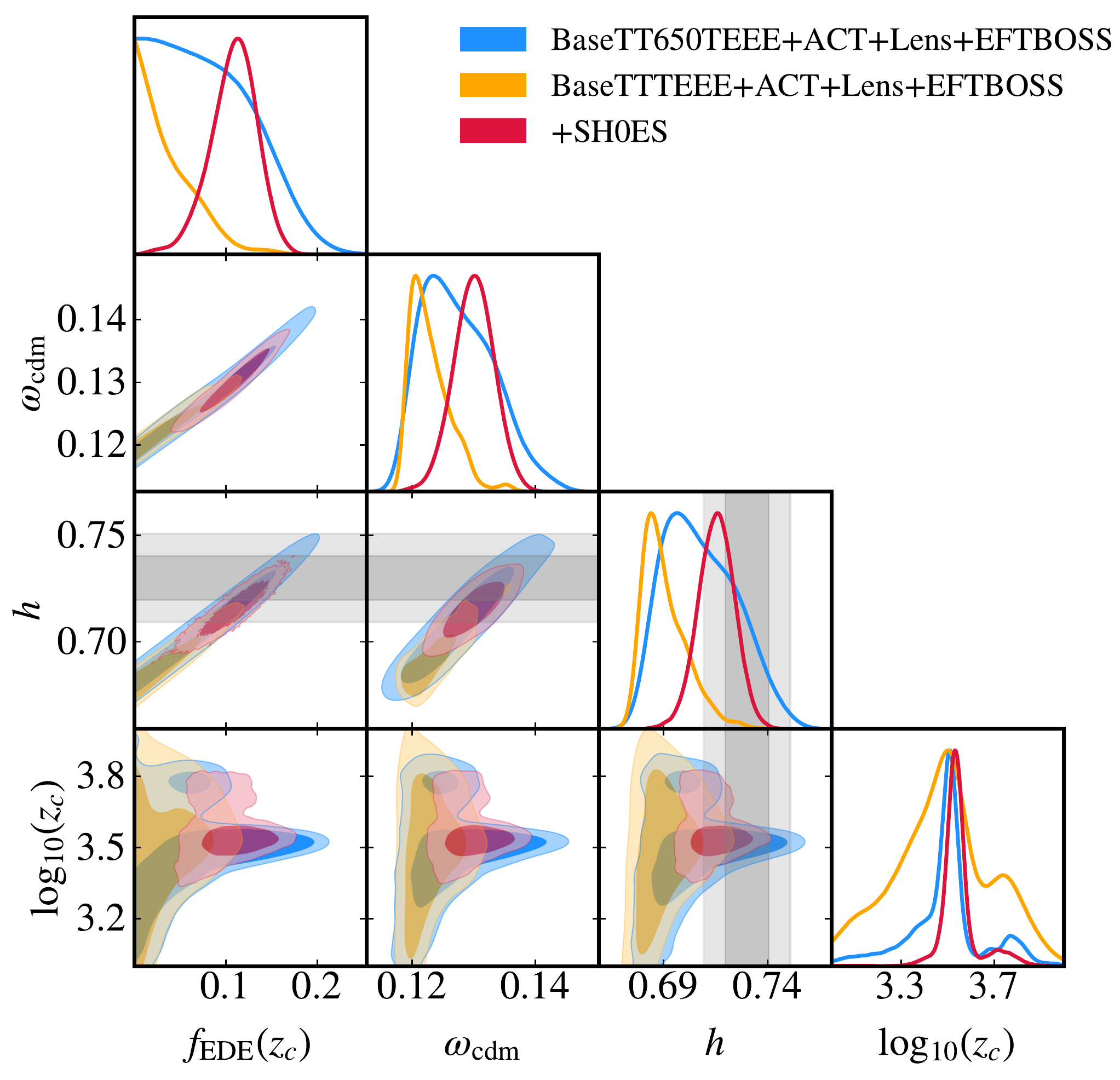}
    \caption{{\it Left Panel:} 2D posterior distributions from BaseTT650TEEE+ACT in combination with $f\sigma_8$, EFTBOSS and \Planck{} lensing. Note that BaseTT650TEEE refers to \Planck TT650TEEE+BAO+Pan18 data. 
    {\it Right Panel:}  2D posterior distributions from ACT+Lens+EFTBOSS in combination with either BaseTT650TEEE, or BaseTTTEEEE  with and without SH0ES. }
    \label{fig:EDE_EFT_ACT}
\end{figure*}
Nonetheless, EFTBOSS data can impact the recent hints of EDE observed in ACT DR4 data. 
Ref.~\cite{Simon:2022adh} showed that they reduce the preference for EDE over $\Lambda$CDM seen when analyzing ACT DR4, alone or in combination with restricted \Planck\ TT data. 
Constraints from combining ACT and EFTBOSS with \Planck\ are shown in Fig.~\ref{fig:EDE_EFT_ACT}.
The combination of \Planck TT650TEEE+Lens+BAO+Pan18+ACT+EFTBOSS leads to a mild constraint on $f_{\rm EDE}(z_c)<0.172$ with $\Delta\chi^2({\rm EDE}-\Lambda{\rm CDM})=-11.1$,  to be compared with $f_{\rm EDE}(z_c)=0.128^{+0.064}_{-0.039}$ without EFTBOSS, with $\Delta\chi^2({\rm EDE}-\Lambda{\rm CDM})=-14.6$.  
When full \Planck{} data are included, the constraints narrow to $f_{\rm EDE}(z_c)<0.110$. 
Nevertheless, when all CMB data are included in combination with EFTBOSS, the Hubble tension is reduced to $1.5\sigma$ in the EDE model, to be compared with $4.7\sigma$ in $\Lambda$CDM. 
The inclusion of the SH0ES prior leads to $f_{\rm EDE}(z_c)=0.108^{+0.028}_{-0.021}$ at $z_c = 3565^{+220}_{-495}$.
Therefore, EFTBOSS data do not rule out EDE as a resolution to the Hubble tension, although they do constrain very high EDE fractions, as seen from analyzing ACT DR4 data. 

A novel way incorporating of EFTBOSS data was recently suggested in Refs.~\cite{Philcox:2020xbv,Farren:2021grl,Philcox:2022sgj}. 
Their constraining power on $h$ predominately comes
from the BAO sensitivity to the sound horizon, and the
same is true of measurements of the CMB. 
It is therefore of interest to develop new analysis methods
that extract information about $h$ from observations which
are based on pre-recombination physics without relying
on the value of the sound horizon. 
By marginalizing over the $r_s$-information within BOSS data and combining with constraints from light-element abundances, CMB lensing and $\Omega_m$ from the Pantheon+ Type Ia supernovae, Ref.~\cite{Philcox:2022sgj} recently obtained the tight constraint $H_0=64.8^{+2.2}_{-2.5}$km s$^{-1}$Mpc$^{-1}$ within the $\Lambda$CDM model. 
As this determination does not depend on $r_s$, it has been suggested that it may constrain models that change the CMB prediction of $H_0$ by adjusting the sound horizon. 
In particular, any mismatch between the value of $H_0$ obtained from analyses that consider information from BAO, and those that marginalize over $r_s$ would be a smoking-gun signal of the inconsistency of the model \cite{Farren:2021grl}.  
With current data however, Ref.~\cite{Smith:2022iax} showed that the consistency tests based on comparing $H_0$ posteriors with and without $r_s$ marginalization are currently inconclusive and in the $r_s$-marginalized analysis, EDE is not in significant tension with the SH0ES determination of $H_0$.
Finally, the galaxy power spectrum is not the only LSS observables affected by the presence of EDE. 
For instance, Ref.~ \cite{Klypin:2020tud} has found that the predicted halo mass function is significantly different in the EDE model. 
The EDE model predicts more halos at any redshift, but the difference is small at $z = 0$, a $1-10\%$ effect for very massive clusters. 
Interestingly, at larger redshifts, the number of halos in EDE is substantially larger than in $\Lambda$CDM. 
For example, the EDE model predicts about 50\% more massive clusters of mass $M = (3-5)\times10^{14}h^{-1}M_\odot$ at $z=1$. 
Such differences will soon be tested by JWST observations \cite{Endsley,Klypin:2020tud}, and in fact, the first publicly available data are, in part, better fit by EDE than \lcdm\ \cite{Boylan-Kolchin:2022kae}. 
On the other hand, the recent work of Ref.~\cite{Goldstein:2023gnw} shows that  Lyman-$\alpha$ data from eBOSS \cite{Chabanier:2018rga}, the XQ-100 \cite{10.1093/mnras/stw3372} and MIKE/HIRES \cite{Viel:2013fqw} quasar samples are in tension with the EDE prediction, leading to the strongest constraints to date on $f_{\rm EDE}$. 
In fact, even under $\Lambda$CDM, Lyman-$\alpha$ data favor significantly different value of the power spectrum tilt and amplitude at small scales, than reconstructed from analysis of CMB data, which (barring a simple statistical fluctuation) may indicate either a systematic effect, or some new physics is at play such as the running of the spectral tilt \cite{Palanque-Delabrouille:2019iyz} or a new type of DM interactions \cite{Hooper:2021rjc}.

\subsection{Age of the Universe tension}

\begin{figure}
    \centering
    \includegraphics[width=0.8\columnwidth]{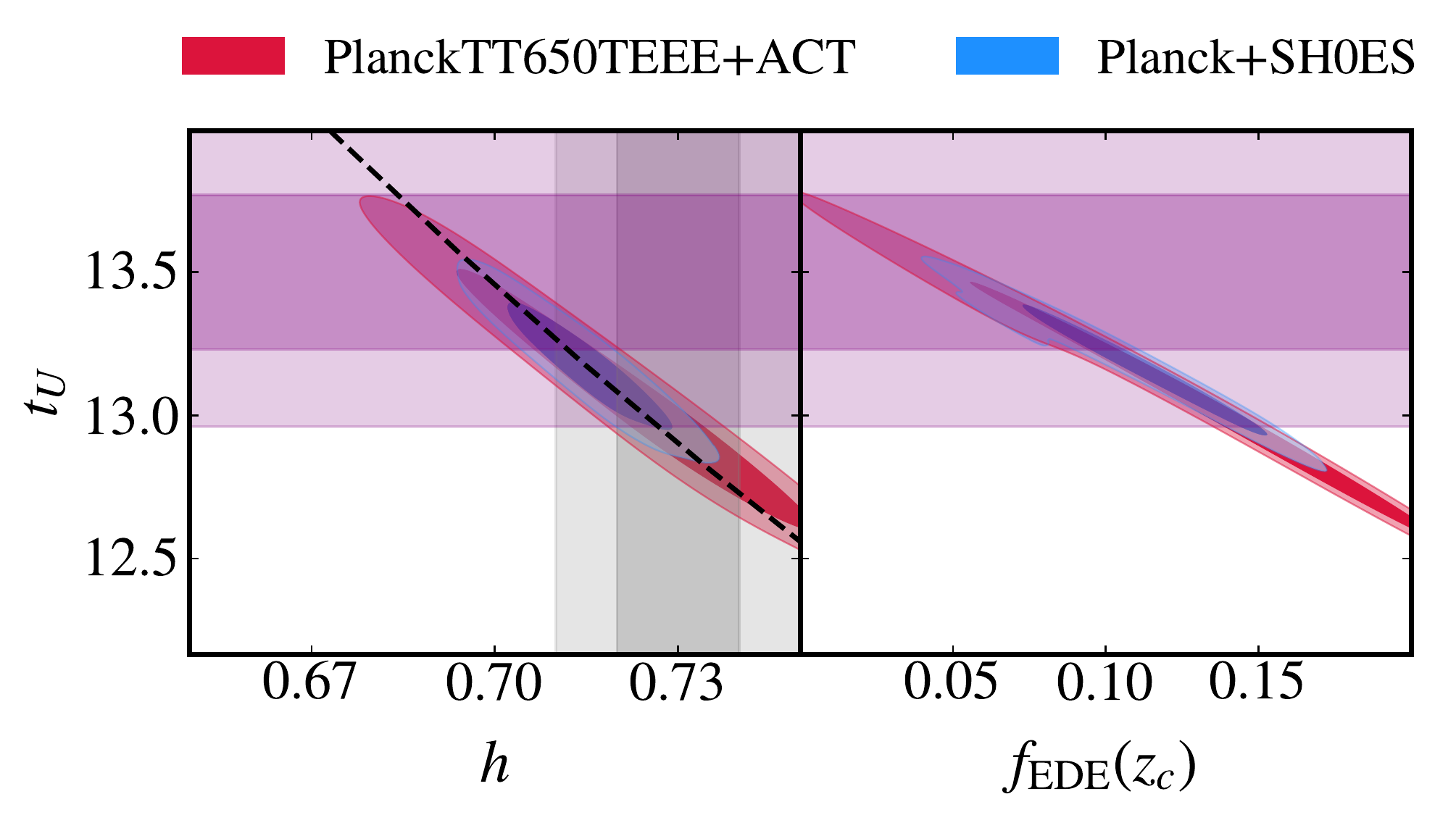}
    \caption{The age of the universe $t_{\rm U}$ vs $\{h,f_{\rm EDE}(z_c)\}$ in the axEDE cosmology. The purple bands mark the recent measurements from GCs, $t_{\rm U}=13.5\pm0.27$ Gyrs \cite{Valcin:2020vav,Valcin:2021jcg,Bernal:2021yli}. The dashed black line shows the prediction from Eq.~\ref{eq:tU}, accounting for the small difference in $\Omega_m$. }
   \label{fig:tU_EDE}
\end{figure}

Another interesting consequence of the presence of Early Dark Energy in the early universe is that the age of the Universe is significantly smaller than $\Lambda$CDM. 
Given the definition of time $t$ in an FLRW universe with scale factor $a$
\begin{equation}\label{eq:t_U}
    t(a)=\int_0^a\frac{da}{aH(a)} \,,
\end{equation}
one finds (assuming flat-$\Lambda$CDM and radiation is sub-dominant) that the age of the Universe today is approximately \cite{Boylan-Kolchin:2021fvy}
\begin{equation}
\label{eq:tU}
    t_{U}\equiv t(a=1) 
  \simeq t_{\rm pl}\bigg(\frac{\Omega_m}{\Omega_{m,{\rm pl}}}\bigg)^{-0.28}\bigg(\frac{h_{\rm pl}}{h}\bigg)\,.
\end{equation}
As EDE reaches high $H_0$ without changing $\Omega_m$ significantly, one immediately sees that the higher $h$ value measured by SH0ES leads to a universe $\sim 7.5\%$ younger than $\Lambda$CDM predicts.
This is illustrated in Fig.~\ref{fig:tU_EDE}. 

This raises a potential issue in EDE cosmologies, and more generally, models that resolve the Hubble tension by adjusting the sound horizon, as it may lead to tensions with the measured ages of old objects such as globular clusters of stars (GCs) \cite{Bernal:2021yli,Boylan-Kolchin:2021fvy,Vagnozzi:2021tjv}. 
Recently, such measurements have been shown to be in slight tension with predictions from the axEDE cosmology fit to {\it Planck}+SH0ES, namely $t_U =13.17_{-0.16}^{+0.14}$ Gyr \cite{Bernal:2021yli,Boylan-Kolchin:2021fvy}. 
Here, we illustrate this potential tension in Fig.~\ref{fig:tU_EDE}, for the axEDE model reconstructed from our \Planck TT650TEEE+ACT and \Planck+SH0ES analyses (two cosmologies for which the Hubble tension is resolved). 
The purple bands mark the recent measurements from GCs, $t_{\rm U}=13.5\pm0.27$ Gyrs \cite{Valcin:2020vav,Valcin:2021jcg,Bernal:2021yli} (see also Ref.~\cite{Moresco:2022phi,Cimatti:2023gil} for a discussion of other measurements). 
One can see that while the predicted age of the Universe is systematically lower than that measured from GCs, the differences are not yet statistically significant. 
The dashed black line shows the prediction from Eq.~\ref{eq:tU}, accounting for the small difference in $\Omega_m$. 
Any model that only modifies the early universe would follow a similar degeneracy line, and therefore can be tested with more accurate measurements of $t_U$.

Similar considerations apply to cosmic chronometers (e.g.~Ref.~\cite{Vagnozzi:2020dfn,Moresco:2022phi}) which are currently too imprecise to probe the EDE scenarios \cite{Schoneberg:2021qvd}, but could provide interesting tests in the future.
In fact, Eq.~\ref{eq:t_U} indicates that the entire time-redshift relation is affected by the EDE model. 
Ref.~\cite{Boylan-Kolchin:2021fvy} points out that it differs from the base \lcdm{} model by at least $\sim 4\%$ at all $z$ and $t$, and therefore this difference represents an important theoretical uncertainty on the connection between $z$ and $t$. 
This can have significant implications for the era of reionization, that occurs at different times in the EDE and \lcdm{} cosmologies. 
As a result, given the formation time of a typical globular cluster, such objects may form {\it before} (in the EDE cosmology) or {\it after} (in \lcdm{}) reionization. 
Yet, current uncertainties on the ages of such objects are too large to unambiguously constrain EDE. 
Given the constraints on $\Omega_m$ from BAO and Pantheon data, this potential issue in the EDE cosmology is very generic to models resolving the Hubble tension solely by modifying the pre-recombination era \cite{Bernal:2021yli}, and may indicate modifications the late-universe dynamics to fully restore cosmic concordance. 
Ref.~\cite{Bernal:2021yli} in fact proposes the use of ternary plots to
simultaneously visualize independent constraints on key quantities related to $H_0$ like $t_U$, $r_s$, and $\Omega_m$. 
These quantities representing an over-constrained problem can be used as a diagnostic tool of consistency, and help find solutions to the $H_0$ tension. 
For further discussion about the age of the Universe tension and future prospects, we refer to Refs.~\cite{Vagnozzi:2021tjv,Boylan-Kolchin:2021fvy,Valcin:2021jcg,Bernal:2021yli}.

\subsection{Impact of EDE for inflation}
\label{sec:inflation}
\begin{figure}[!t]
    \centering
    \includegraphics[width=0.6\columnwidth]{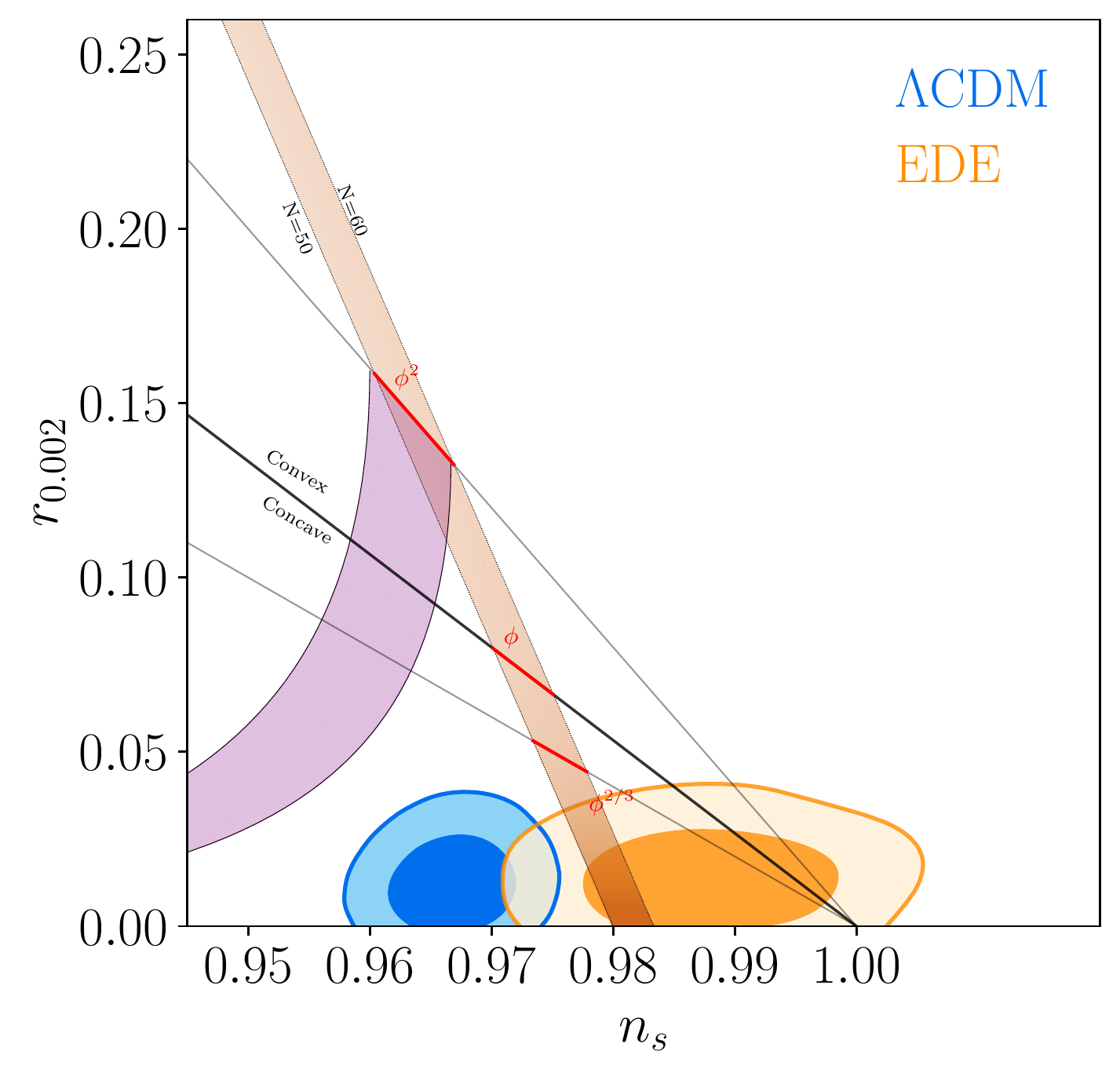}
    \caption{
    Constraints on inflation in the $n_s-r$ plane from the combination of \Planck+Bicep/Keck+BAO data, in the $\Lambda$CDM and EDE models. 
    Constraints are shifted towards larger $n_s\sim 1$ in the EDE model, but the bound on $r$ is unaffected. 
    }
    \label{fig:EDE_inflation}
\end{figure}

The presence of EDE has important consequences for inflation, as the value of $n_s$ reconstructed from the fit to CMB data is significantly larger than in $\Lambda$CDM. 
Under the $\Lambda$CDM model, current \Planck\ data which favor $n_s\sim 0.96$ and $r\lesssim 0.1$ are in perfect agreement with predictions from slow-roll inflation, which suggests $n_s -1 \sim {\cal O}(1)/N$ with $N\sim 60$ the number of e-folds of inflation, such as Starobinsky (or $R^2$)  inflation. 
One the other hand, they disfavor power law (or convex) potentials \cite{Planck:2018jri}.  
However, if the Hubble tension persists, and the preference for EDE models with $n_s\simeq 1$ increases, it can reshape our understanding of cosmic inflation \cite{Takahashi:2021bti,Cruz:2022oqk}. 
We illustrate bounds on $n_s-r$ in the $\Lambda$CDM and EDE models in Fig.~\ref{fig:EDE_inflation}.
All constraints on the standard $n_s-r$ plane are shifted towards larger $n_s$, while the bound on $r$ is essentially unaffected by the presence of EDE \cite{Cruz:2022oqk}.
In the EDE case, convex potentials can generally be rescued, with the important consequence that Starobinsky inflation is disfavored compared to power-law potential, or the curvaton model.

It is important to note that the preference for $n_s \sim 1$ is very generic to models that resolve the Hubble tension by increasing the pre-recombination expansion rate (such as an extra radiation species $\Delta N_{\rm eff}$). 
As discussed in the introduction, it follows from the fact that the angular scale $\theta_D$ corresponding to damping roughly depends on 
\begin{equation}
    \theta_D(z_*) \sim \frac{H_0}{\sqrt{\dot \tau(z_*) H(z_*)}},
    \label{eq:theta_D}
\end{equation}
where $ \dot \tau(z_*)=n_e(z_*) x_e(z_*) \sigma_T/(1+z_*)$, $n_e$ is the electron density, $x_e$ is the ionization fraction, and $\sigma_T$ is the Thomson cross section. 
If the indirect $H_0$ increases by a fraction $f$, with $\theta_s(z_*)$ fixed, $\theta_D(z_*)$ increases by $f^{1/2}$, leading to a suppression of power, which  can be compensated for by increasing $n_s$.
Although details of specific models will change how much the value of $n_s$ must increase, this degeneracy is very generic, and should therefore be considered an important caveat to our current understanding of cosmic inflation (regardless of the cosmological nature of the Hubble tension).
 
\subsection{An extended EDE model to restore cosmic concordance}

The fact that EDE models cannot explain the low-value of $S_8$ (and even slightly worsen the tension) is often presented as a major argument against the existence of an EDE-like component in the pre-recombination era. 
Given that both the $\Lambda$CDM cosmology and the EDE cosmology are in tension with $S_8$ measurements, it is nevertheless reasonable to ask whether there exist EDE extensions that would help in accommodating the low $S_8$ value. 
In fact, as discussed in this review, it is striking that at fixed $\omega_{\rm cdm}$, the EDE leads to a decrease in power at small scales that seemingly goes in the right direction to resolve the $S_8$ tension.
The problem arises from the EDE-$\omega_{\rm cdm}$ degeneracy that counteracts the effect of the EDE on the gravitational potential wells as seen in the CMB.
It is hence possible that a simple extension of the EDE model would mitigate the need for an increase in the CDM density, or alternatively compensate  the associated increase in the amplitude of fluctuations.

One of the least `theoretically costly' explanations of the $S_8$ tension is to invoke massive neutrinos that already lead to a power suppression at small scales which decreases the value of $\sigma_8$. 
Unfortunately, in practice, the sum of neutrino masses $\sum m_\nu\sim0.3$eV required to resolve the tension is excluded by \Planck~data (e.g. \cite{Planck:2018vyg,FrancoAbellan:2020xnr}). Moreover, neither DES nor KiDS prefer a non-zero $\sum m_\nu$ \cite{DES:2021wwk,Heymans:2020gsg}. 
However, constraints on the sum of neutrino masses are substantially broadened in extended cosmologies (e.g. \cite{Chacko:2019nej,Oldengott:2019lke}). 
In the context of EDE, simply adding the neutrino mass as free parameter in an analysis of the EDE model does not change the reconstructed value of $S_8$ \cite{Murgia:2020ryi,Fondi:2022tfp}.
Adding more complexity, Ref.~\cite{Sakstein:2019fmf} suggested that the non-relativistic transition of neutrinos could trigger a phase-transition in EDE. 
Further exploration of any possible connections between EDE and neutrino masses would be interesting. 

Another proposal showed that the existence of a dark sector, where DM is composed of $\sim 5\%$ ultra-light axions (ULAs), can help in relieving the $S_8$ tension without affecting the success of the EDE solution in explaining a high-$H_0$ \cite{Allali:2021azp}. 
Such a dark sector could be the manifestation of the presence of numerous scalar-fields that are ubiquitous in string theory \cite{Svrcek:2006yi,Arvanitaki:2009fg}.
Ref.~\cite{Clark:2021hlo} obtained a similar result considering a decaying DM model instead of the ULA component. 
While these models may seem `ad-hoc', and disfavored from the `Occam's razor' point-of-view, they demonstrate that $H_0$ and $S_8$ solutions can come from different (potentially disconnected) sectors, and therefore that one should not dismiss EDE models on the basis that they cannot simultaneously explain both tensions. 

Nevertheless, there exist attempts of connecting $S_8$ and $H_0$ resolutions within the same theoretical models. 
For instance, Ref.~\cite{McDonough:2021pdg} points out that the Swampland Distance Conjecture (SDC) implies that the near-Planckian scalar-field excursion required in axion-like EDE models would lead to an exponential sensitivity of the mass $m_{\rm DM}$ of DM, if DM is composed of a scalar-field, $m_{\rm DM}(\phi)= m_0\exp(c\phi/M_{\rm pl})$ where $c$ is a free parameter. 
Authors argue that this additional time-dependent mass manifests as an EDE-induced ``fifth force'' that has the potential of reducing the growth of structure and relieving the tension with LSS data. 
A similar model was discussed for EDE and AdS-EDE in Ref.~\cite{Wang:2022nap} and a chameleon-inspired EDE in Ref.~\cite{Karwal:2021vpk}. 

Another EDE-like model designed to resolve both tensions is presented in Ref.~\cite{Alexander:2022own}. 
In this string-theory-inspired model, an
axion, playing the role of the EDE field, is kinetically coupled to a dilaton field, which syphons off energy density
as the axion falls down its potential, allowing it to redshift faster. 
The introduction of the kinetically-coupled dilaton field allows the use of a standard axion potential for solving the Hubble tension. 
In addition, the axion with mass ${\cal O}(10^{-27})$ eV contributes a small fraction ($\sim 1\%$) to dark matter
post-recombination, and naturally suppresses power on scales sensitive to $\sigma_8$.

Finally, we mention Ref.~\cite{Sabla:2022xzj}, where authors studied EDE solutions to both tensions in the generalized Dark Matter (GDM) framework covers more general microphysical descriptions of the EDE component than the simple scalar-field picture described thus far. 
Interestingly, they found that EDE with an anisotropic sound speed can soften both the $H_0$ and $S_8$
tensions while still providing a quality fit to CMB data. 
Such models will be distinguishable from standard scalar-field EDE with future CMB-S4 data. 

The existence of several models based on EDE dynamics that can alleviate both tensions demonstrate that, with current data, one should not dismiss the EDE resolutions of the Hubble tension on the basis that it does not simultaneously address the $S_8$ tension. 
Nevertheless, it is clear that there are irreducible features that are signs of an EDE-like component acting pre-recombination, and future CMB and LSS data will be pivotal in settling the fate of the EDE resolution to the Hubble tension.

\section{Conclusions}
\label{sec:concl}

Early Dark Energy remains one of the most effective models at resolving the Hubble tension \cite{Schoneberg:2021qvd}, the discrepancy between direct and indirect estimates of the Hubble constant, and as such has received a lot of attention in the recent literature. 
In this review, we present the status of this solution, describing both the qualitative requirement that a successful EDE model must have, and the quantitative results obtained for various EDE models that have been studied so far (described in Sec.~\ref{sec:EDE_models}). 
In a nutshell, EDE refers to an additional component of dark energy active at early times (typically manifested by a scalar-field) that quickly dilutes away at a redshift close to that of matter-radiation equality. 
The role of EDE is to decrease the sound horizon by briefly contributing to the Hubble rate in the pre-recombination era. Different models (and different perturbation dynamics) of EDE lead to subtle signatures in the CMB and matter power spectra beyond this simple background description. 
Further details of the phenomenological effects of EDE are presented in Sec.~\ref{sec:EDE_pheno}.

We summarize the results of analyses of several EDE models in light of a combination of cosmological data that includes {\it Planck} data (2015 or 2018), BOSS BAO DR12, Pantheon and SH0ES in Fig.~\ref{fig:EDE_whis} and expound on these in Sec.~\ref{sec:EDE_Planck}.
EDE models are typically able to reduce the Hubble tension to below the $2\sigma$ level, although not all models achieve the same level of success. 
In fact, while all models behave similarly at the background level, data favor very specific dynamics for their perturbations, which can differentiate various models. 
For instance, scalar-field power-law potentials are disfavored over potentials that flatten around the initial field value, such as the modified cosine potential in axion-like models. 
Introducing a non-minimal coupling to the Ricci scalar also improves the results of the quartic potential. 

Despite these successes, Bayesian analyses of the EDE models in light of {\it Planck} data do not indicate a preference for EDE in the absence of information from SH0ES on $H_0$. 
On the other hand, likelihood-profile analyses show a $\sim 2\sigma$ preference for EDE from {\it Planck} data alone. 
This illustrates the impact of prior-volume effects inherent to the Bayesian framework, pivotal in disfavoring EDE models, and therefore cautions against naively interpreting results of Bayesian analyses. 
Such differences between the Bayesian and Frequentist approaches are likely tied to the fairly low constraining power of current data on EDE models, and we anticipate that if the preference for (or against) EDE becomes stronger with higher precision data, both frameworks would provide similar results.

In Sec.~\ref{sec:EDE_ACT_SPT}, we reviewed the recent hints of EDE in ACT data, alone or in combination with restricted {\it Planck} TT data, with $H_0$ values that are in good agreement with the SH0ES determination. 
Although supported by {\it Planck} polarization data, which also weakly favor the EDE model on their own, the addition of {\it Planck} high-$\ell$ temperature data removes the preference for EDE, raising the question of a potential (mild) inconsistency between ACT and {\it Planck} high-$\ell$ TT data. 
Future ACT and SPT data, expected next year, will be crucial to confirm or exclude the current hints of EDE. 
Looking forward, it is expected that high-accuracy measurements at high-$\ell$'s ($\gtrsim 1000$) will be sensitive to the characteristic increase of power induced by the EDE cosmology (mostly due to the reshuffling of $\Lambda$CDM parameters as a consequence of the presence of EDE), while improvements at intermediate $\ell$'s ($\sim 50-500$) can help probe the details of the EDE perturbative dynamics.

If EDE is responsible for the current Hubble tension, Fig.~\ref{fig:EDE_whis} illustrates that one generically expects a higher DM density $\omega_{\rm cdm}$ and primordial tilt $n_s$. 
These shifts in parameters have important consequences on the growth of structure at late times, and lead to a very different matter power spectrum (at the 5$-$10\% level in the range $k\sim 0.01-1 ~h$/Mpc) than predicted in $\Lambda$CDM, which can be used to further probe the model (see Secs.~\ref{sec:S8} and \ref{sec:BOSS}). 
In fact, EDE cosmologies predict a slightly larger $S_8$ parameter than $\Lambda$CDM, which is at odds with recent measurements from weak-lensing surveys. 
We have presented constraints from galaxy weak-lensing and clustering surveys such as KiDS, DES and BOSS on the EDE scenarios, concluding that current surveys generally do not exclude the EDE resolution to the Hubble tension (although they can constrain large EDE contributions), and that current models of EDE cannot explain a low-$S_8$. 
Future constraints with DESI and subsequent surveys will be crucial to test EDE models further. 
In addition, we have argued that the existence of an EDE phase may have important consequences for inflation, as the value of $n_s$ reconstructed in EDE models that resolve the Hubble tension favor curvaton models over Starobinsky inflation \cite{Takahashi:2021bti,Cruz:2022oqk}.
There are additional questions associated with EDE cosmologies described in Sec.~\ref{sec:EDE_challenges}, including a second `cosmic coincidence' problem in the EDE cosmology, a potential `age of the Universe' tension and whether additional dynamics on top of the EDE can explain both the $H_0$ and $S_8$ tensions. 

With this review, we argue that EDE is a promising mechanism to resolve the $H_0$ tension, with consequences for a wide variety of observables that will be firmly tested with next-generation surveys. 
Yet, it is clear that this cannot be `the end of the story' at least in its current form, for various observational and theoretical reasons. 
Connections between inflation, late DE and early DE are still largely unexplored, and may provide an interesting path forward for model-building with undetermined consequences for EDE models. 
Nevertheless, further work is required to find a common solution to recent cosmic tensions, that may or may not be based on some aspects presented in this review, and that will hopefully shed new light on the still unknown fundamental nature of dark matter and dark energy.

\acknowledgments{We thank Th\'eo Simon for providing us with the residuals of the galaxy power spectrum calculated with the EFTofLSS between the EDE cosmology and $\Lambda$CDM and Alexa Bartlett and Yashvi Patel for computing the EDE profile likelihood. We thank Kim Berghaus,  Eoin Colg\'ain, Francis-Yan Cyr-Racine, Fabio Finelli, Laura Herold, Emil Brinch Holm, Meng-Xiang Lin, Evan McDonough, Florian Niedermann, Yun-Son Piao,  Adam Riess, Jeremy Sakstein, Martin Sloth, Mark Trodden for useful comments on the draft. This article is based upon work from COST Action CA21136 Addressing observational tensions in cosmology with systematics and fundamental physics (CosmoVerse) supported by COST (European Cooperation in Science and Technology). This project has received funding from the European Research Council (ERC) under the
European Union’s HORIZON-ERC-2022 (Grant agreement No. 101076865). This work used the Strelka Computing Cluster, which is run by Swarthmore College. TLS is supported by NSF Grant No.~2009377 and the Research Corporation.
TK is supported by NASA ATP Grant 80NSSC18K0694 and by funds provided by the Center for Particle Cosmology at the University of Pennsylvania. The authors acknowledge the use of computational resources from the Excellence Initiative of Aix-Marseille University (A*MIDEX) of the “Investissements d’Avenir” programme. These results have also been made possible thanks to LUPM's cloud computing infrastructure founded by Ocevu labex, and France-Grilles. This project has received support from the European Union’s Horizon 2020 research and innovation program under the Marie Skodowska-Curie grant agreement No.~860881-HIDDeN.
}

\appendix
\section{Appendix: Impact of $f_{\rm EDE}$, $a_c$ and $w_f$ on the power spectra}
\label{app:EDE_pheno}

In this section, we discuss the impacts of varying $f_{\rm EDE}$, $z_c$ and $w_f$ on the CMB power spectra. 
We show relative differences with respect to $\Lambda$CDM keeping $\{\omega_{\rm cdm},\omega_b,A_s,n_s,\tau_{\rm reio}\}$ fixed for all curves. 
Finally, we fix either $h$ to emphasize the dominant background effects of EDE (left panels) or fix $\theta_s$ to show the role of perturbations (right panels). 
We refer to the text under Sec.~\ref{sec:EDE_pheno} for a detailed discussion on the role of the sound speed $c_s^2$. 

\subsection{Impact of $f_{\rm EDE}$}

\begin{figure}
    \centering
    \includegraphics[width=0.99\columnwidth]{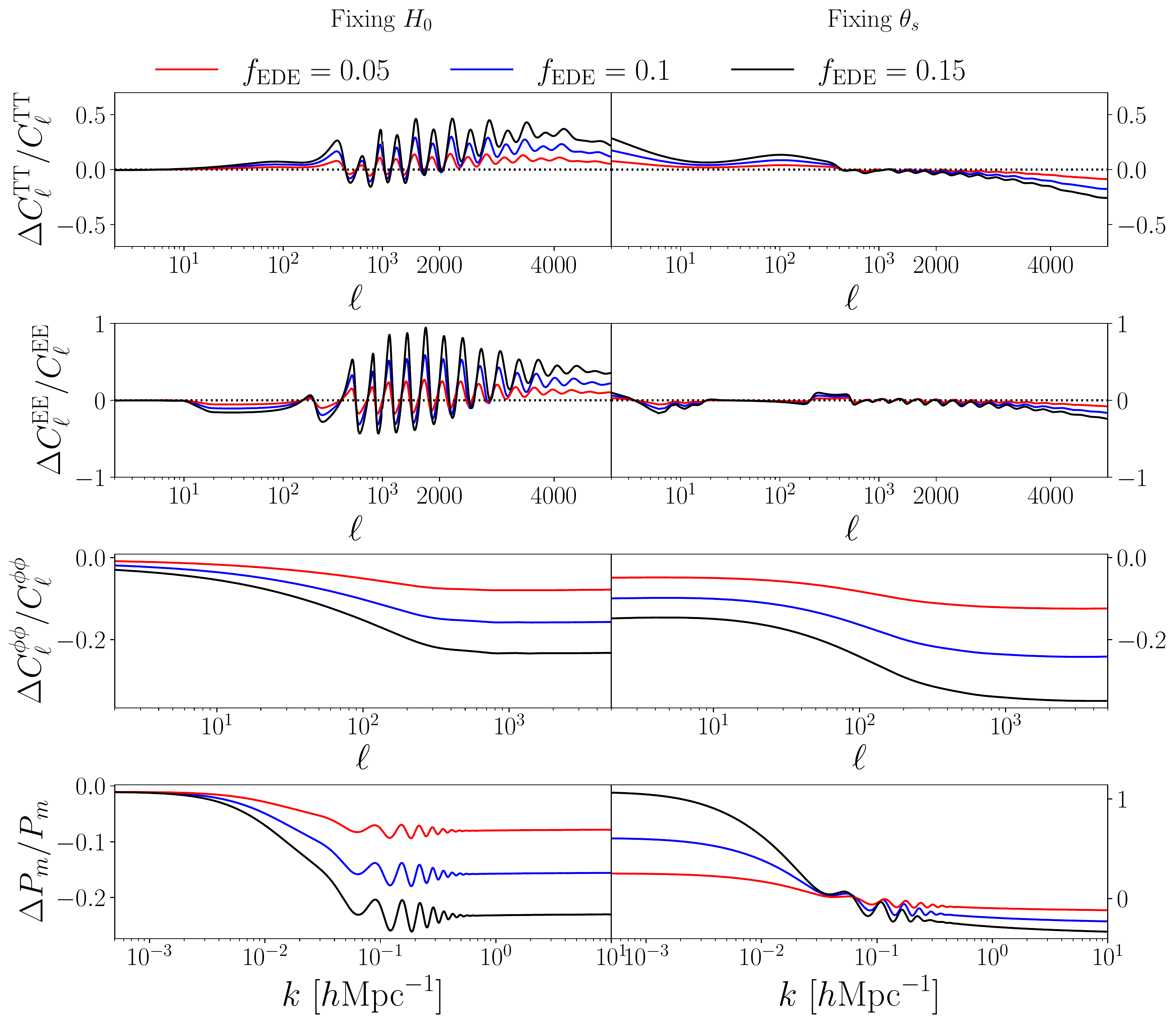}
    \caption{
    Impact of changing $f_{\rm EDE}$ at fixed $\log_{10}(z_c)=3.5$ and $w_f = 1/2$ on the CMB and matter power spectra. 
    We show relative differences with respect to $\Lambda$CDM and choose to keep $\{\omega_{\rm cdm},\omega_b,A_s,n_s,\tau_{\rm reio}\}$ fixed along with either $h$ (left panels) or $\theta_s$ (right panels) in all models.
    Note that the $y$-axis ranges are common across rows for all but the last $\Delta P_m / P_m$ panels that have different ranges as indicated. 
    }
    \label{fig:fede_power_spectra}
\end{figure}

Increasing $f_{\rm EDE}$ at fixed $\log_{10}(z_c)=3.5$ and $w_f = 1/2$ amounts to changing {\it how much} the EDE contributes to $H(z)$ and therefore the amplitude of the reduction of the sound horizon and damping scale. 
This leads to enhanced oscillations in the residuals shown in Fig.~\ref{fig:fede_power_spectra}, left panel, and more power in the damping tail.  
In addition, the EDE contributes more significantly to the acoustic driving of CMB perturbations, and affects the amplitude of the SW term, while leading to a larger residual time-evolution of the gravitational potentials, which affects the amplitude of the eISW term (at $\ell \sim 100$). 
The latter effects are more clearly visible on the right panel. 
In the matter and lensing power spectra, we see the effect of a larger Hubble friction that damps matter perturbations, and leads to a suppression at scales that enter or are within the horizon at $a_c$. 
Note that, once $h$ is adjusted to fix $\theta_s$ (right panel), the small-$k$ branch is largely increased due to the larger $\Omega_m=\omega_{\rm cdm}/h^2$. 

\subsection{Impact of $z_c$}

Increasing $\log_{10}(z_c)$ at fixed $f_{\rm EDE}=0.1$ and $w_f = 1/2$ amounts to changing {\it when} the EDE contributes to $H(z)$. 
This affects the ability of EDE to reduce the sound horizon and damping scale. 
For $\log_{10}(z_c)=4$, one can notice the absence of a shift in the damping tail (left panel). 
The impact of EDE is mostly visible on the acoustic driving of CMB perturbations, and therefore on the amplitude of the SW term (right panel). 
For $\log_{10}(z_c)=3$ on the other hand, EDE does not strongly affect CMB acoustic driving since the universe is essentially matter dominated at that time. 
The effect is therefore dominated by the background impact of the sound horizon and damping scale. 
Pushing $z_c$ to much larger values would lead to EDE impacting smaller and smaller scales. 
Considering much smaller $z_c$, for which the EDE contributes only in the post-recombination era, would affect the angular diameter distance to recombination, and lead to additional decay of the gravitational potential and an ISW effect, similar to the effect of Dark Energy at late-times. 

In the matter and lensing power spectra, the dominant effect is an overall shift of the power suppression,  centered around $k_c = H(z_c)/(1+z_c)$. 
The amplitude of the suppression is also mildly affected, as the EDE contribution before matter-domination leads to shallower suppression at late times. 
Note also that when fixing $\theta_s$, the model with $\log_{10}(z_c)=3$ shows a higher amplitude than when keeping $h$ fixed, compared to other models. 
This is due to the fact that EDE contributes more here than in other models in the post recombination universe, affecting the angular diameter distance, and limiting the required change in $h$ and therefore $\Omega_m$ which scales the overall amplitude of the spectrum \cite{lesgourgues_mangano_miele_pastor_2013}. 
\begin{figure}
    \centering
    \includegraphics[width=0.99\columnwidth]{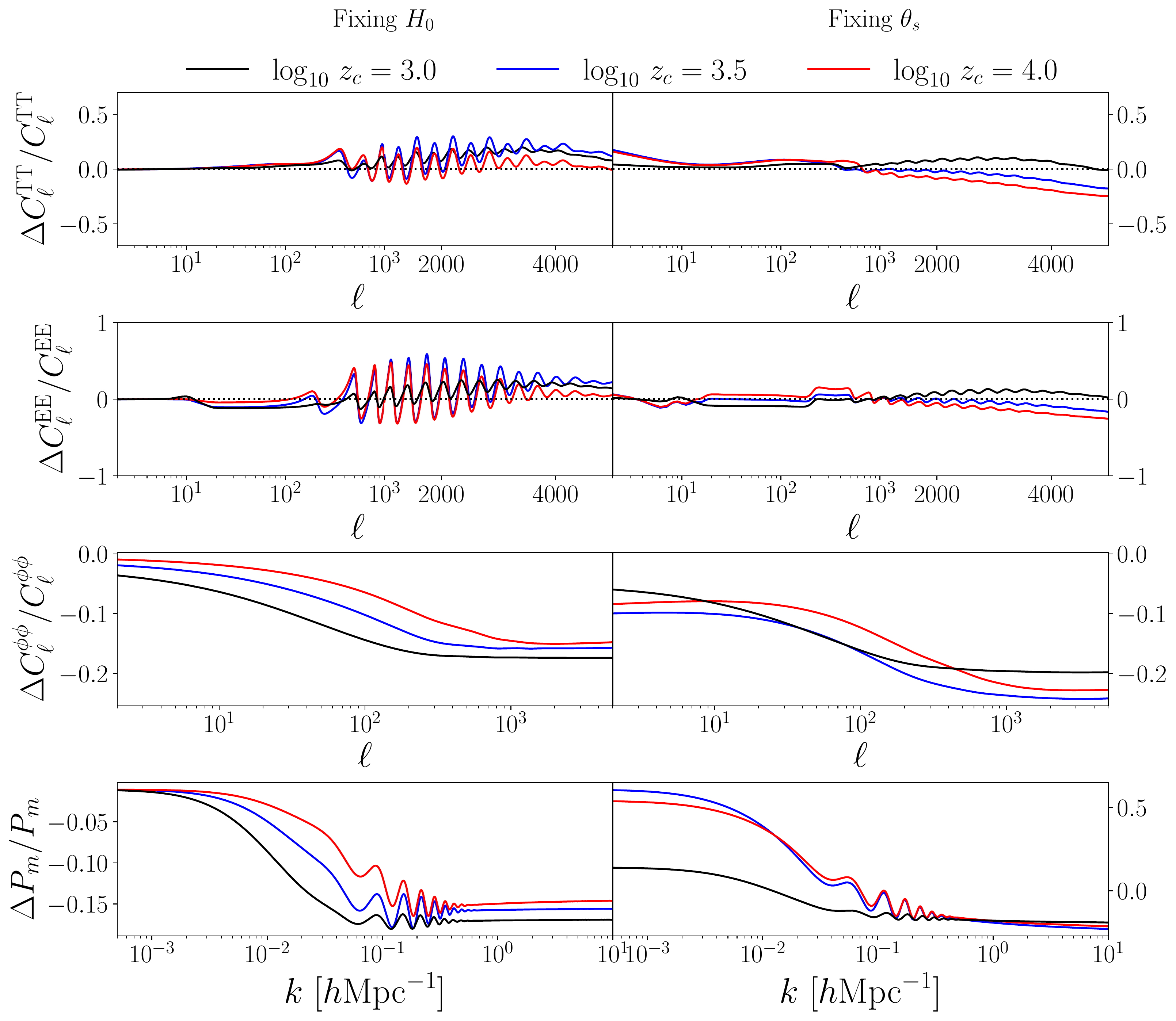}
    \caption{Same as Fig.~\ref{fig:fede_power_spectra} but now changing  $\log_{10}(z_c)$ at fixed $f_{\rm EDE}=0.1$ and $w_f = 1/2$.  }
    \label{fig:log10ac_power_spectra}
\end{figure}

\subsection{Impact of $w_f$}

The impact of changing $w_f$ at fixed $f_{\rm EDE}=0.1$ and  $\log_{10}(z_c)=3.5$ is more subtle as it amounts to changing {\it how long} EDE contributes, and leads to effects that are a combination of those induced by changing $f_{\rm EDE}$ or $\log_{10}(a_c)$. 
Larger $w_f$ means faster EDE dilution and smaller cumulative contribution to $H(z)$. 
The dominant effect on the TT and EE spectra is similar to the effect of larger $f_{\rm EDE}$, namely an overall shift of the peak and more power on small scales, due to a smaller sound horizon and damping scale induced by the larger EDE contribution. 
However, one can see in the matter and lensing power spectra, that both the amplitude and the shape of the power suppression is affected: at fixed $h$ (left panel), a smaller $w_f$ means both that modes are affected longer (larger suppression), and that more modes can be affected  (shift the suppression towards larger scales). Once $\theta_s$ is adjusted (right panel), the change in $h$, and therefore $\Omega_m$, can partly compensate these effects, and even overcome them at large scales.

There are additional subtle effects induced at the level of EDE perturbations, whose growth on super-horizon scales is controlled by $w_f$, that lead to a correlation with $c_s^2$ in the data analysis. 
These are described in more detail in Sec.~\ref{sec:EDE_pheno}.
\begin{figure}
    \centering
    \includegraphics[width=0.99\columnwidth]{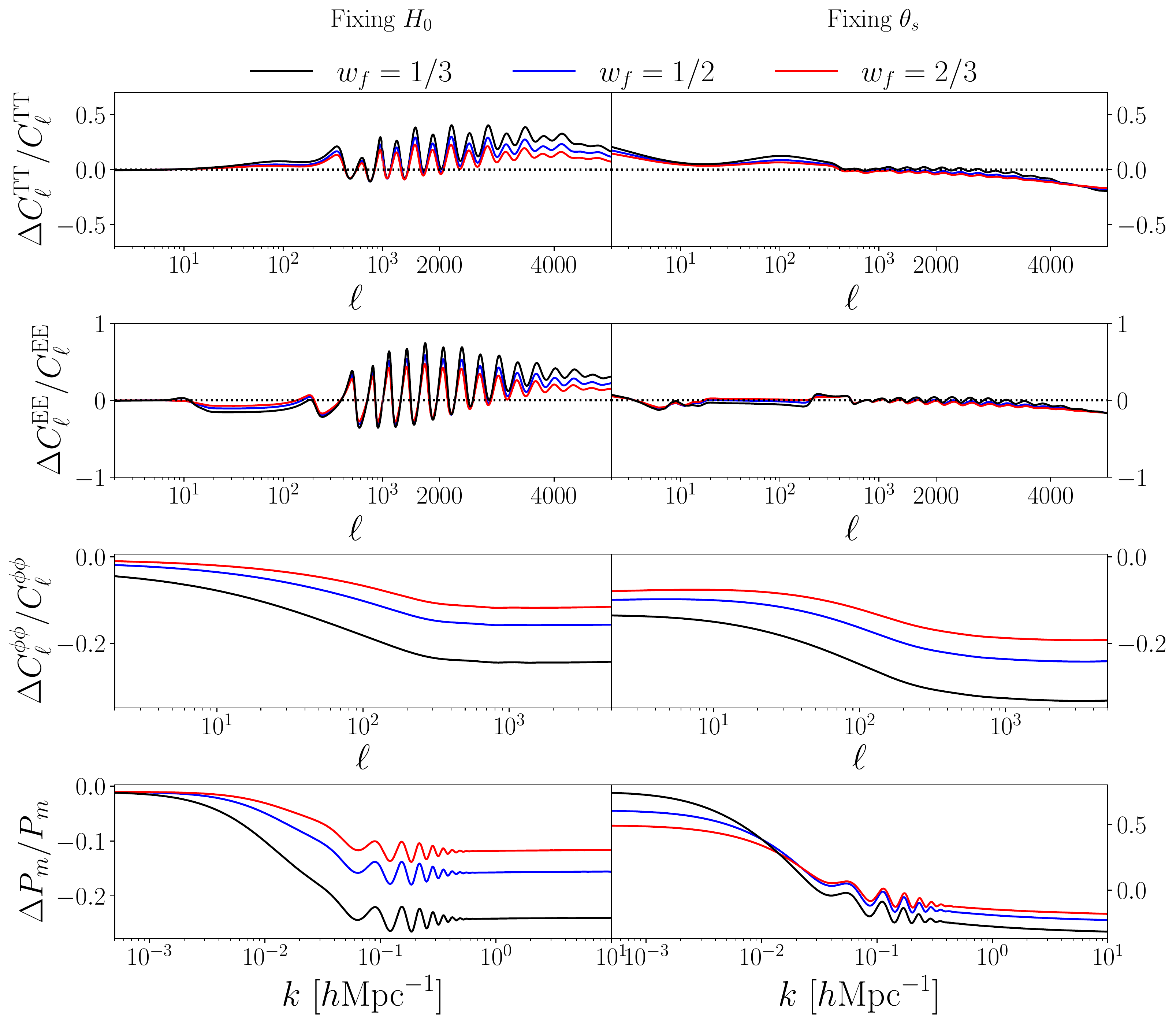}
    \caption{Same as Fig.~\ref{fig:fede_power_spectra} but now changing $w_f$ at fixed $f_{\rm EDE}=0.1$ and  $\log_{10}(z_c)=3.5$.}
    \label{fig:wf_power_spectra}
\end{figure}

\bibliography{biblio}

\end{document}